\documentclass[a4paper,onecolumn,showpacs,notitlepage,floatfix,nofootinbib,superscriptaddress,prd]{revtex4-2}
\pdfoutput=1
\usepackage[utf8]{inputenc}

\usepackage[bookmarks, linktocpage, colorlinks = true, linkcolor = blue, urlcolor  = blue, citecolor = blue, anchorcolor = green, hyperindex = true, hyperfigures]{hyperref}

\usepackage[T1]{fontenc} 
\usepackage{bm} 

\usepackage{amsthm}

\usepackage{mathrsfs} 
\usepackage{subcaption}
\usepackage{mdframed}
\usepackage{enumitem}
\usepackage{xcolor}
\usepackage{subcaption}
\usepackage{mathtools}
\usepackage{tikz-cd}
\usepackage[normalem]{ulem}
\usepackage{amsmath}
\usepackage{amssymb}
\usepackage{amsfonts}

\usepackage{latexsym}

\usepackage{subcaption}
\usepackage{mdframed}
\usepackage{enumitem}
\usepackage{subcaption}
\usepackage{mathtools}
\usepackage{tikz-cd}
\usepackage{pict2e}
\newmdenv[skipabove=6mm]{kotak}   

\DeclareFontEncoding{LS1}{}{}
\DeclareFontSubstitution{LS1}{stix}{m}{n}
\DeclareSymbolFont{symbols2}{LS1}{stixfrak}{m}{n}
\DeclareMathSymbol{\lcangle}{\mathopen}{symbols2}{"E9}
\DeclareMathSymbol{\rcangle}{\mathclose}{symbols2}{"EA}

\makeatletter
\newsavebox{\@brx}
\newcommand{\llcangle}[1][]{\savebox{\@brx}{\(\m@th{#1\lcangle}\)}%
  \mathopen{\copy\@brx\kern-0.5\wd\@brx\usebox{\@brx}}}
\newcommand{\rrcangle}[1][]{\savebox{\@brx}{\(\m@th{#1\rcangle}\)}%
  \mathclose{\copy\@brx\kern-0.5\wd\@brx\usebox{\@brx}}}
\makeatother

\newcommand{\subbe}{\begin{subequations}}
\newcommand{\subee}{\end{subequations}}
\newcommand{\be}{\begin{eqnarray}}
\newcommand{\ee}{\end{eqnarray}}
\newcommand{\nn}{\nonumber}
\newcommand{\nl}{\nonumber \\}
\newcommand{\pd}{\partial}

\newcommand{\appropto}{\mathrel{\vcenter{
  \offinterlineskip\halign{\hfil$##$\cr
    \propto\cr\noalign{\kern2pt}\sim\cr\noalign{\kern-2pt}}}}}

\newtheorem{proposition}{Proposition}

\newtheorem*{fact*}{Fact}

\usepackage{mdframed}
\usepackage{lipsum}

\allowdisplaybreaks

\begin{document}

\title{Exact Sum Rules and Zeta Generating Formulas \\
  from the ODE/IM correspondence
}

\author{Syo Kamata}
\email{skamata11phys@gmail.com}
\affiliation{Department of Physics, The University of Tokyo, 7-3-1 Hongo, Bunkyo-ku, Tokyo 113-0033, Japan}

\begin{abstract}
We develop a spectral-zeta framework for quantum mechanics with the ${\cal PT}$-symmetric potential $V_{{\cal PT}}(x)=x^{2K}(ix)^{\varepsilon}$ $(K,\varepsilon \in {\mathbb N})$ and the Hermitian potential $V_{{\cal H}}(x)=x^{2M}$ $(M \in {\mathbb N}+1)$, based on the fusion relations of the $A_{2M-1}$ T-system.
Using the ODE/IM correspondence, we construct exact sum rules (ESRs) and zeta generating formulas (ZGFs) for the spectral zeta functions (SZFs) $\zeta_n(s)$. In contrast to recursive T-Q relations, the ZGFs provide fixed-source, closed-form mappings between different fusion sectors. For Hermitian $M=2$, our ESRs reproduce exact WKB results, extending them systematically to ${\cal PT}$ sectors and (half-)integer $M$.
Our analysis reveals a phenomenon of \textit{algebraic information loss}, distinct from analytic ambiguity. The structure is governed by a selection rule ${\cal S}_n$, derived from the Chebyshev structure of fusion relations and $\mathbb{Z}_{2M+2}$ Symanzik symmetry. For odd integer $M$, we identify a structural non-invertibility: mapping from \textit{odd} to \textit{even} fusion sectors causes exact coefficient cancellation due to phase interference, rendering the map non-invertible. This implies even-sector data carry strictly less information than odd-sector data, yielding a \textit{no-go} statement for inverse spectral reconstruction. Conversely, for even and half-integer $M$, all relevant sectors form an information-equivalent, mutually invertible family.
Finally, we provide a spectral-zeta formulation of the massless Ai-Bender-Sarkar (ABS) conjecture. By connecting ${\cal PT}$ and Hermitian spectra via ZGFs, we establish a purely spectral-theoretic route to the conjectured relation, avoiding explicit analytic continuation.
\end{abstract}
\maketitle
\tableofcontents
\flushbottom

\section{Introduction}

Non-Hermitian quantum theory has garnered significant interest from both phenomenological and mathematical perspectives. 
A particularly notable subclass is ${\cal PT}$-symmetric quantum mechanics (QM)~\cite{Bender:1998gh}, which is characterized by pseudo-Hermiticity~\cite{Mostafazadeh:2001jk,Mostafazadeh:2008pw}. 
Despite being non-Hermitian, ${\cal PT}$-symmetric QM exhibits several properties traditionally associated with Hermitian systems, such as real and bounded energy spectra, a well-defined ${\cal ``CPT"}$ inner product, and formal duality to Hermitian QMs~\cite{Bender:1998ke,Dorey:2001uw,Shin:2002vu,Mostafazadeh:2002pd,Weigert:2003py,Mostafazadeh:2003gz,Bender:2004zz,Jones:2006qs,Bender:2006wt}.
(See also Refs.~\cite{Bender_2005,Bender:2019,Bender:2023cem} for reviews.)
These features have motivated further exploration of ${\cal PT}$ symmetry in quantum field theories, particularly in contexts beyond the Standard Model~\cite{Bender:2018pbv,Felski:2021evi,Bender:2021fxa,Mavromatos:2021hpe,Grunwald:2022kts,Lawrence:2023woz,Romatschke:2024cld,Mavromatos:2024ozk,Chen:2024ynx,Chen:2024bya}. 
Importantly, the physical realization of ${\cal PT}$-symmetric systems has been established in various experimental platforms, especially in optics and photonics~\cite{ruter2010observation,feng2014single,hodaei2014parity,el2018non,ozdemir2019parity,miri2019exceptional,ashida2020non}.
From a mathematical perspective, ${\cal PT}$-symmetric theories exhibit rich non-perturbative structures that differ markedly from conventional resonance physics, and these topics have been studied using Borel resummation, resurgence theory, and exact quantization conditions (QCs)~\cite{Jentschura:2009jd,taya2021exact,Ai:2022csx,Kamata:2023opn,Kamata:2024tyb,Morikawa:2025grx,Morikawa:2025xjq}.
Classical results of Borel resummation, such as how Stokes data and Voros symbols enter exact quantization and Borel summability, have been studied in Refs.~\cite{Voros1983,DP1,AKT1,Takei2}, and applications to simple Hermitian QMs have been performed in Refs.\cite{Sueishi:2020rug,Sueishi:2021xti,Kamata:2021jrs,Grassi:2018bci,Bucciotti:2023trp,suzuki2023exact,Ture:2024nbi,Misumi:2024gtf}, for example.


A remarkable connection between certain classes of ${\cal PT}$-symmetric QMs and integrable models is known as the ordinary differential equation/integrable model (ODE/IM) correspondence~\cite{Dorey:1998pt,Dorey:1999uk,Dorey:2000ma}. 
A central insight of this correspondence is that the Stokes multipliers, connection coefficients between asymptotic solutions of Schr\"{o}dinger-type equations, satisfy functional relations such as the T-Q and T-system equations, which are hallmark structures in integrable models.
This correspondence was further extended to broader contexts, including quantum KdV theory and conformal field theory (CFT), through the representation theory of the Virasoro algebra and quantum affine algebras~\cite{Bazhanov:1996dr,Bazhanov:1998dq,Bazhanov:2001xm,Dorey:2007ti,Lukyanov:2010rn,Dorey:2012bx,Lukyanov:2013wra,Ito:2015nla,Ito:2016qzt,Ito:2017ypt}. 
In these settings, the spectrum of local integrals of motion in integrable models corresponds to the spectral data of differential operators arising in ${\cal PT}$-symmetric or related quantum mechanical problems.
At the heart of the integrable structure are the fusion hierarchies of transfer matrices, which stem from the representation theory of quantum affine algebras such as $U_q(\widehat{\mathfrak{sl}}_2)$, and give rise to the T- and Y-systems that govern the dynamics of integrable models~\cite{reshetikhin1987spectrum,Reshetikhin:1987,Klumper:1992,Kuniba:1993,Bazhanov:1994ft,Krichever:1997,Kuniba:2010ir}. 
These functional relations re-emerge in the ODE/IM correspondence~\cite{Bazhanov:1994ft}, where they constrain the analytic and algebraic structure of quantum spectral problems via Stokes data and Wronskian-type identities.
(See also Ref.~\cite{Dorey:2007zx} and references therein.)
More recently, exact quantization conditions (QCs) and their resurgent structures, rooted in integrability, have been investigated in Refs.~\cite{Ito:2018eon,Emery:2019znd,imaizumi2020exact,Emery:2020qqu,Gabai:2021dfi,ito2023exact,Ito:2024nlt,Degano:2025mug,ito2025ode}, highlighting the deep interplay between spectral theory, resurgence, and functional relations in QMs.

While the ODE/IM correspondence has been largely explored from the algebraic
structures of integrable models, particularly via the T-/Y-systems, the direct
description of global spectral data has mostly remained implicit in recursive
functional relations. This raises the question of how to encode, in a more
global and quantitative way, how spectral information is distributed across
sectors and constrained by the fusion hierarchy. One natural language for such
questions is provided by spectral zeta functions (SZFs) and their relations.
In this direction, Ref.~\cite{Voros1983} initiated the study of algebraic relations among
SZFs for Hermitian quantum systems, particularly for the
quartic potential, by using exact WKB analysis. In the case of the quartic
oscillator, the QC satisfies the identity
\begin{equation}
\sum_{k=0}^{2} \mathfrak{D}(e^{\frac{2\pi i}{3}k}E) - \prod_{k=0}^{2} \mathfrak{D}(e^{\frac{2\pi i}{3}k}E) + 2 = 0, \nonumber
\end{equation}
which translates to identities among the SZFs
$\zeta(s)=\sum_{\alpha\in\mathbb{N}_0}E_\alpha^{-s}$~\cite{Voros:1986vw,voros1992spectral},
known as \textit{exact sum rules} (ESRs)~\cite{Voros:2012se,Watkins:2011gx,Voros:2022yos}:
\begin{subequations}
  \begin{align}
& \frac{1}{2} \zeta(1)^3 -\frac{3}{2} \zeta(1)\zeta(2) - 3\zeta(3) = 0, \nl
& \frac{1}{240} \zeta(1)^6 - \frac{1}{16} \zeta(1)^4 \zeta(2) + \frac{1}{6} \zeta(1)^3 \zeta(3) + \frac{3}{16} \zeta(1)^2 \zeta(2)^2 - \frac{3}{8} \zeta(1)^2 \zeta(4) \nonumber \\
& \quad - \frac{1}{2} \zeta(1) \zeta(2) \zeta(3) + \frac{3}{5} \zeta(1) \zeta(5) - \frac{1}{16} \zeta(2)^3 - \frac{11}{6} \zeta(3)^2 + \frac{3}{2} \zeta(6) = 0, \nonumber \\
& \vdots. \nn
\end{align} 
\end{subequations}
The present work extends this line of thought within the ODE/IM setting and
systematises it for the $A_{2M-1}$ fusion hierarchy and its
${\cal PT}$-symmetric and Hermitian realisations.
\\ \, \\ \indent
In this paper we study ${\cal PT}$-symmetric quantum mechanics with
$V_{{\cal PT}}(x)=x^{2K}(ix)^{\varepsilon}$ for $K,\varepsilon\in\mathbb{N}$,
together with the Hermitian potential $V_{{\cal H}}(x)=x^{2M}$ for
$M\in\mathbb{N}+1$. Within the ODE/IM correspondence, these problems are
associated with the $A_{2M-1}$ T-system, where $M= K+\varepsilon/2$.
We exploit the fusion relations of this T-system to formulate identities
for SZFs of the corresponding one-dimensional
Schr\"odinger operators. Our aim is to reinterpret the fusion hierarchy
directly in terms of spectral data. Concretely, we present
\begin{itemize}
  \item[(I)] Exact sum rules (ESRs): algebraic relations among SZFs
  at integer arguments within a fixed fusion label $n$;

  \item[(II)] Zeta generating formulas (ZGFs): explicit polynomial
  identities among SZFs that map between different fusion labels,
  expressing the SZFs in any sector in terms of those in a fixed
  \textit{source} sector.
\end{itemize}
Taken together, these constructions provide an explicit and fairly
algorithmic \emph{ODE-side spectral–zeta interpretation} of the
$A_{2M-1}$ fusion hierarchy: they make clear how fusion structures,
together with selection rules, constrain the SZFs and organize the flow
of spectral information between sectors. Unlike the Schr\"{o}dinger
equation, which determines individual eigenvalues $E_\alpha$ sector by
sector, the ZGFs act on the global spectral data of each sector. In this
way our framework complements level-by-level analyses by showing how the
zeta data of one sector is entangled with, and in favourable cases
determines, that of another.
In particular, for odd integer $M$ we find that the ZGFs mapping from odd to even fusion sectors possess a non-trivial algebraic kernel, so that even-sector zeta data cannot reconstruct all integer zeta values in the odd sectors, whereas for even and half-integer $M$ all relevant sectors form an information-equivalent family with mutually invertible ZGFs. We refer to this odd/even pattern as an \emph{algebraic nullification}, or
equivalently an \emph{algebraic information-loss}, phenomenon at the level of SZFs, in order to emphasise that it is generated by exact cancellations in the fusion coefficients rather than by any analytic input. It is, in
particular, distinct from the familiar \textit{analytic ambiguities} of the ODE/IM correspondence, which stem from the choice of boundary conditions and analytic continuation~\cite{Bazhanov:1996dr,Dorey:1998pt,Dorey:2007zx,Kuniba:2010ir}. 
Within each fusion sector the selection rules
${\cal S}_n$ then classify, at fixed fusion label, which integer zeta
moments are algebraically visible or invisible through the fusion
hierarchy, so that the odd/even pattern can be viewed as a local-to-local hierarchy among sectorwise SZF data.
Furthermore, as a concrete application, we show that the same framework
provides a direct spectral–zeta formulation of the massless
Ai–Bender–Sarkar (ABS) conjecture~\cite{Ai:2022csx}: the
${\cal PT}$-symmetric and Hermitian problems are related through Laplace transforms and ZGFs of their SZFs, reformulating the conjectured relation without recourse to explicit analytic continuation.
\\ \, \\ \indent
This paper is structured as follows:
In Sec.~\ref{sec:preparation}, we review the background including our setup, fusion hierarchy, and SZFs.
We also outline the strategy for constructing ESRs and ZGFs.
Sec.~\ref{sec:simple_cases} presents explicit constructions for $M=2,3$, serving as instructive examples.
In Sec.~\ref{sec:general_cases}, we generalize the construction to arbitrary $M \in \frac{1}{2} \mathbb{N} + 1$.
In Sec. \ref{sec:ABS}, we apply our framework to the massless ABS conjecture, demonstrating how ZGFs connect ${\cal PT}$-symmetric and Hermitian spectral data.
Sec.~\ref{sec:summary} concludes with a summary and outlook.
Appendix~\ref{app:review} provides reviews of the derivation of the fusion relations and of the SZFs.


\section{Setup and strategy} \label{sec:preparation}
In this section, we describe our setup and strategy.
In Sec.~\ref{sec:PT_QM}, we introduce the setup of ${\cal PT}$-symmetric QM.
Sec.~\ref{sec:strategy} explains our overall strategy for constructing the ESRs and ZGFs.

\subsection{Setup: The ${\cal PT}$-symmetric QMs and fusion relations} \label{sec:PT_QM}
In this part, we explain our setup, the ${\cal PT}$-symmetric QMs as well as the Hermitian QM.
The ${\cal PT}$-symmetric and Hermitian Schr\"{o}dinger operators, $\widehat{\cal L}_{{\cal PT},{\cal H}}$, are defined as
\be
&& \widehat{\cal L}_{\cal PT}(x,E) :=  - \pd_x^2 + x^{2K}  (i x)^\varepsilon  - E, \qquad K, \varepsilon \in {\mathbb N}, \ E \in {\mathbb R}, 
\label{eq:L_PT} \\
&& \widehat{\cal L}_{\cal H}(x,E) :=  - \pd_x^2 + x^{2 M} - E, \qquad \qquad \ \, M \in {\mathbb N} + 1, \ E \in {\mathbb R}. 
\label{eq:L_Herm}
\ee
The ${\cal PT}$-symmetric and Hermitian domains (or boundary conditions) of the wavefunction, which we denote by $x_{{\cal PT},{\cal H}}$, are given by
\be
&& x_{\cal PT} \in e^{+i \theta(K,\varepsilon)} ( -\infty,0] \cup e^{-i \theta(K,\varepsilon)} [0,+\infty), \qquad \theta(K,\varepsilon) := \frac{\pi \varepsilon}{4 K + 2\varepsilon + 4}, \label{eq:dom_xPT} \\
&& x_{\cal H} \in (-\infty,+\infty), \label{eq:dom_xH}
\ee
and, in the ${\cal PT}$-symmetric operator, $\widehat{\cal L}_{\cal PT}(x,E)$, $\varepsilon$ has a role of a deformation parameter for the domain of wavefunction.
One can see that $\widehat{\cal L}_{\cal PT}(x,E)$ is invariant under the ${\cal PT}$ transform defined as
\be
{\cal P} : (x,i) \mapsto (-x,i), \qquad {\cal T}: (x,i) \mapsto (\bar{x},-i), \label{eq:P_T_trans}
\ee
where $\bar{x}$ is the complex conjugate of $x$.
The energy spectra of the ${\cal PT}$-symmetric QMs in Eq.\eqref{eq:L_PT} are real and bounded~\cite{Bender:1998ke,Dorey:2001uw}.
In the ODE/IM correspondence,  it is convenient to analyze a Hermitian-type Schrödinger operator by introducing a phase angle, $\theta^*$, to the coordinate and the energy in the ${\cal PT}$ symmetric operator, which is obtained as
\be
\widehat{\cal L}(x,E) := e^{2i \theta^*} \widehat{\cal L}_{\cal PT}(e^{i \theta^*} x,e^{-2 i \theta^*} E) = -\pd_x^2  + x^{2 M} - E, \qquad M:= K + \frac{\varepsilon}{2} \in \frac{1}{2} {\mathbb N} + 1, \label{eq:LPTtoLH}
\ee
where $\theta^*$ is given by
\be
\theta^* =
\begin{dcases}
  - \frac{\pi}{2} & \text{for \ $K \in 2{\mathbb N}_0 + 1$} \\
  - \frac{\pi}{2} + \mu
  & \text{for \ $K \in 2{\mathbb N}$} 
\end{dcases} \quad  \text{with} \quad \mu := \frac{\pi}{2M + 2}. \label{eq:theta_star}
\ee
Notice that, in Eq.\eqref{eq:LPTtoLH}, taking $\varepsilon = 0$ corresponds to the Hermitian QM for $M \in {\mathbb N}+1$.
By this phase, $\theta^*$, the ${\cal PT}$-symmetric domains are also modified from Eq.\eqref{eq:dom_xPT} as
\be
&& e^{i \theta^*} x_{\cal PT} \in e^{+i \theta(K,\varepsilon)} ( -\infty,0] \cup e^{-i \theta(K,\varepsilon)} [0,+\infty), \label{eq:dom_xPT_redef}
\ee
and one can simultaneously deal with the ${\cal PT}$-symmetric and Hermitian QMs using $\widehat{\cal L}(x,E)$ by identification of the ${\cal PT}$-symmetric and Hermitian energies $E_{{\cal PT},{\cal H}}$ as
\be
E =
\begin{dcases}
  e^{2i \theta^*} E_{\cal PT} & \\
  E_{\cal H} & \ \ \text{when} \ \ M  \in {\mathbb N + 1}
\end{dcases}. \label{eq:E_PT_H}
\ee

\begin{figure}[t]
\centering
\includegraphics[clip,width=0.75\textwidth]{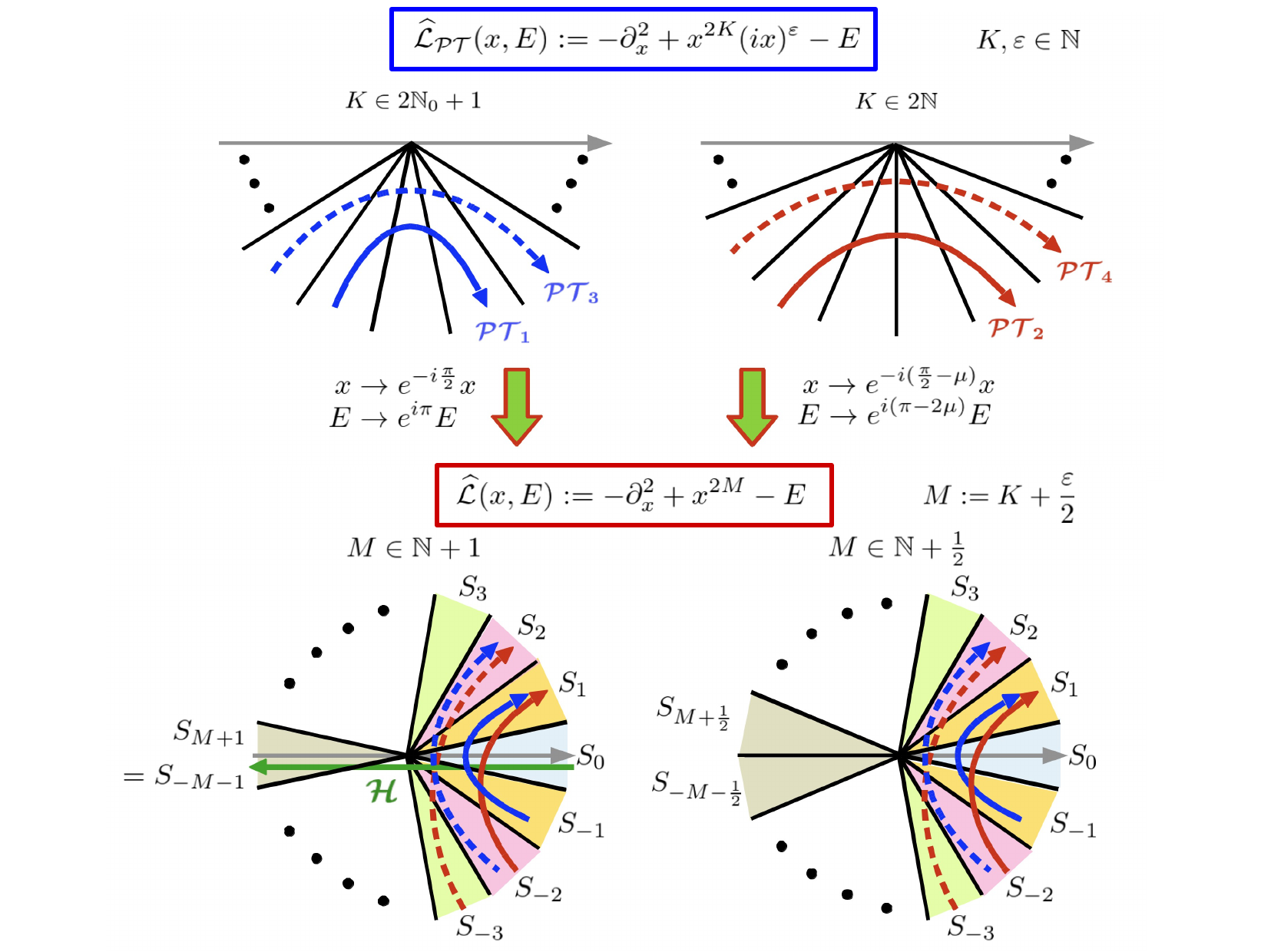}
\caption{The structure of the complex $x$-plane of QMs defined by the ${\cal PT}$-symmetric and the Hermitian-type operators, $\widehat{\cal L}_{\cal PT}(x,E)$ and $\widehat{\cal L}(x,E)$.
  The gray arrow and the black lines denote the real axis and the anti-Stokes lines, respectively.
  The ${\cal PT}$-symmetric QCs are given by analytic continuations represented by the blue and red arrows.
The green arrow denotes the path for the Hermitian QC.}
\label{fig:stokes_sec}
\end{figure}

Then, we split the complex $x$-plane into subdomains $S_k$, called as Stokes sectors, by asymptotic behaviors of the wavefunctions around $|x| = \infty$. 
In the Hermitian-type operator \eqref{eq:LPTtoLH}, those are given by
\be
&& S_k = \{ x \in {\mathbb C} \, | \, \left| \arg(x) -  2 \mu k \right| < \mu \}, \\
&&  2 M + 2 + k \sim  k \in {\mathbb Z} \ \ \Rightarrow \ \   k \in {\mathbb Z}_{2M + 2}. \nn
\ee
Asymptotic behavior of the wavefunctions can be easily seen by the WKB solutions, i.e., $\psi(x) \sim c_\pm x^{-M/2}e^{\pm \frac{x^{M+1}}{M+1}}$ as $|x| \rightarrow \infty$.
If the wavefunction, $\psi(x)$, is convergent as $x \rightarrow +\infty$ on $S_0$, then, $\psi(e^{\pm 2 \mu i}  x)$ on $S_{\pm 1}$ is divergent by taking $x \rightarrow +\infty$.
The boundaries of the Stokes sectors are called as anti-Stokes lines, where the wavefunctions in the leading order asymptotically behave as pure oscillation\footnote{
In mathematical contexts, these are commonly called as Stokes lines.
}.
Our potential in Eq.\eqref{eq:LPTtoLH}, $V(x) = x^{2 M}$, does not have any singular objects, such as poles and branch-cuts, and thus, the Stokes sectors are $(2M + 2)$-periodic, i.e., $S_{k} = S_{k+2M+2}$, without any singular effects.
The QCs of the ${\cal PT}$-symmetric and Hermitian QMs are obtained by analytic continuations starting from and going into $|x| = \infty$ on Stokes sectors as follows:
\be
&& {\cal PT}_{K \in \{ 1, \cdots, \lfloor M - \frac{1}{2} \rfloor \}} : S_{- \lfloor \frac{K+2}{2} \rfloor} \ \rightarrow \ S_{\lfloor \frac{K+1}{2} \rfloor}, \qquad \text{Hermitian} : S_{0} \ \rightarrow \ S_{M+1} \ \ \text{when \ $M \in {\mathbb N} + 1$}, \label{eq:PT_Stokes_Kod} 
\ee
where $\lfloor x \rfloor$ is the floor function.
Fig.~\ref{fig:stokes_sec} shows the transform from $\widehat{\cal L}_{\cal PT}(x,E)$ to $\widehat{\cal L}(x,E)$ in Eq.\eqref{eq:LPTtoLH} and the Stokes sectors.
According to appendix~\ref{sec:ODE_IM}, the quantization conditions (QC) have the fusion relations known as the $A_{2M-1}$ T-system:
\subbe
\be
\text{(i)} &:& C^{(1)}(E)  C^{(n)}(\omega^{n+1} E) = C^{(n-1)}(\omega^{n+2} E) + C^{(n+1)}(\omega^{n} E), \label{eq:fusion_1} \\
\text{(ii)} &:& C^{(n)}(\omega^{-1} E)  C^{(n)}(\omega E) = 1 + C^{(n-1)}(E) C^{(n+1)}(E), \label{eq:fusion_2} \\
\text{(iii)} &:&  C^{(n)}(E) = C^{(2M - n)}(E),\label{eq:Cn_C2Mn}
\ee
\subee
with the initial conditions, 
\be
C^{(-1)}(E) = 0, \qquad C^{(0)}(E) = 1,
\ee
where $C^{(n)}(E)$ are identified as the ${\cal PT}_K$ and Hermitian QCs as
\be
{\cal PT}_{K< \lfloor M + \frac{1}{2}\rfloor} &:&  C^{(K)}(\omega^{(K+1) \bmod 2} E) \mid_{E = e^{2 i \theta^*} E_{\cal PT}} = C^{(K)}(- E_{\cal PT}) = 0, \label{eq:QC_PT} \\
 \text{Hermitian} &:&  C^{(M)}(- E_{\cal H}) = 0. \label{eq:QC_H}
\ee
In the zeta language, we will denote by $\zeta_K(s)$ the SZFs associated with the QC, $C^{(K)}(E)$,
and by $\zeta_M(s)$ those associated with the Hermitian QC, $C^{(M)}(E)$.

In this paper, we mainly use the pair, $(M,K)$, instead of $(K,\varepsilon)$ for identification of the potential and the paths of analytic continuation.
Including the Hermitian QM, the number of the paths to give independent energy spectra is totally $\lfloor M \rfloor$ for $M  = K + \frac{\varepsilon}{2} \in \frac{1}{2} {\mathbb N} + 1$. 
In the below, as fixing $M$, we call the ${\cal PT}$-symmetric (resp. Hermitian) QM defined by the domain labeled by $K$ or the corresponding path of analytic continuation as \textit{${\cal PT}_K$ (resp. $\cal H$) sector}.
In this convention, when $M \in {\mathbb N} + 1$, taking $K=M$ indicates the Hermitian QM, i.e., symbolically ${\cal PT}_M = {\cal H}$.

The QCs \eqref{eq:QC_PT}\eqref{eq:QC_H} have zeros on the negative real axis.
For convenient notations, we redefine the QC as $C^{(n)}(-E) \rightarrow C^{(n)}(E)$ through this paper, which is irrelevant to the fusion relations \eqref{eq:fusion_1}\eqref{eq:fusion_2}.
In addition, we introduce a symbolic notation for a product of $C^{(n)}(\omega^{k} E)$ as 
\be
&& [k_1,\cdots,k_\ell]_{n \in {\mathbb N}} := \prod_{j=1}^{\ell} C^{(n)}(\omega^{k_j} E), \qquad [k_1,\cdots]_{-1} = 0, \qquad [k_1,\cdots]_{0} = 1, \label{eq:simp_1} \\ 
&& k_\ell + 2 M + 2 \sim k_{\ell} \in {\mathbb Z} \ \ \Rightarrow \ \ k_\ell \in
{\cal M} \, \cong \, {\mathbb Z}_{2 M+2},\nn
\ee
where ${\cal M}$ is a set defined as
\be
   {\cal M} := \{ - \lfloor M + 1 \rfloor, - \lfloor M + 1 \rfloor + 1, \cdots, \lfloor M + \tfrac{1}{2} \rfloor - 1,  \lfloor M + \tfrac{1}{2} \rfloor \}, \qquad (M \in \tfrac{1}{2} {\mathbb N} + 1) \label{eq:def_set_M}
\ee
with the floor function, $\lfloor x \rfloor$.
Clearly, $[k_1]_n[k_2]_n = [k_1,k_2]_n = [k_2,k_1]_n$.
Using this notation,
Eqs.\eqref{eq:fusion_1}-\eqref{eq:Cn_C2Mn} (by replacing $E \rightarrow \omega^k E$) are expressed as
\subbe
\be
&& [k]_1 [k+n+1]_n  = [k+n+2]_{n-1} + [k+n]_{n+1}, \label{eq:iden[k]1} \\
&& [k-1, k+1]_n  = 1 + [k]_{n-1} [k]_{n+1}, \label{eq:iden[k]2} \\ 
&& [k]_{n} = [k]_{2M-n}. \label{eq:iden[k]3} 
\ee
\subee
For later discussions, we define a set of $K$ and a subset consisting of its odd (even) numbers as 
\be
   {\cal K} := \{ 1,2, \cdots, \lfloor M \rfloor - 1, \lfloor M \rfloor \}, \qquad {\cal K}_{{\rm od}({\rm ev})} := \{ \text{odd (even) numbers in ${\cal K}$} \}. 
\ee

\subsection{Strategy for formulating the ESRs and the ZGFs} \label{sec:strategy}

For each fusion label $n$, with the associated QC $C_n(E)$ whose zeros are 
$\{E_{n,\alpha}\}_{\alpha\in\mathbb N_0}$, the corresponding SZF
is given by
\be
  \zeta_n(s) := \sum_{\alpha \in {\mathbb N}_0} E_{n,\alpha}^{-s}, \qquad s\in {\mathbb N},
\ee
as recalled in Appendix~\ref{sec:SZF}.
Here, we define a set of the SZFs as
\be
\bm{\zeta}_{K \in {\cal K}} := \{\zeta_K(s) \}_{s \in {\mathbb N}}.
\ee
Our main aim of this paper is to understand the fusion hierarchy in framework of the ODE/IM correspondence from the spectral perspective through $\bm{\zeta}_K$.
For the purpose, we consider the ESRs and the ZGFs defined as follows:
\begin{enumerate}
\item[(I)] \textit{Exact sum rule (ESR)}:
  For fixed $M \in \frac{1}{2}{\mathbb N}+1$ and $K \in {\cal K}$, the ESR is defined as an algebraic relation on $\bm{\zeta}_{K}$ and represented by a polynomial of $\{ \zeta_K(s^\prime) \}_{1 \le s^\prime < s}$ as
\be
\zeta_{K}(s) &=& \sum_{\substack{{\bf m} = \{ m_1,\cdots,m_{s-1} \} \in {{\mathbb N}_0}^{s-1} \\ m_1+2 m_2 + \cdots + (s-1) m_{s-1} = s}} c_{{\bf m}} \prod_{t=1}^{s-1} \zeta_{K}(t)^{m_t}, 
\qquad (c_{{\bf m}} \in {\mathbb R}) \label{eq:def_ESR}
\ee
for some $s \in {\mathbb N}$.
Suppose a set of $s$ to generate non-trivial ESRs and denote it by $G \subseteq {\mathbb N}_0$ by adding $\{ 0 \}$.
We define \textit{selection rule} of the ESRs, which is given by the quotient set as
\be
   {\cal S}_K := {\mathbb N}_0/G, \qquad G = \{ s \in {\mathbb N} \, | \, \text{non-trivial ESR for $\zeta_K(s) $} \} \cup \{ 0 \}. \label{eq:selec_def}
\ee
Here, the equivalence relation for the quotient is defined as: for $ n, n^\prime \in \mathbb{N}_0 $, we write $ n \sim n^\prime $ iff $ n - n^\prime \in G$.

\item[(II)] \textit{Zeta generating formula (ZGF)}:
  For fixed $M \in \frac{1}{2}{\mathbb N}+1$, the ZGF is defined as a mapping from $\bm{\zeta}_{K^\prime \in {\cal K}}$ to $\bm{\zeta}_{K \in {\cal K}}$ and represented by a polynomial of $\{ \zeta_{K^\prime}(s^\prime) \}_{1 \le s^\prime \le s}$ as  
\be
\zeta_{K}(s) &=& \sum_{\substack{{\bf m} = \{ m_1,\cdots,m_s \} \in {{\mathbb N}_0}^s \\ m_1+2 m_2 + \cdots + s m_s = s}} c_{{\bf m}} \prod_{t=1}^s \zeta_{K^\prime}(t)^{m_t}, 
\qquad (c_{\bf m} \in {\mathbb R}) \label{eq:def_ZGF}
\ee
for all $s \in {\mathbb N}$.
We denote the ZGF by $\bm{\zeta}_K = \bm{\zeta}_K(\bm{\zeta}_{K^\prime})$.
For a given $\bm{\zeta}_K = \bm{\zeta}_K(\bm{\zeta}_{K^\prime})$, we say $\bm{\zeta}_K = \bm{\zeta}_K(\bm{\zeta}_{K^\prime})$ is invertible if $\bm{\zeta}_{K^\prime} = \bm{\zeta}_{K^\prime} (\bm{\zeta}_{K})$ is constructable.
\end{enumerate}
Roughly speaking, while the ESRs give the dependence of $\zeta_K(s)$ in the same ${\cal PT}_K$ sector, the ZGFs generate $\bm{\zeta}_K$ in a ${\cal PT}_K$ sector from $\bm{\zeta}_{K^\prime}$ in another ${\cal PT}_{K^\prime}$ sector.
It is notable that, different from the fusion relations, the ZGF is not a recurrence relation found from $K=1$ to higher but a direct mapping between $\bm{\zeta}_K$ and $\bm{\zeta}_{K^\prime \ne K}$\footnote{
In our analysis, instead of $K$ ($K^\prime$), we mainly use $n$ $(n^\prime)$ to specify ${\cal PT}_K$ sectors (or fusion labels) in order to respect the notation in Eqs.\eqref{eq:fusion_1}\eqref{eq:fusion_2}.
}.

We make some remarks on the selection rule.
As we will see in Secs.~\ref{sec:simple_cases} and \ref{sec:general_cases}, adding the identity to $G$, i.e., $s=0$, in the selection rule can be interpreted as finding an algebraic equation in terms of $e^{-\zeta_K^\prime(0)}$.
In addition, looking to the quotient set, rather than $G$ itself, is helpful for comparison with the ${\mathbb Z}_{2M+2}$ Symanzik rotation\footnote{
More precisely, this selection rule is defined to be convenient for comparison with the symmetric structure of a closed form of the QCs, from which the ESRs are derived, as in Eqs.\eqref{eq:C2_iden_PT_Herm_C}\eqref{eq:C2_iden_PT_Herm_C_2}.
}. 
The selection rule ${\cal S}_K$ equips ${\mathbb Z}_N$ group structure if $G = N {\mathbb N}_0$ with $N \in {\mathbb N}$, i.e., ${\mathbb N}_0/(N{\mathbb N}_0) \cong {\mathbb Z}_N$, but it does not always become a group in general.
\\ \, \\ \indent
In order to construct the ESRs and the ZGFs we proceed as follows:
First, for each fusion label $n$ we use the fusion relations to express
$C^{(n)}(E)$ as a function of the rotated fundamental QCs, 
$\{ C^{(1)}(\omega^{k} E)\}_{k \in {\mathbb Z}_{2M + 2}}$.
We then expand $C^{(n)}(E)$ around $E=0$ using
Eq.~\eqref{eq:D_to_zeta} to rewrite the result in terms of the spectral
zeta functions.
Next, we treat separately the cases $n = 2M$ and
$n \in {\cal K} \setminus \{ 1 \}$.
The former yields ESRs written in terms of the PT$_1$ data
$\boldsymbol{\zeta}_1$, while the latter produces ZGFs mapping the
${\cal PT}_1$ sector to the ${\cal PT}_n$ sectors,
\[
  \boldsymbol{\zeta}_{n} = \boldsymbol{\zeta}_n(\boldsymbol{\zeta}_1),
  \qquad n\in {\cal K} \setminus \{1\}.
\]
Whenever a ZGF can be inverted to give an expression of the form
\[
  \boldsymbol{\zeta}_1 = \boldsymbol{\zeta}_1(\boldsymbol{\zeta}_{n'}),
\]
we can rewrite both the ESRs and the remaining ZGFs in terms of the
alternative source sector $n'$: substituting
$\boldsymbol{\zeta}_1(\boldsymbol{\zeta}_{n'})$ into the ESRs and
ZGF for $\boldsymbol{\zeta}_1$ yields ESRs in terms of
$\boldsymbol{\zeta}_{n'}$ and ZGFs of the form
$\boldsymbol{\zeta}_n = \boldsymbol{\zeta}_n(\boldsymbol{\zeta}_{n'})$.
The remaining quantities $\zeta_n^{\prime}(0)$ can, in principle, be
extracted by solving the ${\cal O}(E^{0})$ terms of the ESRs and
ZGFs, but in practice it is more convenient to use the Chebyshev
polynomial method described below.

\begin{figure}[t]
\centering
\includegraphics[clip,width=0.75\textwidth]{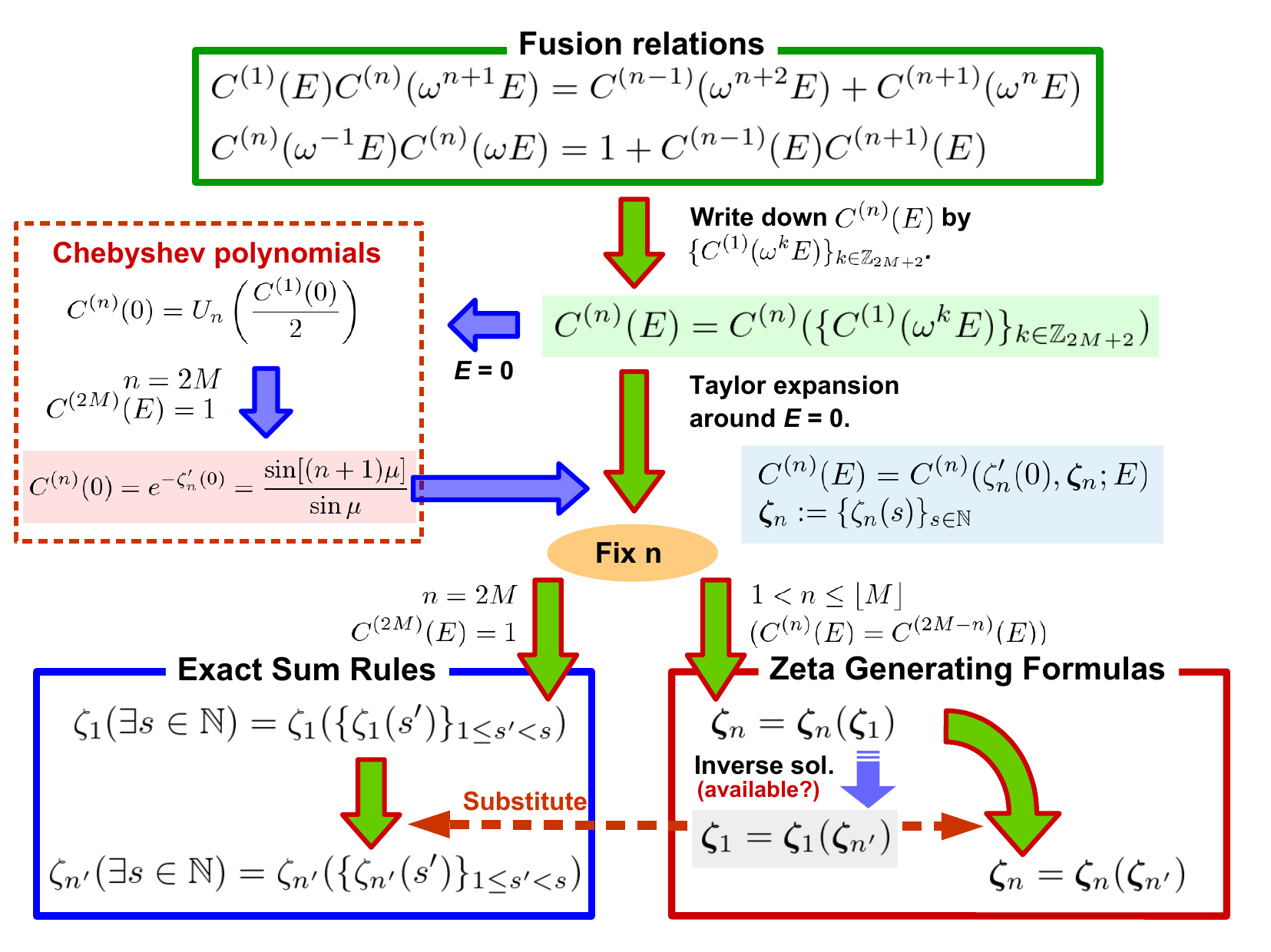}
\caption{The schematic figure of our strategy to formulate the ESRs and the ZGFs by beginning with the fusion relations.
  In this figure, the ESRs and the ZGFs are symbolically denoted by $\zeta_n(s)  = \zeta_n(\{ \zeta_n(s^\prime ) \}_{1 \le s^\prime < s})$ and $\bm{\zeta}_n = \bm{\zeta}_n(\bm{\zeta}_{n^\prime})$.
  The detail is explained in the text.
}
\label{fig:procedure}
\end{figure}

For later reference, we summarize our overall strategy in three steps:
\begin{enumerate}
\item Start from the fusion relations for the QCs, $C^{(n)}(E)$, in the
      $A_{2M-1}$ T-system.
\item Translate these relations into algebraic identities among the
      SZFs via the Taylor expansion around $E=0$.
\item Organize the resulting identities into ESRs (at fixed sector) and
      ZGFs (between sectors), and analyze their selection rules and
      (non-)invertibility.
\end{enumerate}
A key aspect of our strategy is to examine the existence of inverse
mappings of the form
\[
  \boldsymbol{\zeta}_1 = \boldsymbol{\zeta}_1(\boldsymbol{\zeta}_n),
\]
i.e. to what extent the ZGFs can be inverted.
As we will demonstrate in Sec.~\ref{sec:general_cases}, such
invertibility is not guaranteed: it depends in a structural way on the
value of $M$ and on the choice of fusion label $n$.
In particular, for odd $M$ we encounter the algebraic information-loss
phenomenon mentioned in the Introduction, where the inverse mapping
becomes structurally impossible.
The global pattern of which sectors are accessible and mutually
invertible is characterized by the selection rule defined in
Eq.~\eqref{eq:selec_def}.
Our overall procedure is summarized in Fig.~\ref{fig:procedure}.

\section{Construction of the ESRs and the ZGFs: Simple cases} \label{sec:simple_cases}
Before investigating the general cases, we provide demonstrations of formulating the ESRs and the ZGFs in simple cases, $M=2$ and $3$.
When $M$ is small, such as $M = 2,3$, it is convenient to use relations obtained from Eq.\eqref{eq:iden[k]1}\eqref{eq:iden[k]2} by taking $n = M \in {\mathbb N}+1$, which are given by
\subbe
\be
&& [k]_1 [k-M-1]_{M} = [k-M]_{M-1} + [k+M]_{M-1}, \qquad ( M \in {\mathbb N}+1)
\label{eq:iden[k]4} \\ 
&& [k-1, k+1]_{M}  = 1 + [k,k]_{M-1}. \qquad \qquad \qquad \qquad \quad \ \, ( M \in {\mathbb N}+1) \label{eq:iden[k]5}
\ee
\subee
In these cases, thanks to the small $M$, the formulas can be obtained without finding the inverse solutions of $\bm{\zeta}_n = \bm{\zeta}_n (\bm{\zeta}_1)$.
The analysis in this part gives us useful lessons for the generalization discussed in Sec.~\ref{sec:general_cases}.
As we will see in Sec.~\ref{sec:general_cases}, the $M=2$ case already illustrates the structure of ESRs without any obstruction, whereas $M=3$ provides the first example where the information-loss phenomenon and non-invertibility appear.

\subsection{$M=2$} \label{sec:SZF_M2}
We firstly consider the $M=2$ case.
In this case, there exist two sectors, i.e., $K=1$ and $K=2$, and the 1st. and 2nd. sectors correspond to the ${\cal PT}$-symmetric and Hermitian QMs, respectively.
The Hermitian case has been discussed in Ref.~\cite{Voros1983}, and below we reproduce the results in the different derivation.

We firstly derive the ESRs in terms of $\bm{\zeta}_{1,2}$ by using the identity obtained from Eqs.\eqref{eq:simp_1}\eqref{eq:iden[k]3} as
\be
[k]_0 = [k]_{2M} = 1, \label{eq:k0k2M}
\ee
which means taking a path for the analytic continuation going around the origin and imposing the wavefunction to be single-valued for the monodromy.
We write down $[k]_{2M}$ in terms of $[k^\prime]_1$ using the fusion relation \eqref{eq:iden[k]1}, which is given by
\be
&& 0 = \left[ 1 \right]_0 - 1 = \left[ 1 \right]_4 -1 = [-2]_1 \left( [-2,0,2]_1 - [-2]_1 - [0]_1 -[2]_1 \right) \nl
&\Rightarrow \quad& [ -2,0,2 ]_1 - [ -2 ] _1 - [ 0 ]_1 - [ 2 ]_1 = 0. \label{eq:C2_iden_PT}
\ee
Here, we took $k=1$ and $M=2$ in Eq.\eqref{eq:k0k2M} and assumed that $[-2]_1$ is non-zero, which simply excludes the trivial case where the corresponding spectral determinant vanishes identically.

Next, we multiply $[1]_1$ to Eq.\eqref{eq:iden[k]4} with $k=1$, and then, use Eqs.\eqref{eq:iden[k]1}\eqref{eq:iden[k]5} with $n=1$, which leads to
\be
 [1,1]_1 [-2]_2  = [-1,1]_{1} + [1,3]_{1}  &\quad \Rightarrow \quad& ([0,2]_2 - 1) [-2]_2  = ([0]_{2} + 1) + ([2]_{2} + 1) \nl
 &\quad \Rightarrow \quad& [ -2,0,2 ]_2 - [ -2 ]_2 - [ 0 ]_2 - [ 2 ]_2 - 2 = 0. \label{eq:C2_iden_Herm}
\ee
Eqs.\eqref{eq:C2_iden_PT}\eqref{eq:C2_iden_Herm} can be explicitly written as
\subbe
\be
&& \prod_{k=-1}^{1} C^{(1)}(\omega^{2k} E) - \sum_{j=-1}^{1} C^{(1)}(\omega^{2k} E) = 0, \label{eq:C2_iden_PT_Herm_C} \\
&& \prod_{k=-1}^{1} C^{(2)}(\omega^{2k} E) - \sum_{j=-1}^{1} C^{(2)}(\omega^{2k} E) - 2 = 0. \label{eq:C2_iden_PT_Herm_C_2}
\ee
\subee
Remind that, as shown in Eqs.\eqref{eq:QC_PT}\eqref{eq:QC_H}, $C^{(n)}(E) = 0$ with $n=1$ and $2$ are the ${\cal PT}$-symmetric and Hermitian QCs, respectively.
Then, we expand $C^{(1,2)}(\omega^k E)$ in Eqs.\eqref{eq:C2_iden_PT_Herm_C}\eqref{eq:C2_iden_PT_Herm_C_2} around $E=0$ by using Eq.\eqref{eq:D_to_zeta} (and replacing ${\frak D}(E) \rightarrow C^{(n)}(E)$ and $\zeta(s) \rightarrow \zeta_n(s)$).
Here, we use the symbols, $\zeta_{{\cal PT},{\cal H}}(s)$, to explicitly indicate the ${\cal PT}$-symmetric and Hermitian SZFs, respectively.
The $O(E^{0})$ parts of Eqs.\eqref{eq:C2_iden_PT_Herm_C}\eqref{eq:C2_iden_PT_Herm_C_2} correspond to algebraic equations in terms of $C^{(1,2)}(0) = e^{-\zeta^{\prime}_{{\cal PT},{\cal H}}(0)}$, and one can immediately obtain physically reasonable solutions corresponding to the unbroken ${\cal PT}$-symmetry, such that $C^{(n)}(0) \in {\mathbb R}_{>0}$, as
\be
C^{(1)}(0) = \sqrt{3}, \qquad C^{(2)}(0) = 2.  \label{eq:solC0_12}
\ee
In addition, the $O(E^{s \in 3{\mathbb N}})$ parts give the ESRs as 
\subbe
\be
&& \zeta_{{\cal PT}}(1)^3 - 3 \zeta_{{\cal PT}}(1) \zeta_{{\cal PT}}(2) - 4 \zeta_{{\cal PT}}(3) = 0, \label{eq:sym_zeta_PT1} \\
&& \zeta_{\cal PT}(1)^6 - 15 \zeta_{\cal PT}(1)^4 \zeta_{\cal PT}(2) + 40 \zeta_{\cal PT}(1)^3 \zeta_{\cal PT}(3) + 45 \zeta_{\cal PT}(1)^2 \zeta_{\cal PT}(2)^2 - 90 \zeta_{\cal PT}(1)^2 \zeta_{\cal PT}(4) \nl
&& \quad - 120 \zeta_{\cal PT}(1) \zeta_{\cal PT}(2) \zeta_{\cal PT}(3) + 144 \zeta_{\cal PT}(1) \zeta_{\cal PT}(5) - 15 \zeta_{\cal PT}(2)^3 + 90 \zeta_{\cal PT}(2) \zeta_{\cal PT}(4) \nl
&& \quad - 320 \zeta_{\cal PT}(3)^2 + 240 \zeta_{\cal PT}(6) = 0, \\
&& \vdots. \nn 
\ee
\subee
\subbe
\be
&& \zeta_{{\cal H}}(1)^3 - 3 \zeta_{{\cal H}}(1) \zeta_{{\cal H}}(2) - 6 \zeta_{{\cal H}}(3) = 0, \label{eq:sym_zeta_Herm1} \\
&& 
\zeta_{\cal H}(1)^6 - 15 \zeta_{\cal H}(1)^4 \zeta_{\cal H}(2) + 40 \zeta_{\cal H}(1)^3 \zeta_{\cal H}(3) + 45 \zeta_{\cal H}(1)^2 \zeta_{\cal H}(2)^2 - 90 \zeta_{\cal H}(1)^2 \zeta_{\cal H}(4) - 120 \zeta_{\cal H}(1) \zeta_{\cal H}(2) \zeta_{\cal H}(3) \nl
&& \quad + 144 \zeta_{\cal H}(1) \zeta_{\cal H}(5) - 15 \zeta_{\cal H}(2)^3 + 90 \zeta_{\cal H}(2) \zeta_{\cal H}(4) - 440 \zeta_{\cal H}(3)^2 + 360 \zeta_{\cal H}(6) = 0, \label{eq:sym_zeta_Herm2} \\
&& \vdots. \nn
\ee
\subee
and one can find the selection rules as
\be
{\cal S}_{{\cal PT},{\cal H}} =  {\mathbb N}_0/G \cong {\mathbb Z}_3, \qquad G = 3{\mathbb N}_0. \label{eq:sel_rule_M2}
\ee
Notice that these results are consistent with the earlier works~\cite{Voros1983, Romatschke:2024mxr}.


Construction of the ZGFs is more straightforward than the ESRs.
From Eqs.\eqref{eq:iden[k]1}\eqref{eq:iden[k]5} with $n=1$ and $M=2$, one finds 
\subbe
\be
&& [0]_2 =  [-1,1]_1 - 1, \label{eq:C2_iden_PT_Herm_C2_1} \\
&& [-1,1]_2 =  1 + [0,0]_1. \label{eq:C2_iden_PT_Herm_C2_2}
\ee
\subee
Expanding $C^{(1,2)}(E)$ in Eq.\eqref{eq:C2_iden_PT_Herm_C2_1} as Eq.\eqref{eq:D_to_zeta} and using the result of $C^{(1,2)}(0)$ in Eq.\eqref{eq:solC0_12}, one can find the ZGF, $\bm{\zeta}_2 = \bm{\zeta}_2(\bm{\zeta}_1)$, as\footnote{
We thank to Paul Romatschke to mention the relations for $s=1,2,3$.
}
\subbe
\be
\zeta_{\cal H}(1) &=& \frac{3}{2} \zeta_{\cal PT}(1), \label{eq:zetaH_zetaPT_M2_1} \\
\zeta_{\cal H}(2) &=& \frac{3}{4} \zeta_{\cal PT}(1)^2 - \frac{3}{2} \zeta_{\cal PT}(2), \\
\zeta_{\cal H}(3) &=& \frac{3}{4} \zeta_{\cal PT}(1)^3 - \frac{9}{8} \zeta_{\cal PT}(1) \zeta_{\cal PT}(2) - 3 \zeta_{\cal PT}(3), \\
\zeta_{\cal H}(4) &=& \frac{11}{16} \zeta_{\cal PT}(1)^4 - \frac{3}{2} \zeta_{\cal PT}(1)^2 \zeta_{\cal PT}(2) - 2 \zeta_{\cal PT}(1) \zeta_{\cal PT}(3) + \frac{3}{8} \zeta_{\cal PT}(2)^2 - \frac{3}{2} \zeta_{\cal PT}(4), \\
\zeta_{\cal H}(5) &=& \frac{5}{8} \zeta_{\cal PT}(1)^5 - \frac{55}{32} \zeta_{\cal PT}(1)^3 \zeta_{\cal PT}(2) - \frac{5}{2} \zeta_{\cal PT}(1)^2 \zeta_{\cal PT}(3) + \frac{15}{16} \zeta_{\cal PT}(1) \zeta_{\cal PT}(2)^2 \nl
&& - \frac{15}{16} \zeta_{\cal PT}(1) \zeta_{\cal PT}(4) + \frac{5}{4} \zeta_{\cal PT}(2) \zeta_{\cal PT}(3) + \frac{3}{2} \zeta_{\cal PT}(5), \label{eq:zetaH_zetaPT_M2_5} \\
\vdots. && \nn
\ee
\subee
This ZGF is invertible because the coefficient in the last term, $\zeta_{\cal H}(s) = \cdots + c \zeta_{\cal PT}(s)$, is non-zero for any $s \in {\mathbb N}$. 
The inverse ZGF, $\bm{\zeta}_1 = \bm{\zeta}_1 (\bm{\zeta}_2)$, is obtained as
\subbe
\be
\zeta_{\cal PT}(1) &=& \frac{2}{3} \zeta_{\cal H}(1), \\
\zeta_{\cal PT}(2) &=& \frac{2}{9} \zeta_{\cal H}(1)^2 - \frac{2}{3} \zeta_{\cal H}(2), \\
\zeta_{\cal PT}(3) &=& \frac{1}{54} \zeta_{\cal H}(1)^3 + \frac{1}{6} \zeta_{\cal H}(1) \zeta_{\cal H}(2) - \frac{1}{3} \zeta_{\cal H}(3), \\
\zeta_{\cal PT}(4) &=& - \frac{1}{81} \zeta_{\cal H}(1)^4 + \frac{2}{27} \zeta_{\cal H}(1)^2 \zeta_{\cal H}(2) + \frac{8}{27} \zeta_{\cal H}(1) \zeta_{\cal H}(3) + \frac{1}{9} \zeta_{\cal H}(2)^2 - \frac{2}{3} \zeta_{\cal H}(4), \\
\zeta_{\cal PT}(5) &=& \frac{5}{972} \zeta_{\cal H}(1)^5 + \frac{5}{162} \zeta_{\cal H}(1)^3 \zeta_{\cal H}(2) - \frac{5}{81} \zeta_{\cal H}(1)^2 \zeta_{\cal H}(3) - \frac{5}{108} \zeta_{\cal H}(1) \zeta_{\cal H}(2)^2 \nl
&& - \frac{5}{18} \zeta_{\cal H}(1) \zeta_{\cal H}(4) - \frac{5}{27} \zeta_{\cal H}(2) \zeta_{\cal H}(3) + \frac{2}{3} \zeta_{\cal H}(5), \\
\vdots. && \nn
\ee
\subee
\, \\ \indent
We make some comments on the ESRs and the ZGFs for $M=2$.
First, the selection rule in Eq.\eqref{eq:sel_rule_M2} relates to the ${\mathbb Z}_{2M+2}$ Symanzik rotation.
It can be seen from that Eqs.\eqref{eq:C2_iden_PT_Herm_C}\eqref{eq:C2_iden_PT_Herm_C_2} are invariant under the ${\mathbb Z}_3 (\cong {\mathbb Z}_6/\langle 2 \rangle)$ shift symmetry, namely $w^{k} \mapsto w^{k+c}$ (or $[k_1,\cdots,k_\ell]_{1,2} \mapsto [k_1+c,\cdots,k_\ell+c]_{1,2} $) with $c \in \{ 0,2,4\} \cong {\mathbb Z}_{3}$. 
In these equations, the $E$-expansion has the similar role to the ${\mathbb Z}_3$ Fourier expansion, and thus, only the $O(E^{s \in 3{\mathbb N}})$ parts make sense.
In consequence, the selection rules coincide with the shift symmetry of Eqs.\eqref{eq:C2_iden_PT_Herm_C}\eqref{eq:C2_iden_PT_Herm_C_2}.
Second, we used Eq.\eqref{eq:C2_iden_PT_Herm_C2_1} to construct the ZGF, but Eq.\eqref{eq:C2_iden_PT_Herm_C2_2} should give the same ZGF.
However, the resulting forms are slightly different from Eqs.\eqref{eq:zetaH_zetaPT_M2_1}-\eqref{eq:zetaH_zetaPT_M2_5}:
\subbe
\be
\zeta_{\cal H}(1) &=& \frac{3}{2} \zeta_{\cal PT}(1), \label{eq:zetaH_zetaPT_M2_1_2} \\
\zeta_{\cal H}(2) &=& \frac{3}{4} \zeta_{\cal PT}(1)^2 - \frac{3}{2} \zeta_{\cal PT}(2), \\
\zeta_{\cal H}(3) &=& \frac{3}{16} \zeta_{\cal PT}(1)^3 + \frac{9}{16} \zeta_{\cal PT}(1) \zeta_{\cal PT}(2) - \frac{3}{4} \zeta_{\cal PT}(3), \\
\zeta_{\cal H}(4) &=& -\frac{1}{16} \zeta_{\cal PT}(1)^4 + \frac{3}{4} \zeta_{\cal PT}(1)^2 \zeta_{\cal PT}(2) + \zeta_{\cal PT}(1) \zeta_{\cal PT}(3) + \frac{3}{8} \zeta_{\cal PT}(2)^2 - \frac{3}{2} \zeta_{\cal PT}(4), \\
\zeta_{\cal H}(5) &=& \frac{5}{32} \zeta_{\cal PT}(1)^5 + \frac{5}{32} \zeta_{\cal PT}(1)^3 \zeta_{\cal PT}(2) - \frac{5}{8} \zeta_{\cal PT}(1)^2 \zeta_{\cal PT}(3) - \frac{15}{32} \zeta_{\cal PT}(1) \zeta_{\cal PT}(2)^2 \nl
&& - \frac{15}{16} \zeta_{\cal PT}(1) \zeta_{\cal PT}(4) - \frac{5}{8} \zeta_{\cal PT}(2) \zeta_{\cal PT}(3) + \frac{3}{2} \zeta_{\cal PT}(5), \label{eq:zetaH_zetaPT_M2_5_2} \\
\vdots. && \nn
\ee
\subee
The differences come from $\zeta_{{\cal PT},{\cal H}}(s)$ with $s \in 3{\mathbb N}$, and taking into account of the ESRs \eqref{eq:sym_zeta_PT1}-\eqref{eq:sym_zeta_Herm2} yields the same results.

\subsection{$M=3$} \label{sec:SZF_M3}
Next, we consider the $M=3$ case.
The procedure for the ESRs and the ZGFs is the same to the $M=2$ case discussed in Sec.~\ref{sec:SZF_M2}.
Similar to Eqs.\eqref{eq:C2_iden_PT}\eqref{eq:C2_iden_Herm} for $M=2$, one can find a closed form in terms of $[k]_{1,2,3}$.
Hence, we show the results briefly, and then, describe the crucial difference from the $M=2$ case.

The ESRs can be constructed from Eq.\eqref{eq:k0k2M}.
By recursively solving the fusion relation \eqref{eq:iden[k]1}, one can express $[2]_6$ using $[k]_1$ as
\be
&& 0 = [2]_0 - 1 = [2]_{6} - 1  = [-2]_2  \left( [-3,-1,1,3]_1 - [-3,-1]_1 - [-3,3]_1 - [-1,1]_1 - [1,3]_1 + 2 \right) \nl
&\Rightarrow \quad&  [-3,-1,1,3]_1 - [-3,-1]_1 - [-3,3]_1 - [-1,1]_1 - [1,3]_1 + 2 = 0. \label{eq:Phi4_1_M3}
\ee
By applying Eq.\eqref{eq:iden[k]1} with $n=1$ to Eq.\eqref{eq:Phi4_1_M3}, one can find 
\be
  [-4,0]_2  - [-2]_2  - [2]_2  - 1 = 0. \label{eq:Phi4_2_M3} 
\ee
Using Eq.\eqref{eq:iden[k]5} with $M=3$ to Eq.\eqref{eq:Phi4_2_M3} yields
\be
\sqrt{[-3,3]_3 - 1} \sqrt{[-1,1]_3 - 1} - \sqrt{[-3,-1]_3 - 1} - \sqrt{[1,3]_3 - 1} -1 = 0. \label{eq:Phi4_3_M3}
\ee
Notice that, from Eqs.\eqref{eq:QC_PT}\eqref{eq:QC_H}, $C^{(1,2)}(E) = 0$ and $C^{(3)}(E) = 0$ correspond to the ${\cal PT}_{1,2}$ and Hermitian QCs, respectively.
Let us denote $\zeta_{1,2,3}(s)$ by $\zeta_{{\cal PT}_1,{\cal PT}_2,{\cal H}}(s)$ explicitly.
By taking $E=0$ in Eqs.\eqref{eq:Phi4_1_M3}-\eqref{eq:Phi4_3_M3}, one can find  $C^{(1,2,3)}(0) = e^{- \zeta^{\prime}_{{\cal PT}_1, {\cal PT}_2,{\cal H}}(0)}$ as
\be
C^{(1)}(0) 
= \sqrt{2 + \sqrt{2}}, \qquad
C^{(2)}(0) = \sqrt{3 + 2 \sqrt{2}},
\qquad C^{(3)}(0)  = \sqrt{4 + 2\sqrt{2}}, \label{eq:CnE0_M3}
\ee
where we took the real positive solutions.
By using Eq.\eqref{eq:D_to_zeta}, one can find the ESRs for $\zeta_{{\cal PT}_1}(s)$ and $\zeta_{\cal H}(s)$  from Eqs.\eqref{eq:Phi4_1_M3}\eqref{eq:Phi4_3_M3} as\footnote{
Eq.\eqref{eq:Phi4_3_M3} gives non-zero coefficients in $O(E^{s})$ with $s \in  2{\mathbb N} + 2$.
However, those of $O(E^{s})$ with $s \in 4{\mathbb N}$ and $s \in 4{\mathbb N} + 2$ are dependent on each other and give the same relations.
}
\subbe
\be
&& 2 {\zeta}_{{\cal PT}_1}(1)^4 - 8 {\zeta}_{{\cal PT}_1}(1) {\zeta}_{{\cal PT}_1}(3) - 3 \sqrt{2} \zeta_{{\cal PT}_1}(4) = 0, \\
&& \zeta_{{\cal PT}_1}(1)^8 - 56 \zeta_{{\cal PT}_1}(1)^5 \zeta_{{\cal PT}_1}(3) + 210 \zeta_{{\cal PT}_1}(1)^4 \zeta_{{\cal PT}_1}(4) - 336 \zeta_{{\cal PT}_1}(1)^3 \zeta_{{\cal PT}_1}(5) + 280 \zeta_{{\cal PT}_1}(1)^2 \zeta_{{\cal PT}_1}(3)^2 \nl
&& \quad - 840 \zeta_{{\cal PT}_1}(1) \zeta_{{\cal PT}_1}(3) \zeta_{{\cal PT}_1}(4) + 720 \zeta_{{\cal PT}_1}(1) \zeta_{{\cal PT}_1}(7) + 336 \zeta_{{\cal PT}_1}(3) \zeta_{{\cal PT}_1}(5) - (315 \sqrt{2}+315) \zeta_{{\cal PT}_1}(4)^2 \nl
&& \quad + 315 \sqrt{2} \zeta_{{\cal PT}_1}(8) = 0, \\
&& \vdots, \nn 
\ee
\subee
and
\subbe
\be
&& (248 - 175 \sqrt{2}) \zeta_{\cal H}(1)^4 - (8 - 4 \sqrt{2}) \zeta_{\cal H}(1) \zeta_{\cal H}(3) - 3 \zeta_{\cal H}(4) = 0, \\
&& \frac{20469808 - 14474335 \sqrt{2}}{420} \zeta_{\cal H}(1)^8
- \frac{59578 \sqrt{2} - 84256}{15} \zeta_{\cal H}(1)^5 \zeta_{\cal H}(3) - \frac{4305 \sqrt{2} - 6088}{2} \zeta_{\cal H}(1)^4 \zeta_{\cal H}(4) \nl
&& \quad - \frac{992 - 700 \sqrt{2}}{5} \zeta_{\cal H}(1)^3 \zeta_{\cal H}(5) - \frac{592-418 \sqrt{2}}{3} \zeta_{\cal H}(1)^2 \zeta_{\cal H}(3)^2 - (128 - 90 \sqrt{2}) \zeta_{\cal H}(1) \zeta_{\cal H}(3) \zeta_{\cal H}(4) \nl
&& \quad  + \frac{24 - 12 \sqrt{2}}{7} \zeta_{\cal H}(1) \zeta_{\cal H}(7) - \frac{66 - 45\sqrt{2}}{4} \zeta_{\cal H}(4)^2
+  \frac{8 - 4 \sqrt{2}}{5} \zeta_{\cal H}(3) \zeta_{\cal H}(5) + \frac{3}{2} \zeta_{\cal H}(8) = 0, \\
&& \vdots, \nn
\ee
\subee
respectively.
Similarly, the ESR for $\zeta_{{\cal PT}_2}(s)$ is obtained from Eq.\eqref{eq:Phi4_2_M3} as
\subbe
\be
&& \zeta_{{\cal PT}_2}(1)^2 - (2 + \sqrt{2}) \zeta_{{\cal PT}_2}(2) = 0, \\
&& \zeta_{{\cal PT}_2}(1)^4 - 6 \zeta_{{\cal PT}_2}(1)^2 \zeta_{{\cal PT}_2}(2) + 8 \zeta_{{\cal PT}_2}(1) \zeta_{{\cal PT}_2}(3) - (3 + 6 \sqrt{2}) \zeta_{{\cal PT}_2}(2)^2 + 6 \sqrt{2} \zeta_{{\cal PT}_2}(4) = 0, \\
&& \zeta_{{\cal PT}_2}(1)^6 - 15 \zeta_{{\cal PT}_2}(1)^4 \zeta_{{\cal PT}_2}(2) + 40 \zeta_{{\cal PT}_2}(1)^3 \zeta_{{\cal PT}_2}(3) + 45 \zeta_{{\cal PT}_2}(1)^2 \zeta_{{\cal PT}_2}(2)^2 - 90 \zeta_{{\cal PT}_2}(1)^2 \zeta_{{\cal PT}_2}(4) \nl
&& \quad - 120 \zeta_{{\cal PT}_2}(1) \zeta_{{\cal PT}_2}(2) \zeta_{{\cal PT}_2}(3) - (75 + 60 \sqrt{2}) \zeta_{{\cal PT}_2}(2)^3 + (270 + 180 \sqrt{2}) \zeta_{{\cal PT}_2}(2) \zeta_{{\cal PT}_2}(4) \nl
&& \quad + 144 \zeta_{{\cal PT}_2}(1) \zeta_{{\cal PT}_2}(5) + 40 \zeta_{{\cal PT}_2}(3)^2 - (240 + 120 \sqrt{2}) \zeta_{{\cal PT}_2}(6) = 0, \\
&& \vdots. \nn
\ee
\subee
From these results, the selection rules are obtained as
\be
&& {\cal S}_{K} = {\mathbb N}_{0}/G \cong
\begin{cases}
  {\mathbb Z}_4, \quad G = 4 {\mathbb N}_0 &  \ \ \text{for} \ \ {\cal PT}_1 \ \text{and}  \ {\cal H} \ \ (K = 1 \ \text{and} \ 3) \\
  {\mathbb Z}_2, \quad G = 2 {\mathbb N}_0 & \ \ \text{for} \ \ {\cal PT}_2 \ \ (K = 2)
\end{cases}. \label{eq:sel_rule_M3}
\ee

For formulating the ZGFs, we express $[0]_3$ in terms of $[k]_1$ using Eqs.\eqref{eq:iden[k]1} with $n=2$ and \eqref{eq:C2_iden_PT_Herm_C2_1}.
It is derived as
\be
\left[ 0 \right]_3 &=& [-2,0,2]_1 - [-2]_1 - [2]_1. \label{eq:O3_0} 
\ee
From Eqs.\eqref{eq:D_to_zeta}\eqref{eq:C2_iden_PT_Herm_C2_1}\eqref{eq:O3_0} and the solutions of $C^{(n)}(E=0)$ in Eq.\eqref{eq:CnE0_M3}, one can obtain the ZGFs, $\bm{\zeta}_{2,3} = \bm{\zeta}_{2,3}(\bm{\zeta}_{1})$, as
\subbe
\be
\zeta_{{\cal PT}_2}(1) &=&  2 \zeta_{{\cal PT}_1}(1), \\
\zeta_{{\cal PT}_2}(2) &=&  (4 - 2\sqrt{2}) \zeta_{{\cal PT}_1}(1)^2, \\
\zeta_{{\cal PT}_2}(3) &=& (10 - 6 \sqrt{2}) \zeta_{{\cal PT}_1}(1)^3 - 2 \zeta_{{\cal PT}_1}(3), \\
\zeta_{{\cal PT}_2}(4) &=& \frac{76  - 50 \sqrt{2}}{3} \zeta_{{\cal PT}_1}(1)^4 - \frac{16  - 8 \sqrt{2}}{3} \zeta_{{\cal PT}_1}(1) \zeta_{{\cal PT}_1}(3) - 2 \sqrt{2} \zeta_{{\cal PT}_1}(4) = 0, \\
\zeta_{{\cal PT}_2}(5) &=& \frac{197 - 135 \sqrt{2}}{3} \zeta_{{\cal PT}_1}(1)^5 - \frac{50- 30 \sqrt{2}}{3} \zeta_{{\cal PT}_1}(1)^2 \zeta_{{\cal PT}_1}(3) - (5 \sqrt{2}-5) \zeta_{{\cal PT}_1}(1) \zeta_{{\cal PT}_1}(4) \nl
&& -2 \zeta_{{\cal PT}_1}(5) = 0, \\
\vdots, && \nn 
\ee
\subee
and
\subbe
\be
\zeta_{\cal H}(1) &=& (1+ \sqrt{2}) \zeta_{{\cal PT}_1}(1), \\
\zeta_{\cal H}(2) &=& 2 \zeta_{{\cal PT}_1}(1)^2 - \zeta_{{\cal PT}_1}(2), \\
\zeta_{\cal H}(3) &=&  \sqrt{2} \zeta_{{\cal PT}_1}(1)^3 +  (1 + \sqrt{2}) \zeta_{{\cal PT}_1}(3), \\
\zeta_{\cal H}(4) &=&  \frac{4 - 2 \sqrt{2}}{3} \zeta_{{\cal PT}_1}(1)^4 + \frac{8 + 8 \sqrt{2}}{3} \zeta_{{\cal PT}_1}(1) \zeta_{{\cal PT}_1}(3) + (3 + 2\sqrt{2}) \zeta_{{\cal PT}_1}(4), \\
\zeta_{\cal H}(5) &=& - \frac{10 - 3 \sqrt{2}}{6} \zeta_{{\cal PT}_1}(1)^5 + \frac{20 +  15 \sqrt{2}}{3} \zeta_{{\cal PT}_1}(1)^2 \zeta_{{\cal PT}_1}(3) + \frac{10 + 5\sqrt{2}}{2} \zeta_{{\cal PT}_1}(1) \zeta_{{\cal PT}_1}(4) \nl
&& + (1 + \sqrt{2} )\zeta_{{\cal PT}_1}(5), \\
\vdots, && \nn
\ee
\subee
respectively.
\\ \, \\ \indent
We make comments on differences from the $M=2$ case.
There are two crucial observations.
The first is that the number of the ESRs of $\bm{\zeta}_2$ are twice those of $\bm{\zeta}_{1,3}$.
The second is that the ZGFs are not always invertible, i.e., $\bm{\zeta}_{3} = \bm{\zeta}_{3}(\bm{\zeta}_{1})$ is invertible but $\bm{\zeta}_{2} = \bm{\zeta}_{2}(\bm{\zeta}_{1})$ is not.
The first observation regarding the number of ESRs is explained by the selection rules in Eq.~\eqref{eq:sel_rule_M3}: the non-trivial ESRs for $\bm{\zeta}_{1,3}$ come from the $O(E^s)$ parts with $s \in 4\mathbb{N}$, whereas those for $\bm{\zeta}_2$ come from $s \in 2\mathbb{N}$.
The second observation regarding non-invertibility arises because $\zeta_{{\cal PT}_2}(s)$ does not contain the linear term of $\zeta_{{\cal PT}_1}(s)$ for $s \in 4\mathbb{N}_0+2$.
The absence of this linear term is not accidental. This is the first concrete manifestation of the\textit{algebraic nullification} that we will prove generally in Sec.~\ref{sec:general_cases}.
The vanishing of coefficients here anticipates the structural non-invertibility for odd $M$, which is governed by the selection rules.
Another remarkable point is the compatibility with the $\mathbb{Z}_{2M+2}$ Symanzik rotation in Eqs.~\eqref{eq:Phi4_1_M3}--\eqref{eq:Phi4_3_M3}.
Eq.~\eqref{eq:Phi4_1_M3} is invariant under the $\mathbb{Z}_{4} (\cong \mathbb{Z}_{8}/\langle 2 \rangle)$ symmetry as $[k_1,\cdots,k_\ell]_{1} \mapsto [k_1+c,\cdots,k_\ell+c]_{1}$ with $c \in \{0,2,4,6\}$, and Eq.~\eqref{eq:Phi4_2_M3} is $\mathbb{Z}_{2} (\cong \mathbb{Z}_{8}/\langle 4 \rangle)$ symmetric as $[k_1,\cdots,k_\ell]_{2} \mapsto [k_1+c,\cdots,k_\ell+c]_{2}$ with $c \in \{0,4\}$.
Hence, these are consistent with the selection rules in Eq.~\eqref{eq:sel_rule_M3}.
In contrast, Eq.~\eqref{eq:Phi4_3_M3} is $\mathbb{Z}_2$ symmetric but ${\cal S}_{{\cal H}} \cong \mathbb{Z}_4$, which implies that the selection rule does not always naively match the symmetric structure of a closed form of the QCs with a fixed fusion label, $n$.
The symmetric structure is possibly a subgroup (or subset) of the selection rule in general.

This $M=3$ example thus already exhibits the qualitative pattern that will be proved in Sec.~\ref{sec:general_cases}: odd sectors form an information-complete family, whereas even sectors suffer from algebraic information loss and cannot be used to reconstruct the full spectrum.

\section{Construction of the ESRs and the ZGFs: General $M \in \frac{1}{2}{\mathbb N} + 1$} \label{sec:general_cases}
\begin{figure}[t]
    \centering
    \includegraphics[width=0.45\textwidth]{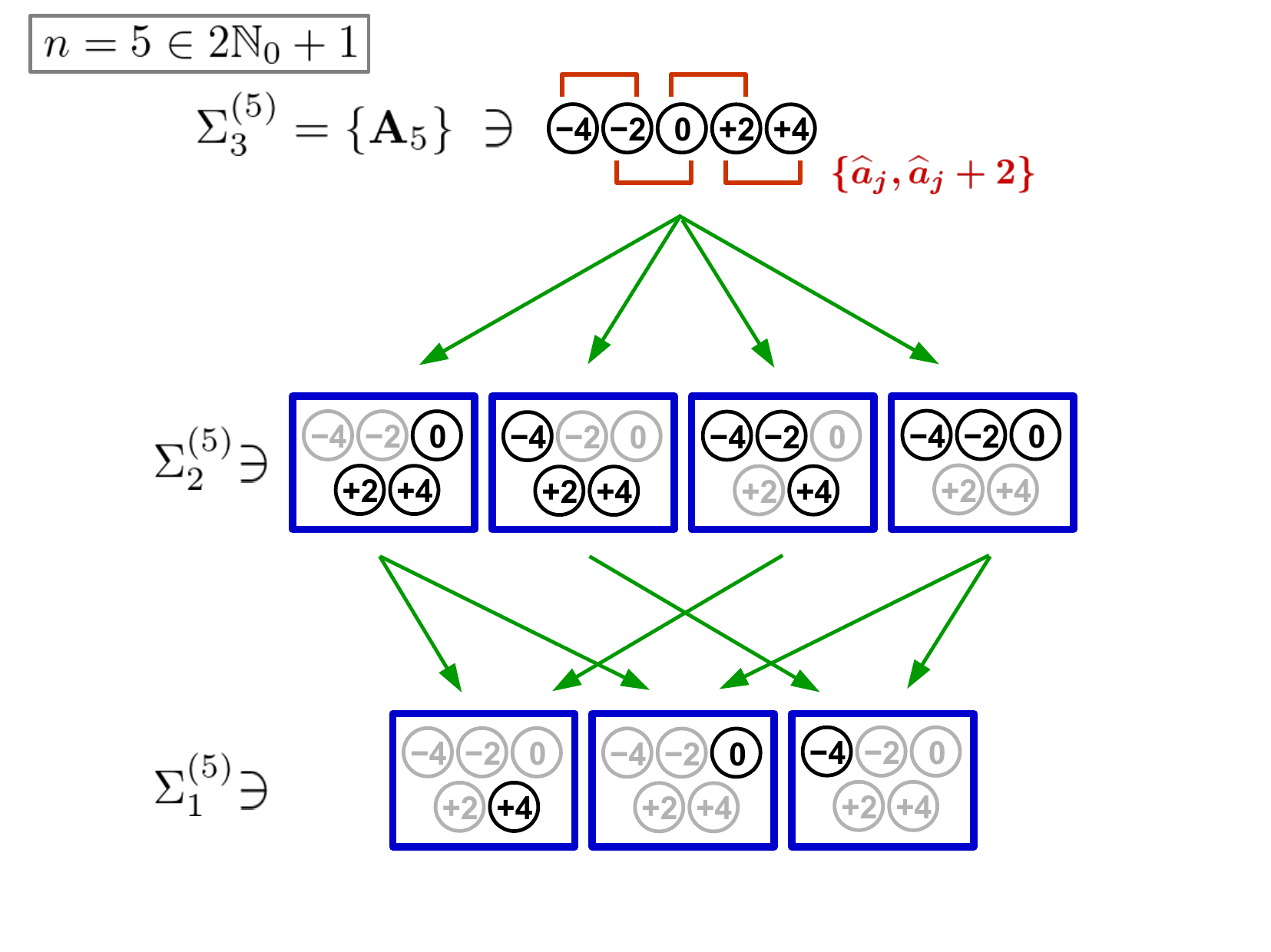}
    \includegraphics[width=0.45\textwidth]{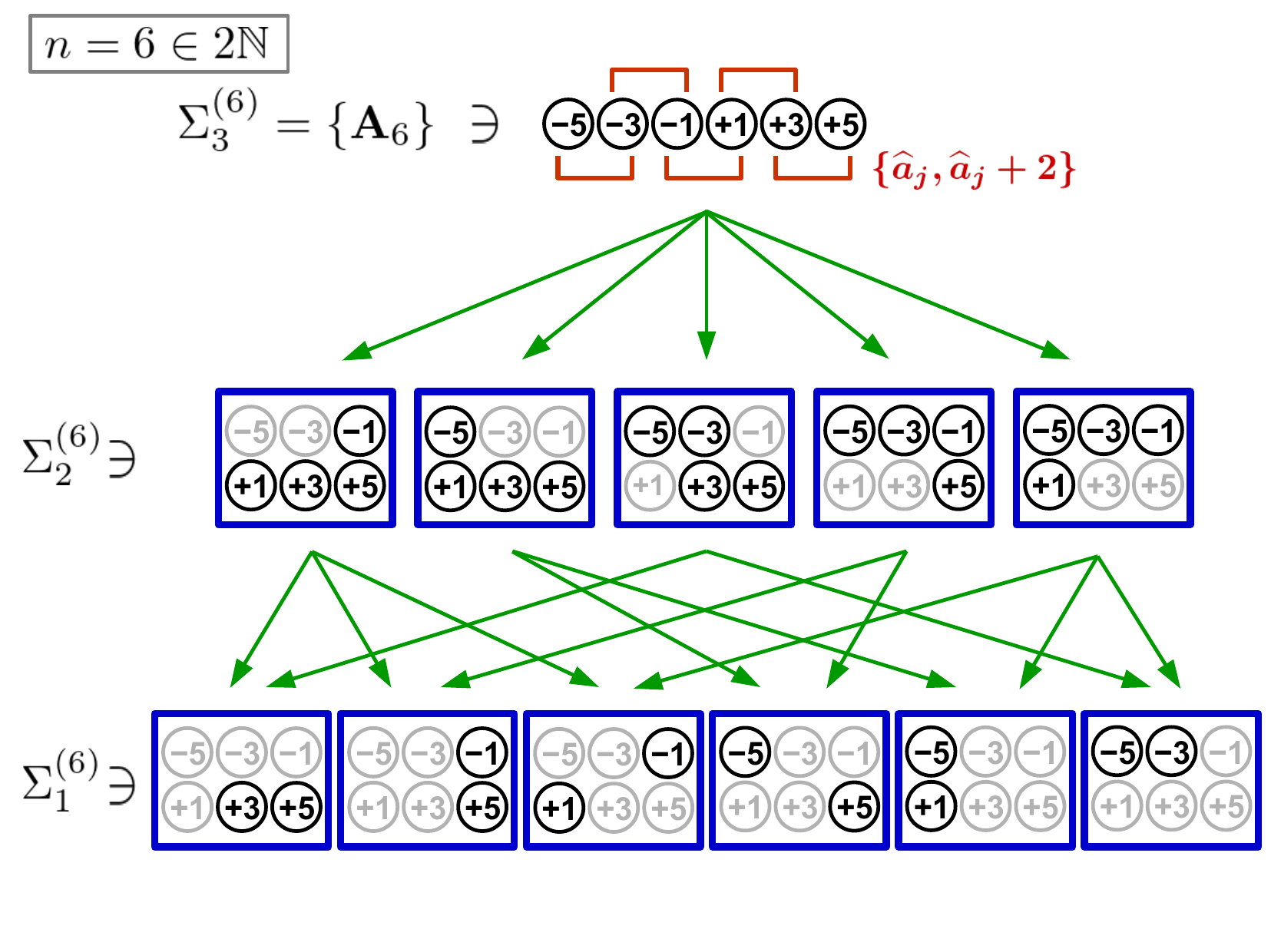}
    \caption{
      The schematic figures of $\Sigma_{k}^{(n)}$ for $n=5$ (Left) and $6$ (Right).
      The black circled numbers are elements of the sets in the family, $\Sigma_{k}^{(n)}$.
      $\Sigma^{(n)}_{\lfloor \frac{n+1}{2} \rfloor}$ only contains ${\bf A}_n$, and the sets which belong to $\Sigma^{(n)}_k $ can be recursively determined by subtracting the pair, $\{\widehat{a}_j,\widehat{a}_j + 2\}$, from those in $\Sigma^{(n)}_{k+1}$.}
    \label{fig:Sigma}
\end{figure}

In this section, we generalize the previous analyses to arbitrary $M \in \frac{1}{2} {\mathbb N}+1$.
We perform the analyses along Fig.~\ref{fig:procedure}, and the procedure is almost the same to Sec.~\ref{sec:simple_cases}.
One of the main differences from the simple cases is that the closed forms in terms of $C^{(n)}(\omega^k E)$ with fixed $n$, such as Eqs.\eqref{eq:C2_iden_PT}\eqref{eq:C2_iden_Herm} and \eqref{eq:Phi4_1_M3}-\eqref{eq:Phi4_3_M3}, can not be obtained.
Therefore, one has to find the ESRs for higher $n$ by the inverse ZGFs obtained from $\bm{\zeta}_n = \bm{\zeta}_n(\bm{\zeta}_1)$.
In order to do it, one needs to write down $C^{(n)}(E) (=[0]_n)$ in terms of $C^{(1)}(\omega^k E) (=[k]_1)$ from the fusion relations \eqref{eq:iden[k]1}\eqref{eq:iden[k]2}.
It is expressed by
\be
[0]_n =
\begin{dcases}
  \sum_{k=1}^{\frac{n+1}{2}} (-1)^{\frac{n+1}{2}-k} \sum_{\{ a_1,\cdots,a_{2k-1}\} \in \Sigma^{(n)}_{k}} [a_1,\cdots,a_{2k-1}]_1
  & \text{for $n \in 2{\mathbb N}_0 + 1$} \\
  \sum_{k=1}^{\frac{n}{2}} (-1)^{\frac{n}{2}-k} \sum_{\{ a_1,\cdots,a_{2k} \} \in \Sigma^{(n)}_{k}}  [a_1,\cdots,a_{2k}]_1  + (-1)^{\frac{n}{2}} & \text{for $n \in 2{\mathbb N}$} 
\end{dcases}, \label{eq:Cn_prods_C1_simp}
\ee
where $\Sigma^{(n)}_{k}$ is a family of sets defined as\footnote{
In Eqs.\eqref{eq:def_SigmaM}\eqref{eq:def_til_SigmaM_2}\eqref{eq:def_hatSigmaM}\eqref{eq:def_hattil_SigmaM_2}, we assume that $\bigcup_{j=1}^{0} \{ \cdots \} = \emptyset$.}
\be
\Sigma^{(n)}_{k \in \{1,\cdots, \lfloor \frac{n+1}{2} \rfloor  \}} &:=& \{ {\bf a} \subseteq {\bf A}_n \, | \, {\bf a} = {\bf A}_{n} \setminus \bigcup_{j=1}^{\lfloor \frac{n+1}{2} \rfloor - k} \{ \widehat{a}_{j} , \widehat{a}_{j} + 2 \} \nl
&& \text{with} \, \,  \widehat{a}_j \in {\bf A}_n \, \, \text{and} \, \, \widehat{a}_{j_1} + 2 < \widehat{a}_{j_2} \le n - 3 \, \, \text{for any} \, \, j_1 < j_2 \}, \label{eq:def_SigmaM}
\ee     
with ${\bf A}_n$ given by
\be
{\bf A}_n &:=& \{ -n+1, -n+3, \cdots, n-3, n-1 \}, \qquad |{\bf A}_n| = n. \label{eq:def_set_An}
\ee
Physically, $\Sigma^{(n)}_{k}$ represents the collection of all possible phase interference patterns appearing in the expansion of the n-th quantization condition, where $k$ counts the number of surviving terms after cancellations.
The schematic figures of $\Sigma^{(n)}_k$ for $n=5,6$ are shown in Fig.~\ref{fig:Sigma}.
The first five specific forms for $n  \in {\mathbb N} + 1$ can be written down as
\subbe
\be
\left[ 0 \right]_2 &=& [-1,1]_1 - 1, \label{eq:O2} \\
   \left[ 0 \right]_3 &=& [-2,0,2]_1 - [-2]_1 - [2]_1, \label{eq:O3} \\
   \left[ 0 \right]_4 &=& [-3,-1,1,3]_1 - [-3,-1]_1 - [-3,3]_1 - [1,3]_1 + 1, \label{eq:O4} \\
   \left[ 0 \right]_5 &=& [-4,-2,0,2,4]_1 - [-4,-2,0]_1 - [-4,-2,4]_1 - [-4,2,4]_1 - [0,2,4]_1 + [-4]_1 + [0]_1 + [4]_1, \label{eq:O5} \\
   \left[ 0 \right]_6 &=& [-5,-3,-1,1,3,5]_1 - [-5,-3,-1,1]_1 - [-5,-3,-1,5]_1 - [-5,-3,3,5]_1 - [-5,1,3,5]_1 \nl
&& - [-1,1,3,5]_1 + [-5,-3]_1 + [-5,1]_1 + [-5,5]_1 + [-1,1]_1 + [-1,5]_1 + [3,5]_1 - 1, \label{eq:O6} \\
\vdots. && \nn
\ee
\subee

As we saw in Sec.~\ref{sec:simple_cases}, structures of the ESRs and the ZGFs characterized by the selection rules in general depend on $M$ and $n \in {\cal K}$.
In the below, we firstly provide a method using the Chebyshev polynomials to easily determine $C^{(n)}(E=0) = e^{-\zeta^\prime_n(0)}$ in Sec.~\ref{sec:SZF_genM}, and then, discuss  the $M \in {\mathbb N} + 1$ and $M \in {\mathbb N} + \frac{1}{2}$ cases in Secs.~\ref{sec:SZF_genM} and \ref{sec:half_M}, respectively.

\subsection{Determination of $\zeta_n^{\prime}(0)$ via the Chebyshev polynomials} \label{sec:sym_iden_genM}

For construction of the ESRs and the ZGFs, one has to determine $C^{(n)}(E=0) = e^{-\zeta_n^{\prime}(0)}$.
It can be in principle found by solving an algebraic equations of $C^{(1)}(0)$ given by the condition \eqref{eq:k0k2M} and substituting the solution, $e^{-\zeta_1^{\prime}(0)}$, into Eq.\eqref{eq:def_SigmaM}.
Finding the solution is not so easy when $M$ is large, but our method using the Chebyshev polynomials can solve the technical difficulty. 
In addition to that, the solutions from the structure of the Chebyshev polynomials indirectly affect the selection rules, which we will discuss in Secs.~\ref{sec:SZF_genM} and \ref{sec:half_M}.

We define the following mapping:
\be
  [k]_n \longmapsto U_n(z), \qquad
  z := \frac{C^{(1)}(E=0)}{2}, \label{eq:k_U_homo} 
\ee
where $U_n(z)$ is the Chebyshev polynomial of the second kind (or Vieta–Fibonacci polynomial), defined by
$U_0(z)=1$, $U_1(z)=2z$, $U_{n+1}(z)=2zU_n(z)-U_{n-1}(z)$ for $n\in\mathbb{N}$.
Here, the mapping simply means that we evaluate the QCs at $E=0$ and identify all labels $[k]_n$ with the same $n$.
Under this identification, the algebraic structure of the fusion relations is preserved.
In fact, Eqs.\eqref{eq:iden[k]1}\eqref{eq:iden[k]2} reduce to the well-known identities:
\subbe
\be
\text{\eqref{eq:iden[k]1}} &\ \ \mapsto \ \ & 2 z U_n(z) = U_{n-1}(z) + U_{n+1}(z), \label{eq:iden[k]1_Un} \\
\text{\eqref{eq:iden[k]2}} &\ \ \mapsto \ \ & U_n(z) U_n(z) = 1 + U_{n-1}(z) U_{n+1}(z). \label{eq:iden[k]2_Un} 
\ee
In addition, Eq.\eqref{eq:iden[k]3} is transformed as
\be
\text{\eqref{eq:iden[k]3}} &\ \ \mapsto \ \ & U_n(z) = U_{2M-n}(z). \label{eq:iden[k]3_Un}
\ee
\subee
From $U_n(\cos \theta) = \frac{\sin[(n + 1) \theta]}{\sin \theta}$, one can find
\be
C^{(1)}(0) = 2 \cos [(2 j + 1) \mu]. \qquad (j \in {\mathbb Z}) \label{eq:C10_sol}
\ee
By requiring $C^{(n)}(0) = U_{n}(\frac{C^{(1)}(0)}{2}) > 0$ for all $n \in \{ 0, \cdots, 2 M - 1 \}$ that follows from the real positive spectrum in unbroken ${\cal PT}$ symmetry, $j$ is uniquely determined as $j = 0$ in Eq.\eqref{eq:C10_sol}\footnote{
This $j$ corresponds to the angular momentum, $\ell$,  when one includes the singular term, $\frac{\ell(\ell + 1)}{x^{2}}$, in the potential.
In such a case, the condition is generally broken.
}.
This solution is consistent with the result obtained by directly solving the ODE with $E=0$ in, e.g., Ref.~\cite{Dorey:2007zx}.
From this solution and Eq.\eqref{eq:k_U_homo}, one can find that
\be
e^{-\zeta_n^{\prime}(0)} = C^{(n)}(0)= U_n\left( \frac{C^{(1)}(0)}{2} \right) = \frac{\sin [(n + 1) \mu]}{\sin \mu}, \qquad (n \in {\cal K}) \label{eq:sol_of_Cn0}
\ee
and it also agrees with the result for the Hermitian case, i.e., $n = M$ for $M \in {\mathbb N}+1$ in Ref.~\cite{Voros1983}.

\subsection{$M \in {\mathbb N} + 1$} \label{sec:SZF_genM}

\begin{figure}[t]
    \centering
    \includegraphics[width=0.45\textwidth]{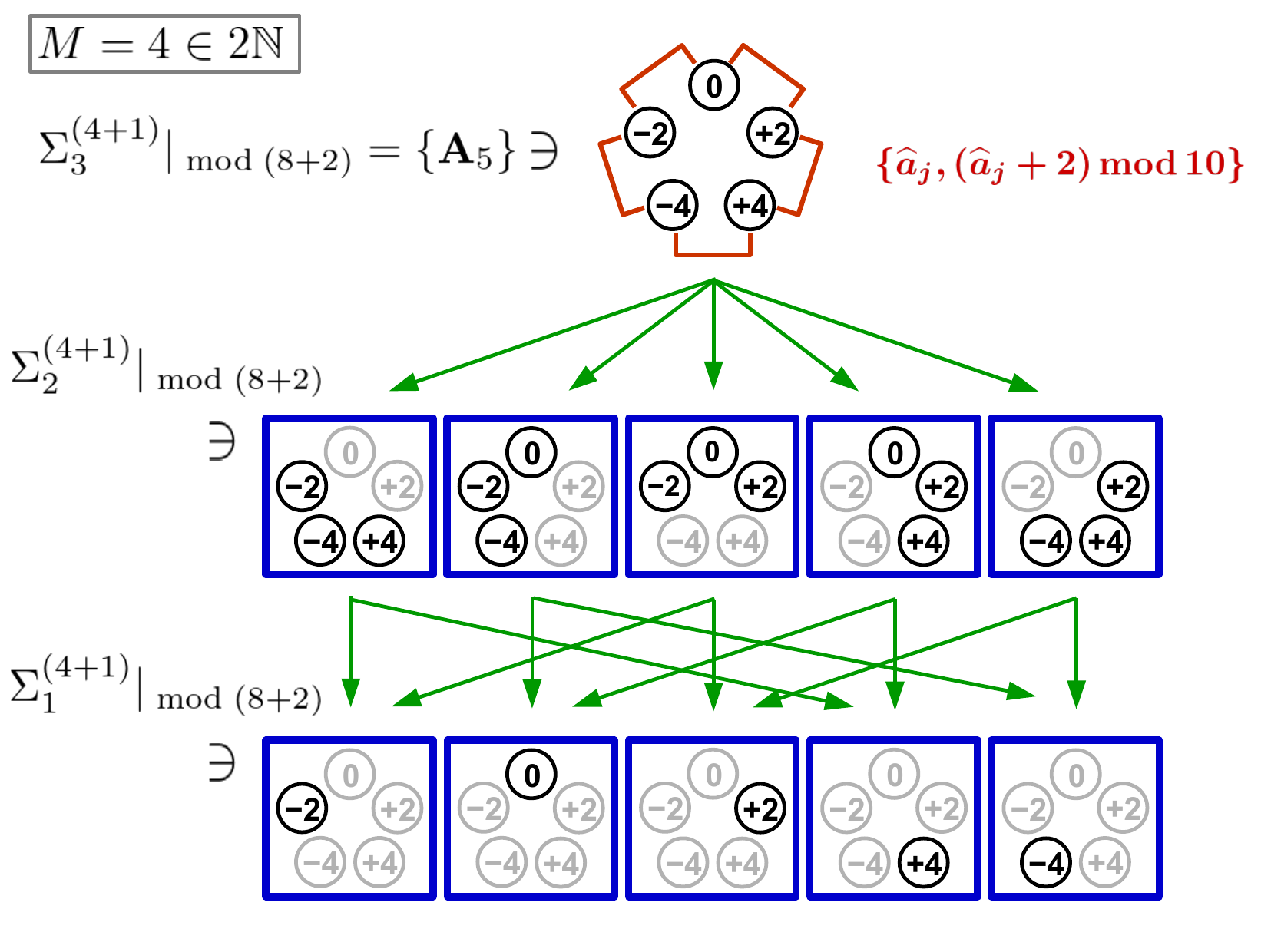}
    \includegraphics[width=0.45\textwidth]{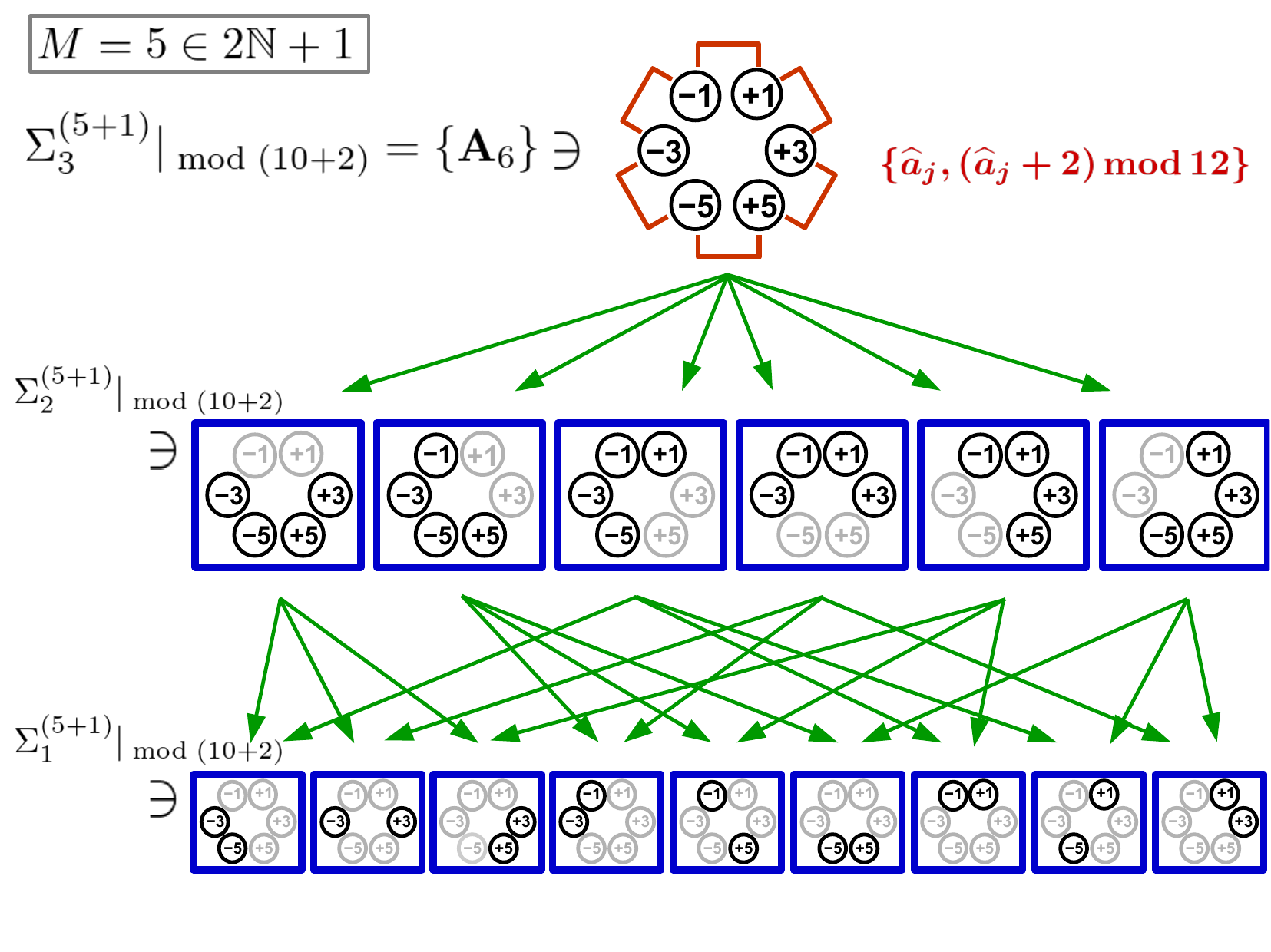}
    \caption{The schematic figures of ${\Sigma}_{k}^{(M+1)}|_{\bmod (2 M + 2)}$ for $M=4$ (Left) and $5$ (Right).
      The black circled numbers are elements of the sets in the family, ${\Sigma}_{k}^{(M+1)}|_{\bmod (2 M + 2)}$. 
      ${\Sigma}^{(M + 1)}_{k}|_{\bmod (2 M + 2)}$ is defined from $\Sigma^{(n)}_k$ in Eq.\eqref{eq:def_SigmaM} by imposing the modulo identification, i.e., $\bmod (2 M + 2)$.}
    \label{fig:Sigma_til}
\end{figure}

Let us consider the $M \in {\mathbb N} + 1$ case by beginning with the derivation of the ESRs of $\bm{\zeta}_1$.
From Eq.\eqref{eq:k0k2M}, one can generalize Eqs.\eqref{eq:C2_iden_PT}\eqref{eq:Phi4_1_M3} as
\be
&&   [M-1]_{2M} - 1 = [-2]_{M-1} \Phi^{(1)}_{2M+2}, \label{eq:gen_iden_PT_pre} \\ \nl
&& \Phi_{2M+2}^{(1)} :=
\begin{dcases}
  \sum_{k=1}^{\frac{M+2}{2}} (-1)^{\frac{M+2}{2}-k} \sum_{\{ a_1,\cdots,a_{2k-1} \} \in {\Sigma}^{(M+1)}_{k}|_{\bmod (2 M + 2)}} [a_1, \cdots, a_{2k-1} ]_1 & \ \ \text{for $M \in 2{\mathbb N}$} \\
\sum_{k=1}^{\frac{M+1}{2}} (-1)^{\frac{M+1}{2}-k} \sum_{\{ a_1,\cdots,a_{2k} \} \in {\Sigma}_{k}^{(M+1)}|_{\bmod (2 M + 2)}} [a_1,\cdots, a_{2k}]_1 + (-1)^{\frac{M+1}{2}} 2 & \ \ \text{for $M \in 2{\mathbb N} + 1$} 
\end{dcases}, \label{eq:Phi_M+1_sym}
\ee
where the family of sets, $\widetilde{\Sigma}_{k}^{(M+1)}$, is defined from Eq.\eqref{eq:def_SigmaM} by imposing the modulo identification to the elements in ${\bf A}_{n=M+1}$ as
\be
   {\Sigma}^{(M+1)}_{k \in \{ 1, \cdots, \lfloor \frac{M + 2}{2} \rfloor \}}|_{\bmod (2 M + 2)} &:=& \{ {\bf a} \subseteq {\bf A}_{M+1} \, | \, {\bf a} = {\bf A}_{M+1} \setminus  \bigcup_{j=1}^{\lfloor \frac{M + 2}{2} \rfloor - k} \{ \widehat{a}_{j} , (\widehat{a}_{j} + 2) \bmod (2 M + 2) \} \nl
&& \text{with} \  \widehat{a}_j \in {\bf A}_{M+1}, \ |\omega^{\widehat{a}_{j_1} - \widehat{a}_{j_2}}| > |\omega^{2}| \ \text{and} \ \widehat{a}_{j_1}  < \widehat{a}_{j_2} \ \text{for any} \ j_1 < j_2 \}. \label{eq:def_til_SigmaM_2}
\ee
Fig.~\ref{fig:Sigma_til} shows the schematic figures of ${\Sigma}^{(M+1)}_k|_{\bmod (2 M + 2)}$ for $M=4,5$.
The remarkable fact is that $[M-1]_{2M}-1$ in Eq.\eqref{eq:gen_iden_PT_pre} can be always factorized into the two parts, $[-2]_{M-1}$ and $\Phi_{2M+2}$ for any $M \in {\mathbb N} + 1$.
The former part, $[-2]_{M-1}$, is irrelevant because it is generally non-zero, and the later part should correspond to the ESRs as
\be
\Phi^{(1)}_{2M+2} = 0. \label{eq:gen_iden_PT}
\ee
For $M=2,\cdots,5$, Eq.\eqref{eq:def_til_SigmaM_2} can be specifically written down as
\subbe
\be
\Phi^{(1)}_{4+2} &=& [-2,0,2]_1 - [-2]_1 - [0]_1 - [2]_1, \label{eq:Phi3_C1} \\
\Phi^{(1)}_{6+2} &=& [-3,-1,1,3]_1 - [-3,-1]_1 - [-3,3]_1 - [-1,1]_1 - [1,3]_1 + 2, \label{eq:Phi4_C1} \\
\Phi^{(1)}_{8+2} &=& [-4,-2,0,2,4]_1 - [-4,-2,0]_1 - [-4,-2,4]_1 - [-4,2,4]_1 - [-2,0,2]_1 - [0,2,4]_1 \nl
&& + [-4]_1 + [-2]_1 + [0]_1 + [2]_1 + [4]_1, \label{eq:Phi5_C1} \\
\Phi^{(1)}_{10+2} &=& [-5,-3,-1,1,3,5]_1 - [-5,-3,-1,1]_1 - [-5,-3,-1,5]_1 - [-5,-3,3,5]_1 - [-5,1,3,5]_1 \nl
&& - [-3,-1,1,3]_1 - [-1,1,3,5]_1 + [-5,-3]_1  + [-5,1]_1 + [-5,5]_1 + [-3,-1]_1 \nl
&& + [-3,3]_1 + [-1,1]_1 + [-1,5]_1 + [1,3]_1 + [3,5]_1 - 2, \label{eq:Phi6_C1} \\
\vdots. && \nn
\ee
\subee
From Eq.\eqref{eq:D_to_zeta}, one can write down $[a_1,\cdots,a_k]_{n \in {\cal K}}$ as 
\be
   [a_1,\cdots,a_k]_{n \in {\cal K}} &=& e^{- k \zeta_n^{\prime}(0)} \exp \left[ - \sum_{s \in {\mathbb N}} \frac{\zeta_n(s)}{s} \Omega(\{ a_1,\cdots,a_k\},s) E^s \right] \nl
   &=& e^{-k \zeta_n^{\prime}(0)}  + e^{-k \zeta_n^{\prime}(0)} \sum_{m \in {\mathbb N}} \sum_{\substack{{\bf t} \in {\mathbb N}_0^m  \\ |{\bf t}| > 0}} (-1)^{|{\bf t}|} \left[ \prod_{s=1}^{m} \frac{\zeta_n(s)^{t_s}}{t_s ! s^{t_s}} \Omega(\{a_1,\cdots,a_k \}, s)^{t_s} \right] E^{({\bf m}, {\bf t})}, \label{eq:[a_1]_expand_zeta}
\ee
where 
\be
\Omega(\{ a_1, \cdots, a_k \}, s) &:=& \sum_{j=1}^{k} \omega^{s a_j}, \label{eq:def_Omega}
\ee
and substituting Eq.\eqref{eq:[a_1]_expand_zeta} with $n=1$ into Eq.\eqref{eq:Phi_M+1_sym} yields the ESRs of $\bm{\zeta}_1$.
As one can see from Eqs.\eqref{eq:Phi3_C1}-\eqref{eq:Phi6_C1}, these expressions are ${\mathbb Z}_{M+1} (\cong {\mathbb Z}_{2M+2}/\langle 2 \rangle)$ invariant under the constant shift, i.e., $[k_1,\cdots,k_\ell]_1 \mapsto [k_1+c,\cdots,k_\ell+c]_1$ with $c \in \{0,2, \cdots, 2M \}$.
This property directly propagates to the selection rule of $\bm{\zeta}_1$ through $\Omega(\{ a_1, \cdots, a_k \}, s)$.
Thanks to the factorization in Eq.\eqref{eq:gen_iden_PT_pre}, the sum of $[k_1,\cdots,k_\ell]_1$ appearing in $\Phi^{(1)}_{2M+2}$ is ${\mathbb Z}_{M+1}$ invariant for each fixed $\ell$, and the ESR is non-trivial iff the corresponding sum of $\sum_{\bf a} \Omega(\{ a_1, \cdots, a_k \}, s)$ does not vanish.
Consequently, the selection rule is
\be
{\cal S}_{1} \cong {\mathbb Z}_{M+1} \ \ \text{for} \ \ M \in {\mathbb N} + 1. 
\ee

It is notable that the factorizability of $[M-1]_{2M}-1$ in Eq.\eqref{eq:gen_iden_PT_pre} can be also seen from the mapping to the Chebyshev polynomial discussed in Sec.~\ref{sec:sym_iden_genM}.
The mapping \eqref{eq:k_U_homo} reduces Eq.\eqref{eq:gen_iden_PT_pre} to the well-known identity of the Chebyshev polynomials as
\be
&& \text{\eqref{eq:gen_iden_PT_pre}} \ \ \mapsto \ \ U_{2M}(z) - 1 = 2 U_{M-1}(z) T_{M+1}(z), \qquad (M \in {\mathbb N} + 1) \label{eq:prod_form_Cheby}
\ee
where $T_n(z)$ is the Chebyshev polynomial of the first kind (or the Vieta–Lucas polynomial) defined as
\be
T_0(z) = 1, \qquad T_1(z) = z, \qquad T_{n+1}(z) = 2z T_{n}(z) - T_{n-1}(z), \quad (n \in {\mathbb N}) \label{eq:cheby_pols}
\ee
and $\Phi^{(1)}_{2M+2}\mid_{E=0}$ can be identified as
\be
\Phi^{(1)}_{2M+2} \mid_{E=0} = 2 T_{M+1}\left( \frac{C^{(1)}(0)}{2} \right). \qquad (M \in {\mathbb N} + 1)
\label{eq:Phi1_Ezero}
\ee
Since $T_{M+1}(\cos \mu) = \cos [(M+1) \mu] = 0$, one can consistently have $\Phi^{(1)}_{2M+2} \mid_{E=0} = 0$, where we used Eq.\eqref{eq:C10_sol} with $j=0$.

\, \\ \indent
For constructing the ESRs for higher $n$, one needs to obtain the ZGFs, $\bm{\zeta}_n = \bm{\zeta}_n(\bm{\zeta}_1)$, and their inverse solutions.
The ZGFs can be obtained by substituting Eq.\eqref{eq:[a_1]_expand_zeta} into Eq.\eqref{eq:Cn_prods_C1_simp}, which is given by
\be
&& \sum_{m \in {\mathbb N}} \sum_{\substack{{\bf t} \in {\mathbb N}_0^m \\ |{\bf t}| > 0 \\ ({\bf m}, {\bf t}) = N}} (-1)^{|{\bf t}|} \left[ \prod_{s=1}^{m} \frac{\zeta_n(s)^{t_s}}{t_s ! s^{t_s}} - \sum_{k=1}^{\lfloor \frac{n+1}{2} \rfloor} \Xi_{n,k} \sum_{{\bf a} \in \Sigma^{(n)}_{k}} \prod_{s=1}^{m} \frac{\zeta_1(s)^{t_s}}{t_s ! s^{t_s}} \Omega({\bf a} ,s)^{t_s} \right] 
= 0, \qquad (n \in {\cal K}) \label{eq:Cn_prods_C1_simp_zeta2} \\
&&  \Xi_{n,k} := \frac{(-1)^{\lfloor \frac{n+1}{2} \rfloor -k}}{e^{- \zeta_n^\prime(0)}}
\left( e^{- \zeta_1^\prime(0)} \right)^{2k - n \bmod 2} \quad \text{with} \quad e^{- \zeta_n^\prime(0)} = \frac{\sin [ (n+1) \mu ]}{\sin \mu}.
\nn 
\ee
Here, we fixed the order of the energy as $O(E^{({\bf m}, {\bf t}) = N})$ in Eq.\eqref{eq:[a_1]_expand_zeta} to derive Eq.\eqref{eq:Cn_prods_C1_simp_zeta2}.
Remind that the constant part in $O(E^{0})$ becomes zero by Eq.\eqref{eq:sol_of_Cn0}.
It is also notable that the ESR of $\bm{\zeta}_1$ for any $M \in \frac{1}{2} {\mathbb N} + 1$ can be also expressed by Eq.\eqref{eq:Cn_prods_C1_simp_zeta2} by taking $n=2M$, $\zeta_{n = 2M}(s) = 0$, and $\zeta^\prime_{n = 2M}(0) = 1$.
One can find $\zeta_n(s)$ as a polynomial of $\{ \zeta_1(s^\prime) \}_{1 \le s^\prime \le s}$ by recursively solving  Eq.\eqref{eq:Cn_prods_C1_simp_zeta2} from $N=1$ to higher.
The first five orders can be specifically written as
\subbe
\be
\zeta_n(1) &=& \sum_{k=1}^{\lfloor \frac{n+1}{2} \rfloor} \Xi_{n,k} \sum_{{\bf a} \in \Sigma^{(n)}_{k}} \zeta_1(1) \Omega({\bf a} ,1), \label{eq:zeta_n1_zeta1} \\
 \zeta_n(2) &=& \zeta_n(1)^2  - \sum_{k=1}^{\lfloor \frac{n+1}{2} \rfloor} \Xi_{n,k} \sum_{{\bf a} \in \Sigma^{(n)}_{k}} \left[ \zeta_1(1)^2 \Omega({\bf a} ,1)^2 - \zeta_1(2) \Omega({\bf a} ,2) \right], \\
 \zeta_n(3) &=& - \frac{1}{2} \zeta_n(1)^{3} + \frac{3}{2} \zeta_n(1)^{2} \zeta_n(2) \nl
 && + \sum_{k=1}^{\lfloor \frac{n+1}{2} \rfloor} \Xi_{n,k} \sum_{{\bf a} \in \Sigma^{(n)}_{k}} \left[ \frac{1}{2} \zeta_1(1)^{3} \Omega({\bf a},1)^3 - \frac{3}{2} \zeta_1(1) \zeta_1(2) \Omega({\bf a},1) \Omega({\bf a},2) + \zeta_1(3) \Omega({\bf a},3) \right], \\
 \zeta_n(4) &=&  \frac{1}{6} \zeta_n(1)^{4} - \zeta_n(1)^{2} \zeta_n(2) + \frac{4}{3} \zeta_n(1) \zeta_n(3) + \frac{1}{2} \zeta_n(2)^{2} \nl
 && - \sum_{k=1}^{\lfloor \frac{n+1}{2} \rfloor} \Xi_{n,k} \sum_{{\bf a} \in \Sigma^{(n)}_{k}} \left[ \frac{1}{6} \zeta_1(1)^{4} \Omega({\bf a},1)^4 - \zeta_1(1)^{2} \zeta_1(2) \Omega({\bf a},1)^2 \Omega({\bf a},2) + \frac{1}{2} \zeta_1(2)^2 \Omega({\bf a},2)^2 \right. \nl
 && \left. + \frac{4}{3} \zeta_1(1) \zeta_1(3) \Omega({\bf a},1) \Omega({\bf a},3) - \zeta_1(4) \Omega({\bf a},4) \right], \label{eq:zeta_n4_zeta1} \\
 \zeta_n(5) &=& - \frac{1}{24} \zeta_n(1)^{5} + \frac{5}{12} \zeta_n(1)^{3} \zeta_n(2) - \frac{5}{6} \zeta_n(1)^{2} \zeta_n(3) - \frac{5}{8} \zeta_n(1) \zeta_n(2)^2 + \frac{5}{4} \zeta_n(1) \zeta_n(4) + \frac{5}{6} \zeta_n(2)\zeta_n(3) \nl
&& + \sum_{k=1}^{\lfloor \frac{n+1}{2} \rfloor} \Xi_{n,k} \sum_{{\bf a} \in \Sigma^{(n)}_{k}} \left[ \frac{1}{24} \zeta_1(1)^{5} \Omega({\bf a},1)^5
 - \frac{5}{12} \zeta_1(1)^{3} \zeta_1(2) \Omega({\bf a},1)^3 \Omega({\bf a},2) \right. \nl
&& \left. + \frac{5}{6} \zeta_1(1)^{2} \zeta_1(3) \Omega({\bf a},1)^2 \Omega({\bf a},3) + \frac{5}{8} \zeta_1(1) \zeta_1(2)^2 \Omega({\bf a},1) \Omega({\bf a},2)^2 - \frac{5}{4} \zeta_1(1) \zeta_1(4) \Omega({\bf a},1) \Omega({\bf a},4) \right. \nl
&& \left. - \frac{5}{6} \zeta_1(2)\zeta_1(3) \Omega({\bf a},2) \Omega({\bf a},3) + \zeta_1(5) \Omega({\bf a},5) \right], \label{eq:zeta_n5_zeta1} \\
 \vdots. && \nn 
\ee
\subee
If $\sum_k \Xi_{n,k}\sum_{\bf a} \Omega({\bf a},s)$ in the last terms of the r.h.s. is non-zero for all $s \in {\mathbb N}$, then the inverse ZGF, $\bm{\zeta}_1 = \bm{\zeta}_1(\bm{\zeta}_n)$, exist, and thus, the ESRs of $\bm{\zeta}_n$ are obtained by substituting the inverse ZGF into the ESRs of $\bm{\zeta}_1$.
In such a case, the selection rules of the resulting ESRs are the same to that of $\bm{\zeta}_1$, because the number of independent $\zeta_n(s)$ in $\bm{\zeta}_n$ are the same and can be expressed as the selection rule, as we discussed in Sec.~\ref{sec:SZF_M3}.
Thus, it should be ${\cal S}_n \cong {\mathbb Z}_{M+1}$ as far as its inverse ZGF exists.

However, the existence of the inverse ZGF depends on the values of $n$ and $M$, and it indeed does not exist when $(M,n) \in (2{\mathbb N} + 1) \times {\cal K}_{\rm od}$.
Let us see this fact by focusing on $\sum_{\bf a} \Omega({\bf a},s)$.
This part is evaluated as
\be
\sum_{{\bf a} \in \Sigma^{(n)}_{k}} \Omega({\bf a},s) =
\begin{dcases}
  \sum_{j=0}^{\frac{n-1}{2}} c_{j} \cos \left[ 4j s \mu \right]  & \text{for $n \in {\cal K}_{\rm od}$} \\
  \sum_{j=1}^{\frac{n}{2}} c_{j} \cos \left[ (4j-2) s \mu \right] & \text{for $n \in {\cal K}_{\rm ev}$} 
\end{dcases}, \label{eq:sum_Omega}
\ee
where the real coefficients, $c_j$, are determined as the number of $\pm 2 j$ (resp. $\pm (2j - 1)$) in ${\bf a}$ for $n \in {\cal K}_{\rm od}$ (resp. $n \in {\cal K}_{\rm ev}$). 
Eq.\eqref{eq:sum_Omega} with any $s \in {\mathbb N}$ can not be simultaneously zero for all $k$ when $(M,n) \in 2{\mathbb N} \times {\cal K}$ or $(2{\mathbb N}+1) \times {\cal K}_{\rm od}$, but it can vanish by taking certain $s \in {\mathbb N}$ when $(M,n) \in (2{\mathbb N}+1) \times {\cal K}_{\rm ev}$.
The (sufficient) condition for the vanishing $\sum_k \Xi_{n,k} \sum_{\bf a} \Omega({\bf a},s)$ can be found from $\cos [ (4j-2)s^* \mu] = 0$ and derived as
\footnote{%
More precisely, in Eq.~\eqref{eq:s_vanish} we focus on the possibility that the inner sum
$\sum_{{\bf a} \in \Sigma^{(n)}_k}\Omega({\bf a},s)$ vanishes for a fixed $k$.
In principle, one could also imagine cancellations where
$\sum_k \Xi_{n,k}\sum_{{\bf a} \in\Sigma^{(n)}_k}\Omega({\bf a},s)$ vanishes while each inner sum is non-zero.
One can check that this second possibility does not occur for $M\in\mathbb{N}+1$, but it can occur for $M\in\mathbb{N}+\tfrac12$.
This difference will be crucial in the discussion of Sec.~\ref{sec:half_M}.}

\be
s^* \in (M+1) \left( {\mathbb N}_0 + \frac{1}{2} \right) \ \ \text{for} \ \ M \in 2 {\mathbb N} + 1. \label{eq:s_vanish}
\ee
In addition, for such an $s=s^*$, one can also find that
\be
\sum_{{\bf a} \in \Sigma^{(n)}_{k }} \Omega({\bf a},s_1) \cdots \Omega({\bf a},s_\ell) \Omega({\bf a},s^*) = 0 \ \ \text{for} \ \ n \in {\cal K}_{\rm ev}, \ M \in 2{\mathbb N} + 1, \ \ \left( s_{t \in \{ 1,\cdots,\ell \} } \in {\mathbb N} \right) \label{eq:zero_omega_mult}
\ee
and thus, $\zeta_1(s^*)$ does not appear in $\zeta_{n}(s)$ with any $n \in {\cal K}_{\rm ev}$ and $s \in {\mathbb N}$.
These observations mean that, although $\bm{\zeta}_{n} = \bm{\zeta}_{n} (\bm{\zeta}_{1})$ can be always formulated for any $n \in {\cal K}$, $\zeta_{1}(s^*)$ does not relate to $\zeta_{n}(s^*)$ when $n \in {\cal K}_{\rm ev}$ and $M \in 2 {\mathbb N} + 1$.
Thus, $\bm{\zeta}_{n \in {\cal K}_{\rm od}} = \bm{\zeta}_{n}(\bm{\zeta}_{1})$ are invertible, but $\bm{\zeta}_{n \in {\cal K}_{\rm ev}} = \bm{\zeta}_n (\bm{\zeta}_1)$ are not.
In consequence, $\bm{\zeta}_n = \bm{\zeta}_n(\bm{\zeta}_{n^\prime})$ are invertible if $(n,n^\prime) \in {\cal K}_{\rm od} \times{\cal K}_{\rm od}$, but they are not if $(n,n^\prime) \in {\cal K}_{\rm ev} \times {\cal K}_{\rm od}$ for $M \in 2{\mathbb N} + 1$.
This situation was indeed observed in Sec.~\ref{sec:SZF_M3}.
The ESRs for $n \in {\cal K}_{\rm od}$ can be obtained from the inverse ZGF and consequently has the same selection rule, i.e.,
\be
   {\cal S}_{n} \cong {\mathbb Z}_{M+1} \ \ \text{for} \ \ n \in {\cal K}_{\rm od}, \ M \in 2{\mathbb N} + 1.
\ee
Notice that, when $M \in 2{\mathbb N}$, this non-invertibility does not happen for any $n,n^\prime \in {\cal K}$, and the selection rule of the ESRs is
\be
   {\cal S}_{n} \cong {\mathbb Z}_{M+1} \ \ \text{for} \ \ n \in {\cal K}, \ M \in 2 {\mathbb N}.
\ee

\, \\  \indent
The question is, for $M \in 2{\mathbb N} + 1$, if there are any other ZGFs as $\bm{\zeta}_{n \in {\cal K}_{\rm ev}} = \bm{\zeta}_{n}(\bm{\zeta}_{n^\prime})$ that has the invertibility.
We below assume $M \in 2{\mathbb N}+1$ and show that such a ZGF exists for $n^\prime \in {\cal K}_{\rm ev}$.
One can see this fact by multiplying $[-n+3,-n+5,\cdots,n-5,n-3]_1$ to $[0]_{n \in {\cal K}_{\rm ev}}$ in Eq.\eqref{eq:Cn_prods_C1_simp} and expressing it by $[k_1,\cdots,k_\ell]_{2}$ using Eq.\eqref{eq:O2} as\footnote{
In Eqs.\eqref{eq:[0]n_2}\eqref{eq:Phi_M+1_sym2}, we assume that $\prod_{\widehat{a} \in \widehat{\bf a}} ([\widehat{a}]_2 + 1) = 1$ if $\widehat{\bf a} = \emptyset$.}
\be
[0]_{n \in {\cal K}_{\rm ev}} &=& \sum_{k=0}^{\frac{n}{2}-1} (-1)^{k} \sum_{ \widehat{\bf a} \in \widehat{\Sigma}^{(n)}_{k}} \frac{\prod_{j=1}^{\frac{n}{2}} ([-n + 4 j - 2]_2 + 1)}{\prod_{\widehat{a} \in \widehat{\bf a}} ([\widehat{a}]_2 + 1)}  + (-1)^{\frac{n}{2}}, \label{eq:[0]n_2} \\
\widehat{\Sigma}^{(n)}_{k \in \{ 0,\cdots, \frac{n}{2} - 1 \}} &:=& \{ {\bf a} \subseteq {\bf A}_n \, | \, {\bf a} = \bigcup_{j=1}^{k} \{ \widehat{a}_{j} + 1 \} \nl
&& \text{with} \  \widehat{a}_j \in {\bf A}_n \ \text{and} \ \widehat{a}_{j_1} + 2 < \widehat{a}_{j_2} \le n - 3 \ \text{for any} \ j_1 < j_2 \}. \label{eq:def_hatSigmaM}
\ee
Here, the family of sets, $\widehat{\Sigma}^{(n)}_{k}$, is defined from the intermediate values of the pairs, $\{ \widehat{a}_j, \widehat{a}_j + 2 \}$, in the definition of $\Sigma_{k}^{(n)}$ in Eq.\eqref{eq:def_SigmaM}.
See also Fig.~\ref{fig:Sigma}.
For example, for $n=4,6,8$, one can find
\subbe
\be
\left[ 0 \right]_4 &=& \frac{1}{[0]_2 + 1} \left( [-2,0,2]_2 - [-2]_2  - [2]_2 - 1 \right), \label{eq:O4_2} \\ 
\left[ 0 \right]_6 &=& \frac{1}{([-2]_2 + 1)([2]_2 + 1)} \left(  [-4,-2,0,2,4]_2 - [-4,-2,0]_2 - [-4,-2,4]_2 - [-4,2,4]_2 - [0,2,4]_2 + [0]_2 \right. \nl
&& \left. - [-4,-2]_2 - [-4,4]_2 - [2,4]_2 + 1 \right), \label{eq:O6_2} \\
\left[ 0 \right]_8 &=& \frac{1}{([-4]_2 + 1)([0]_2 + 1)([4]_2 + 1)}
\left(  [-6,-4,-2,0,2,4,6]_2 - [-6,-4,-2,0,2]_2 - [-6,-4,-2,0,6]_2  \right. \nl
&& \left. - [-6,-4,-2,4,6]_2 - [-6,-4,2,4,6]_2 - [-6,0,2,4,6]_2 - [-2,0,2,4,6]_2+ [-6,-4,2]_2 + [-6,0,2]_2 \right. \nl
&& \left. + [-6,0,6]_2 + [-2,0,2]_2 + [-2,0,6]_2 + [-2,4,6]_2 + [-6]_2 + [0]_2 + [6]_2 \right. \nl
&& \left. - [-6,-4,-2,0]_2 - [-6,-4,-2,6]_2 - [-6,-4,4,6]_2 - [-6,2,4,6]_2 - [0,2,4,6]_2 \right. \nl
&& \left. + [-6,-4]_2 + [-6,0]_2 + [-6,2]_2 + [-6,6]_2 + [-2,0]_2 + [-2,6]_2 + [0,2]_2 + [0,6]_2 + [4,6]_2 \right). \label{eq:O8_2}
\ee
\subee
By expanding $[k_1,\cdots,k_\ell]_n$ using Eq.\eqref{eq:[a_1]_expand_zeta}, one can find the expression of ZGFs for $n \in {\cal K}_{\rm ev}$, i.e., $\bm{\zeta}_{n \in {\cal K}_{\rm ev}} = \bm{\zeta}_{n}(\bm{\zeta}_{2})$.
The remarkable fact is that the resulting ZGFs, $\bm{\zeta}_{n \in {\cal K}_{\rm ev}} = \bm{\zeta}_{n}(\bm{\zeta}_{2})$, are invertible.
For example, $\bm{\zeta}_4 = \bm{\zeta}_4(\bm{\zeta}_2)$ for $M=5$ is given by
\subbe
\be
\zeta_{{\cal PT}_{4}}(1) &=& 2 \zeta_{{\cal PT}_{2}}(1), \label{eq:zeta24(1)_M5} \\
\zeta_{{\cal PT}_{4}}(2) &=& (12 - 6 \sqrt{3}) \zeta_{{\cal PT}_{2}}(1)^2 - (4 - 2 \sqrt{3}) \zeta_{{\cal PT}_{2}}(2), \\
\zeta_{{\cal PT}_{4}}(3) &=& \frac{21 \sqrt{3} - 33}{2} \zeta_{{\cal PT}_{2}}(1)^3 + \frac{27 - 15 \sqrt{3}}{2} \zeta_{{\cal PT}_{2}}(1) \zeta_{{\cal PT}_{2}}(2) - (7 - 3 \sqrt{3}) \zeta_{{\cal PT}_{2}}(3), \\
\zeta_{{\cal PT}_{4}}(4) &=& (246 - 141 \sqrt{3}) \zeta_{{\cal PT}_{2}}(1)^4 - (156 - 90 \sqrt{3})\zeta_{{\cal PT}_{2}}(1)^2 \zeta_{{\cal PT}_{2}}(2) + (26 - 15\sqrt{3}) \zeta_{{\cal PT}_{2}}(2)^2 \nl
&& - (4-2 \sqrt{3}) \zeta_{{\cal PT}_{2}}(4), \\
\zeta_{{\cal PT}_{4}}(5) &=& \frac{1780 \sqrt{3} - 3075}{4}\zeta_{{\cal PT}_{2}}(1)^5
- \frac{2540 \sqrt{3} - 4395}{6} \zeta_{{\cal PT}_{2}}(1)^3 \zeta_{{\cal PT}_{2}}(2) \nl
&&  - (165 - 95 \sqrt{3}) \zeta_{{\cal PT}_{2}}(1)^2 \zeta_{{\cal PT}_{2}}(3)
- \frac{625 - 360 \sqrt{3}}{4} \zeta_{{\cal PT}_{2}}(1) \zeta_{{\cal PT}_{2}}(2)^2 \nl
&&+ \frac{10 \sqrt{3} - 15}{2} \zeta_{{\cal PT}_{2}}(1) \zeta_{{\cal PT}_{2}}(4) + \frac{165 - 95 \sqrt{3}}{3} \zeta_{{\cal PT}_{2}}(2) \zeta_{{\cal PT}_{2}}(3) + 2 \zeta_{{\cal PT}_{2}}(5), \label{eq:zeta24(5)_M5} \\
\zeta_{{\cal PT}_{4}}(6) &=&  \frac{213999 - 123499 \sqrt{3}}{40} \zeta_{{\cal PT}_{2}}(1)^6 - \frac{42405 - 24477 \sqrt{3}}{8} \zeta_{{\cal PT}_{2}}(1)^4 \zeta_{{\cal PT}_{2}}(2) \nl
&& - (217 \sqrt{3} - 375) \zeta_{{\cal PT}_{2}}(1)^3 \zeta_{{\cal PT}_{2}}(3) - \frac{7731 \sqrt{3} - 13383}{8} \zeta_{{\cal PT}_{2}}(1)^2 \zeta_{{\cal PT}_{2}}(2)^2 \nl
&& + \frac{207 \sqrt{3} - 351}{4} \zeta_{{\cal PT}_{2}}(1)^2 \zeta_{{\cal PT}_{2}}(4) + (123 \sqrt{3} - 213) \zeta_{{\cal PT}_{2}}(1) \zeta_{{\cal PT}_{2}}(2) \zeta_{{\cal PT}_{2}}(3) \nl
&& - \frac{18 \sqrt{3} - 18}{5} \zeta_{{\cal PT}_{2}}(1) \zeta_{{\cal PT}_{2}}(5) - \frac{1173 - 677 \sqrt{3}}{8} \zeta_{{\cal PT}_{2}}(2)^3 \nl
&& + \frac{111 + 63 \sqrt{3}}{4} \zeta_{{\cal PT}_{2}}(2) \zeta_{{\cal PT}_{2}}(4) + (33 - 19 \sqrt{3}) \zeta_{{\cal PT}_{2}}(3)^2 + (5 -\sqrt{3} ) \zeta_{{\cal PT}_{2}}(6), \label{eq:zeta24(6)_M5} \\
\vdots, && \nn
\ee
\subee
and one can indeed see that $\zeta_{{\cal PT}_4}(s)$ contains the linear term of $\zeta_{{\cal PT}_2}(s)$ for any $s \in {\mathbb N}$.
Therefore, the ZGFs with any $n,n^\prime \in {\cal K}_{\rm ev}$, i.e., $\bm{\zeta}_{n \in {\cal K}_{\rm ev}} = \bm{\zeta}_{n} (\bm{\zeta}_{n^\prime \in {\cal K}_{\rm ev}})$, are constructable from the inverse ZGFs, $\bm{\zeta}_{2} = \bm{\zeta}_{2} (\bm{\zeta}_{n^\prime \in {\cal K}_{\rm ev}})$.


In order to more understand the invertibility of the ZGFs for $(M,n) \in (2{\mathbb N}+1)\times {\cal K}_{\rm ev}$, we look to their selection rule.
Again, we assume that $M \in 2{\mathbb N}+1$.
Once one finds the expression of $[0]_{n \in {\cal K}_{\rm ev}}$ using $[k_\ell]_2$ in Eq.\eqref{eq:[0]n_2}, in principle the ESRs of $\bm{\zeta}_2$ can be obtained by taking the condition in Eq.\eqref{eq:k0k2M}.
However, since these computations are quite complicated, we formulate it by employing $\Phi^{(1)}_{2M+2} = 0$ in Eq.\eqref{eq:Phi_M+1_sym}.
We define $\Phi_{2M+2}^{(2)}$ from $\Phi_{2M+2}^{(1)}$ such that
\be
\Phi_{2M+2}^{(2)}([k_\ell]_{2} = [k_\ell-1,k_\ell + 1]_1 - 1) := \Phi_{2M+2}^{(1)}([k_\ell]_{1}) = 0, \qquad (M \in 2{\mathbb N}+1) \label{eq:Phi_M+1_sym2_cond}
\ee 
and taking account of the periodicity of $\omega$, i.e., $[k_\ell + 2 M + 2]_n = [k_\ell]_n$, one can find $\Phi^{(2)}_{2M+2}$ as
\be
\Phi^{(2)}_{2M+2} &=&
\sum_{k=0}^{\frac{M-1}{2}} (-1)^{k} \sum_{\widehat{{\bf a}} \in  \widehat{{\Sigma}}_{k}^{(M+1)}|_{\bmod (2 M + 2)}} 
\frac{\prod_{j=0}^{\frac{M-1}{2}} ([-M- 1 + 4 j]_2 + 1)}{\prod_{\widehat{a} \in \widehat{\bf a}} ([\widehat{a}]_2 + 1)} + (-1)^{\frac{M+1}{2}} 2, \label{eq:Phi_M+1_sym2} \\
\widehat{{\Sigma}}^{(M+1)}_{k \in \{ 0,\cdots,\frac{M-1}{2} \}}|_{\bmod (2 M + 2)} &:=& \{ {\bf a} \subseteq {\bf A}_{M+1} \, | \, {\bf a} = \bigcup_{j=1}^{k} 
\{ (\widehat{a}_{j} + 1) \bmod (2 M + 2) \}
\nl
&& \text{with} \  \widehat{a}_j \in {\bf A}_{M+1}, \ |\omega^{\widehat{a}_{j_1} - \widehat{a}_{j_2}}| > |\omega^{2}| \ \text{and} \ \widehat{a}_{j_1}  < \widehat{a}_{j_2} \ \text{for any} \ j_1 < j_2 \}.
\label{eq:def_hattil_SigmaM_2}
\ee
Here, similar to ${\Sigma}^{(n)}_{k}|_{\bmod (2 M + 2)}$ in Eq.\eqref{eq:def_hatSigmaM}, the family of sets, $\widehat{\Sigma}^{(M+1)}_{k}|_{\bmod (2 M + 2)}$, is defined from the intermediate values of the pairs, $\{ \widehat{a}_j, (\widehat{a}_j + 2) \bmod (2 M + 2) \}$, in the definition of ${\Sigma}_{k}^{(M+1)}|_{\bmod (2 M + 2)}$ in Eq.\eqref{eq:def_til_SigmaM_2}.
See also Fig.~\ref{fig:Sigma_til}.
The specific forms for $M=3,5,7$ can be written down as\footnote{
To make the forms manifest for the invariance of the ${\mathbb Z}_{\frac{M+1}{2}}$ shift,  Eqs.\eqref{eq:Phi4_C2}-\eqref{eq:Phi8_C2} have been rewritten by slight modification to the denominators in Eq.\eqref{eq:Phi_M+1_sym2} using
\be
\prod_{j=0}^{\frac{M-1}{2}} ([-M-1 + 4 j]_2 + 1) = \prod_{j=0}^{\frac{M-1}{2}} ([-M + 1 + 4 j]_2 + 1). \qquad (M \in 2{\mathbb N} + 1)
\ee
}
\subbe
\be
\Phi^{(2)}_{6+2} &=& [-4,0]_2 - [-2]_2 - [2]_2 -1, \label{eq:Phi4_C2} \\
\Phi^{(2)}_{10+2} &=& [-6,-2,2]_2 - [-4,0]_2 - [-4,4]_2 - [0,4]_2 - [-4]_2 - [0]_2 - [4]_2 \nl
&& + \frac{([-6]_2 + 1)([-2]_2 + 1)}{[-4]_2 + 1} + \frac{([-6]_2 + 1)([2]_2 + 1)}{[4]_2 + 1} + \frac{([-2]_2 + 1)([2]_2 + 1)}{[0]_2 + 1} - 1, \label{eq:Phi6_C2} \\
\Phi^{(2)}_{14+2} &=& [-8,-4,0,4]_2 +[-6,2]_2 +[-2,6]_2 \nl
&& - \frac{[0,4]_2 ([-8]_2 + 1)([-4]_2 + 1)}{[-6]_2 + 1}  - \frac{[-4,0]_2 ([-8]_2 + 1)([4]_2 + 1)}{[6]_2 + 1} \nl
&& - \frac{[-8,4]_2([-4]_2 + 1)([0]_2 + 1)}{[-2]_2+1} - \frac{[-8,-4]_2([0]_2 + 1)([4]_2 + 1)}{[2]_2+1} \nl
&& +\frac{[4]_2([-8]_2+1)([-4]_2+1)([0]_2+1)}{([-6]_2 + 1)([-2]_2 + 1)} + \frac{[-4]_2([-8]_2+1)([0]_2+1)([4]_2+1)}{([2]_2 + 1)([6]_2 + 1)} \nl
&& + \frac{[0]_2([-8]_2+1)([-4]_2+1)([4]_2+1)}{([-6]_2 + 1)([6]_2 + 1)} + \frac{[-8]_2([-4]_2+1)([0]_2+1)([4]_2+1)}{([-2]_2 + 1)([2]_2 + 1)} - 1. \label{eq:Phi8_C2}
\ee
\subee
These forms are invariant under the ${\mathbb Z}_{\frac{M+1}{2}}$ shift, $[k_1,\cdots,k_\ell]_2 \mapsto [k_1+c,\cdots,k_{\ell} + c]_2$ with $c \in \{ 0, 4,\cdots,2M-6, 2M-2 \}$. and thus, these symmetric structures are ${\mathbb Z}_{\frac{M+1}{2}} (\cong {\mathbb Z}_{2 M + 2}/\langle 4 \rangle)$.
Meanwhile, by specific computations for Eqs.\eqref{eq:[a_1]_expand_zeta}\eqref{eq:Phi_M+1_sym2} or from the discussion below Eq.\eqref{eq:def_Omega}, one can obtain the selection rule of $\bm{\zeta}_2$ as ${\cal S}_2 \cong {\mathbb Z}_{\frac{M+1}{2}}$.
Since the ZGFs, $\bm{\zeta}_{n \in {\cal K}_{\rm ev}} = \bm{\zeta}_{n}(\bm{\zeta}_{2})$, are invertible, the selection rules of $\bm{\zeta}_{n \in {\cal K}_{\rm ev}}$ are given by
\be
 {\cal S}_{n} \cong {\mathbb Z}_{\frac{M+1}{2}} \ \ \text{for} \ \  n \in {\cal K}_{\rm ev}, \ M \in 2{\mathbb N} + 1.
\ee
For $M=5$, the first two ESRs of $\bm{\zeta}_2$ can be obtained from Eqs.\eqref{eq:[a_1]_expand_zeta}\eqref{eq:Phi6_C2} as
\subbe
\be
&& (9 + 5 \sqrt{3}) \zeta_{{\cal PT}_2}(1)^{3} + (57 + 33 \sqrt{3}) \zeta_{{\cal PT}_2}(1) \zeta_{{\cal PT}_2}(2) - (142 + 82 \sqrt{3}) \zeta_{{\cal PT}_2}(3) = 0, \\
&& \frac{509 + 297 \sqrt{3}}{40} \zeta_{{\cal PT}_2}(1)^{6} + \frac{3505 + 2021 \sqrt{3}}{8} \zeta_{{\cal PT}_2}(1)^{4} \zeta_{{\cal PT}_2}(2) + \frac{1109 + 641 \sqrt{3}}{3} \zeta_{{\cal PT}_2}(1)^3 \zeta_{{\cal PT}_2}(3) \nl
&& \quad + \frac{3381 + 1953 \sqrt{3}}{8} \zeta_{{\cal PT}_2}(1)^{2} \zeta_{{\cal PT}_2}(2)^2 - \frac{3549 + 2049 \sqrt{3}}{4} \zeta_{{\cal PT}_2}(1)^2 \zeta_{{\cal PT}_2}(4) \nl
&& \quad + (47 + 27 \sqrt{3}) \zeta_{{\cal PT}_2}(1) \zeta_{{\cal PT}_2}(2) \zeta_{{\cal PT}_2}(3) - \frac{10278 + 5934 \sqrt{3}}{5} \zeta_{{\cal PT}_2}(1) \zeta_{{\cal PT}_2}(5) + \frac{161 + 93 \sqrt{3}}{8} \zeta_{{\cal PT}_2}(2)^3 \nl
&& \quad - \frac{795 + 459 \sqrt{3}}{4} \zeta_{{\cal PT}_2}(2) \zeta_{{\cal PT}_2}(4) - \frac{227 + 131 \sqrt{3}}{3} \zeta_{{\cal PT}_2}(3)^2 - (989 + 571 \sqrt{3}) \zeta_{{\cal PT}_2}(6) = 0, \\ 
\vdots. && \nn
\ee
\subee
Substituting Eqs.\eqref{eq:zeta24(1)_M5}-\eqref{eq:zeta24(6)_M5} into these results lead to the ESRs of $\bm{\zeta}_4$ as
\subbe
\be
&& \frac{201 + 116 \sqrt{3}}{22} \zeta_{{\cal PT}_4}(1)^{3} + \frac{348 + 201 \sqrt{3}}{22} \zeta_{{\cal PT}_4}(1) \zeta_{{\cal PT}_4}(2) - \frac{866 + 500 \sqrt{3}}{11} \zeta_{{\cal PT}_4}(3) = 0, \\
&& \frac{186258 + 107537 \sqrt{3}}{9680} \zeta_{{\cal PT}_4}(1)^{6} + \frac{147883 + 85380 \sqrt{3}}{1936} \zeta_{{\cal PT}_4}(1)^{4} \zeta_{{\cal PT}_4}(2) - \frac{5969 + 3446 \sqrt{3}}{726} \zeta_{{\cal PT}_4}(1)^3 \zeta_{{\cal PT}_4}(3) \nl
&& \quad + \frac{4848 + 2799 \sqrt{3}}{1936} \zeta_{{\cal PT}_4}(1)^{2} \zeta_{{\cal PT}_4}(2)^2 - \frac{9987 + 5766 \sqrt{3}}{88} \zeta_{{\cal PT}_4}(1)^2 \zeta_{{\cal PT}_4}(4) \nl
&& \quad + \frac{16013 + 9245 \sqrt{3}}{242} \zeta_{{\cal PT}_4}(1) \zeta_{{\cal PT}_4}(2) \zeta_{{\cal PT}_4}(3) - \frac{17298 + 9987 \sqrt{3}}{55} \zeta_{{\cal PT}_4}(1) \zeta_{{\cal PT}_4}(5) - \frac{3329 + 1922 \sqrt{3}}{176} \zeta_{{\cal PT}_4}(2)^3 \nl
&& \quad - \frac{17298 + 9987 \sqrt{3}}{88} \zeta_{{\cal PT}_4}(2) \zeta_{{\cal PT}_4}(4) - \frac{5366 + 3098 \sqrt{3}}{363} \zeta_{{\cal PT}_4}(3)^2 - \frac{3329 + 1922 \sqrt{3}}{11} \zeta_{{\cal PT}_4}(6) = 0, \\
\vdots. && \nn
\ee
\subee

Remind that $\bm{\zeta}_n = \bm{\zeta}_n(\bm{\zeta}_{n^\prime})$ with $(n,n^\prime)\in {\cal K}_{\rm od} \times{\cal K}_{\rm ev}$ do not exist for $M \in 2{\mathbb N} + 1$.
This can be also seen from the selection rules as $ {\mathbb Z}_{M+1} \not\subset {\mathbb Z}_{\frac{M+1}{2}}$, which means that $\bm{\zeta}_{n \in {\cal K}_{\rm ev}}$ is too less information to reproduce $\bm{\zeta}_{n \in {\cal K}_{\rm od}}$ because the smaller order of the selection rule implies stricter constraints over $\bm{\zeta}_n$.

\, \\ \indent
Let us briefly summarize our results in this part.
Taking the procedure shown in Fig.~\ref{fig:procedure}, we have formulated the ESRs and the ZGFs for $M \in {\mathbb N} + 1$ by beginning with the fusion relations \eqref{eq:iden[k]1}\eqref{eq:iden[k]2}.
The structure of the ESRs and the ZGFs in general depends on $M$ and $n$ and can be characterized by selection rules, which are some ${\mathbb Z}_N$ group originated from the ${\mathbb Z}_{2M + 2}$ Symanzik rotation.
The ESRs, $\zeta_n(s) = \zeta_n(\{ \zeta_n(s^\prime) \}_{1 \le s^\prime < s})$, are constructable for $s$ such that
\be
&& s \in G = N {\mathbb N}_0, \qquad (s \in {\mathbb N}_0) \nl
&& N = 
\begin{dcases}
  M+1 
  & \ \  \text{for \ \ $(M,n) \in 2{\mathbb N} \times{\cal K}$ \ or \ $(2{\mathbb N} + 1) \times{{\cal K}_{\rm od}}$}  \\
  \frac{M+1}{2}   & \ \  \text{for \ \ $(M,n) \in  (2{\mathbb N} + 1) \times{{\cal K}_{\rm ev}}$}
\end{dcases}. \label{eq:s_mod_N_0}
\ee
The selection rules are given by the quotient group as ${\cal S}_n = {\mathbb N}_0/(N {\mathbb N}_0) \cong {\mathbb Z}_N  \subset {\mathbb Z}_{2M + 2}$, which coincides with the shift symmetry of $\Phi^{(1,2)}_{2M + 2}$ in Eqs.\eqref{eq:Phi_M+1_sym}\eqref{eq:Phi_M+1_sym2_cond}.
For $M \in 2{\mathbb N}$, the ZGFs, $\bm{\zeta}_{n} = \bm{\zeta}_{n}(\bm{\zeta}_{n^\prime})$, can be constructed for any $n,n^\prime \in {\cal K}$ and are all invertible.
For $M \in 2{\mathbb N} + 1$, although the ZGFs are constructable for any $(n,n^\prime) \in {\cal K} \times {\cal K}_{\rm od}$, those are invertible when $n \in {\cal K}_{\rm od}$ and one way-mappings when $n \in {\cal K}_{\rm ev}$.
The ZGFs for $(n,n^\prime) \in {\cal K}_{\rm ev} \times {\cal K}_{\rm ev}$ constitute of another closed family equipping the invertibility.
We summarize these observations as the following proposition:
\begin{proposition}[Structural non-invertibility]
For $M \in 2\mathbb{N}+1$, the ZGF mapping
${\cal Z}_{n \rightarrow n^\prime} : \bm{\zeta}_{n} \mapsto \bm{\zeta}_{n^\prime}$ with $n  \in {\cal K}_{\rm od}$ and $n^\prime \in {\cal K}_{\rm ev}$
is structurally non-invertible.
\end{proposition}
\begin{proof}
The ZGF relates the SZFs in different fusion sectors
via coefficients built from the sums
$\Sigma_\Omega(s) = \sum_{\bf } \Omega({\bf a},s)$.
As derived in Eqs.\eqref{eq:sum_Omega}-\eqref{eq:zero_omega_mult}, for odd $M$ there exists an integer
$s^* \in (M+1)(\mathbb{N}_0+\tfrac{1}{2})$,
belonging to the\textit{null} class singled out by the selection rule, such that
\be
  \Sigma_{\Omega}(s^*) = 0, \nn
\ee
and, more generally, any coefficient that contains $\Omega({\bf a},s^*)$ vanishes identically.
Therefore all terms proportional to $\zeta_{n}(s^*)$  are eliminated from the ZGF, i.e., the even–sector data $\zeta_{n^\prime}$ are completely independent of the value of $\zeta_{n}(s^*)$.
Equivalently, at fixed fusion label the map
${\cal Z}_{n \rightarrow n^\prime} :\bm{\zeta}_{n}\mapsto \bm{\zeta}_{n^\prime}$, viewed on the sectorwise SZF spaces reduced by the ESRs, has a non-trivial kernel generated by the \textit{null} zeta moments singled out by the selection rule ${\cal S}_{n,n^\prime}$. In this sense ${\cal S}_{n,n^\prime}$ characterizes, for a given local fusion sector, which integer zeta moments are algebraically invisible under the odd$\to$even ZGF.
In particular, two distinct odd–sector datasets that differ only in $\zeta_{n}(s^*)$ are mapped to the same even–sector data, so ${\cal Z}_{n \rightarrow n^\prime}$ is not injective and hence not invertible.
Consequently, the spectral data in the even sector cannot uniquely reconstruct the odd sector (a spectral no-go theorem).
\end{proof} 
\noindent\textit{Remark.}
The above non-invertibility is purely algebraic, i.e., it uses only the $A_{2M-1}$ fusion
relations and the $\mathbb{Z}_{2M+2}$ Symanzik rotation phases entering the sums
$\Sigma_\Omega(s)$, and is therefore insensitive to further details of the potential. In particular, any ODE/IM realization with the same T-system and Symanzik rotation exhibits the same odd/even information-loss pattern at the level of SZFs.
\\ \, \\ \noindent 
This general statement matches what we already observed explicitly in the $M=3$ example of Sec.~\ref{sec:SZF_M3}.
\\ \, \\ \indent
These results can be schematically depicted as
\be
\begin{aligned}
& \text{For $M \in 2{\mathbb N}$}: & & \text{For $M \in 2{\mathbb N} + 1$}:  \\
& \large
  \begin{tikzcd} 
    \bm{\zeta}_{n \in {\cal K}} \arrow[leftrightarrow]{rrr}{{\cal S}_n \cong {\mathbb Z}_{M+1}}[swap]{\bm{\zeta}_{n} = \bm{\zeta}_{n} (\bm{\zeta}_{n^\prime})} & & & \bm{\zeta}_{n^{\prime} \in {\cal K}}
\end{tikzcd} & \qquad \qquad &
\large  
\begin{tikzcd}
  \bm{\zeta}_{n \in {\cal K}_{\rm od}} \arrow[leftrightarrow]{rrr}{{\cal S}_n \cong {\mathbb Z}_{M+1}}[swap]{\bm{\zeta}_{n} = \bm{\zeta}_{n}(\bm{\zeta}_{n^\prime})} & & & \bm{\zeta}_{n^{\prime} \in {\cal K}_{\rm od}} \\ \\
  \bm{\zeta}_{\widetilde{n} \in {\cal K}_{\rm ev}} \arrow[leftrightarrow]{rrr}{{\cal S}_{\widetilde{n}} \cong {\mathbb Z}_{\frac{M+1}{2}}}[swap]{\bm{\zeta}_{\widetilde{n}} = \bm{\zeta}_{\widetilde{n}}(\bm{\zeta}_{\widetilde{n}^\prime})} \arrow[Leftarrow]{uu}{\bullet/\langle 2 \rangle}[swap]{\bm{\zeta}_{\widetilde{n}} = \bm{\zeta}_{\widetilde{n}}(\bm{\zeta}_{n})} & & & \bm{\zeta}_{\widetilde{n}^{\prime} \in {\cal K}_{\rm ev}}
\end{tikzcd} 
\end{aligned} \label{eq:sch_fig_Mint} 
\ee
where the arrows with ``$\bm{\zeta}_{n_1} = \bm{\zeta}_{n_1}(\bm{\zeta}_{n_2})$'' denote the existence of the ZGFs from $\bm{\zeta}_{n_2}$ to $\bm{\zeta}_{n_1}$.
Since the ${\cal PT}_K$ and Hermitian SZFs corresponds to the label with $n = K < M$ and $n = M$, respectively, one can always access the ${\cal PT}_K$ SZFs from the Hermitian SZFs, but the opposite is not always true.

\subsection{$M \in {\mathbb N}+\frac{1}{2}$} \label{sec:half_M}

\begin{figure}[t]
\centering
\includegraphics[clip,width=0.65\textwidth]{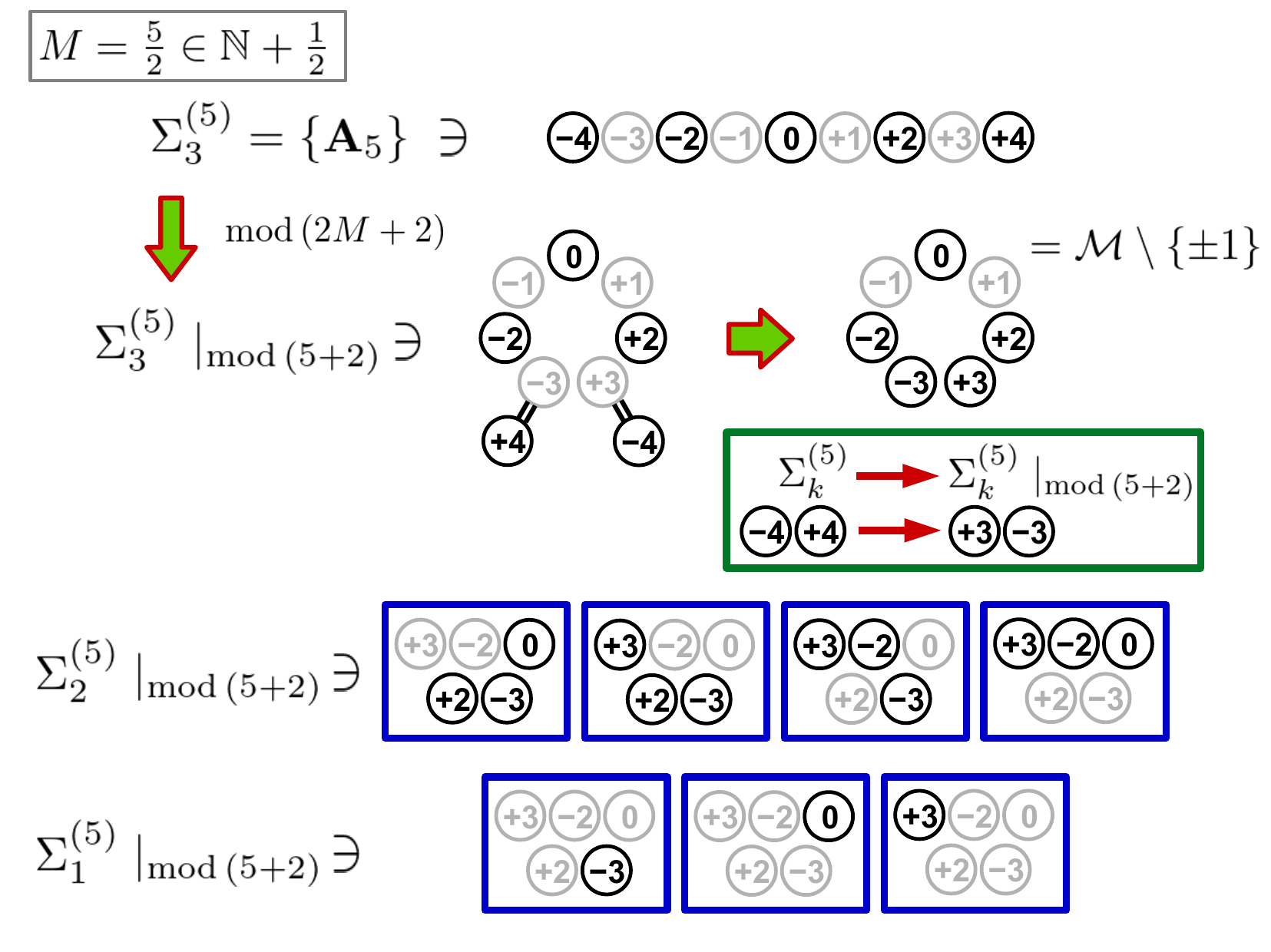}
\caption{The construction of $\Sigma^{(2M)}_{k}\mid_{\bmod (2M+2)}$ from $\Sigma^{(2M)}_{k}$ for $M=\frac{5}{2}$.
  After finding $\Sigma^{(2M)}_{k}$ for all $k$, imposing the ${\mathbb Z}_{2M+2}$ modulo identification to elements of the sets in the family yields $\Sigma^{(2M)}_{k}\mid_{\bmod (2M+2)}$.
  In the $M = \frac{5}{2}$ case, the elements, $\{\mp 4 \}$, in the sets are identified as $\{ \mp 4 \} \sim \{\pm 3\}$.
  The sets in the family $\Sigma_k^{(2M)} \mid_{2M+2}$ are in general subsets of ${\cal M} \setminus \{ \pm 1\}$. 
See also the left panel of Fig.~\ref{fig:Sigma} for comparison.
}
\label{fig:Sigma_mod}
\end{figure}

We consider the $M \in {\mathbb N} + \frac{1}{2}$ case.
The procedure for formulating the ESRs and the ZGFs is the same as the $M \in {\mathbb N}+1$ case, as shown in Fig.~\ref{fig:procedure}.
For this reason, we mainly describe remarkable differences from the $M \in {\mathbb N}+1$ case.

The essential difference can be readily seen in the process to find the ESRs.
In contrast to Eq.\eqref{eq:gen_iden_PT_pre}, Eq.\eqref{eq:k0k2M} for $M \in {\mathbb N} + \frac{1}{2}$ is unfactorizable.
As a result, the ESRs of $\bm{\zeta}_1$ for $M \in {\mathbb N} + \frac{1}{2}$, i.e., $\Phi^{(1)}_{2M+2} = 0$, can be directly derived from Eq.\eqref{eq:Cn_prods_C1_simp} as
\be
\Phi^{(1)}_{2M+2} :=
\sum_{k=1}^{M + \frac{1}{2}} (-1)^{M + \frac{1}{2}-k} \sum_{\{ a_1,\cdots,a_{2k-1}\} \in \Sigma^{(2M)}_{k} \mid_{\bmod (2M + 2)}} [a_1,\cdots,a_{2k-1}]_1 - 1 \ \ \text{for} \ \ M \in {\mathbb N} + \frac{1}{2}, \label{eq:Phi_M+1_halfM}
\ee
where $\Sigma^{(2M)}_{k} \mid_{\bmod (2M + 2)}$ denotes a family of sets generated from Eq.\eqref{eq:def_SigmaM} by imposing the ${\mathbb Z}_{2M+2}$ modulo identification, as
\be
   {\Sigma}^{(2M)}_{k \in \{ 1, \cdots,  M + \frac{1}{2} \} } \mid_{\bmod (2M+2)} &:=& \{ {\bf a}  \subseteq {\cal M} \setminus \{ \pm 1 \}\, | \, {\bf a} = \bigcup_{\widehat{a} \in \widehat{\bf a}} \{ \widehat{a} \ {\rm mod} \ (2M+2) \}  \ \text{with} \ \widehat{\bf a} \in {\Sigma}^{(2M)}_{k} \}, 
   \label{eq:def_Sigma_M_mod}
\ee
and ${\cal M}$ is given in Eq.\eqref{eq:def_set_M}.
Fig.~\ref{fig:Sigma_mod} shows the construction of $\Sigma^{(2M)}_{k} \mid_{\bmod (2M+2)}$ for $M=\frac{5}{2}$.
For $M=\frac{3}{2}, \frac{5}{2}$, and $\frac{7}{2}$, $\Phi^{(1)}_{2 M + 2}$ can be specifically expressed by
\subbe
\be
\Phi^{(1)}_{3 + 2} &=& [-2,0,2]_1 - [-2]_1  - [2]_1 - 1, \label{eq:Phi3/2} \\
 \Phi^{(1)}_{5 + 2} &=&  [-3,-2,0,2,3]_1 - [-3,-2,3]_1 - [-3,2,3]_1 - [-3,0,2]_1 - [-2,0,3]_1 + [-3]_1 + [0]_1 + [3]_1 - 1, \label{eq:Phi5/2} \\
 \Phi^{(1)}_{7 + 2} &=&  [-4,-3,-2,0,2,3,4]_1 - [-4,-3,-2,0,3]_1 - [-4,-3,2,3,4]_1
 - [-4,-3,-2,3,4]_1  \nl
 && - [-4,-2,0,2,3]_1  - [-3,-2,0,2,4]_1 - [-3,0,2,3,4]_1 + [-4,-3,3]_1 + [-4,-2,3]_1 + [-4,2,3]_1 \nl
 && + [-3,-2,0]_1 + [-3,-2,4]_1  + [-3,0,3]_1 + [-3,2,4]_1 + [-3,3,4]_1  + [-2,0,2]_1 + [0,2,3]_1   \nl
 &&  -[-3]_1 -[-2]_1 - [2]_1 -[3]_1 - 1. \label{eq:Phi7/2} 
\ee
\subee
One can see that  Eqs.\eqref{eq:Phi3/2}-\eqref{eq:Phi7/2} are invariant only under the trivial shift, i.e., $[k_1,\cdots,k_\ell]_1 \mapsto [k_1 + c,\cdots,k_\ell+c]_1$ with $c \in \{ 0 \} \subset {\mathbb Z}_{2M + 2}$.
Hence, the ${\mathbb Z}_N$ shift symmetry is broken for $M \in {\mathbb N} + \frac{1}{2}$.
Notice that unfactorizability of Eq.\eqref{eq:k0k2M} for $M \in {\mathbb N} + \frac{1}{2}$ can be also seen from the mapping to the Chebyshev polynomials in Eq.\eqref{eq:prod_form_Cheby} works when $M \in {\mathbb N} + 1$ but does not when $M \in {\mathbb N} + \frac{1}{2}$.

The situation for the selection rule is more complicated than that of the $M \in {\mathbb N} + 1$ case, and another feature arises by including the contribution from $C^{(n \in {\cal K})}(0) = e^{- \zeta_n^\prime(0)}$, which is implicitly related to the structure of the Chebyshev polynomials.
After finding $C^{(n \in {\cal K})}(0) = e^{-\zeta^{\prime}_n(0)}$ as Eq.\eqref{eq:sol_of_Cn0}, one can construct the ESRs by expanding $[k_1,\cdots,k_\ell]_1$ as Eq.\eqref{eq:[a_1]_expand_zeta}.
Similar to Eq.\eqref{eq:Cn_prods_C1_simp_zeta2}, the expanded forms contain $\sum_{k} \Xi_{n,k} \sum_{{\bf a}} \Omega({\bf a},s)$, and from this part the condition to $s \in {\mathbb N}$ for yielding independent ESRs is obtained as
\be
\sum_{k=1}^{M + \frac{1}{2}} (-1)^{k} e^{ - 2k \zeta_1^{\prime}(0)} \sum_{{\bf a} \in \Sigma^{(2M)}_{k} \mid_{\bmod {(2M + 2)}}} \Omega({\bf a},s) \ne 0 \quad \text{with} \quad e^{ - \zeta_1^{\prime}(0)} = 2 \cos \mu, \label{eq:Phi_M+1_halfM_zero_zeta}
\ee
which means that the linear term of $\zeta_1(s)$ exists in $O(E^s)$.
By expressing $e^{-2 k \zeta_1^\prime(0)}$ by $\omega$ as
\be
e^{-2 k \zeta_1^\prime(0)} &=& (2 \cos \mu)^{2k} =
\begin{pmatrix}
  2k \\
  k
\end{pmatrix}
+ \sum_{\ell = 0}^{k-1}
\begin{pmatrix}
  2k \\
  \ell
\end{pmatrix}  ( \omega^{k-\ell} + \omega^{-k + \ell}),
\label{eq:zeta1_omega}
\ee
one can find that Eq.\eqref{eq:Phi_M+1_halfM_zero_zeta} is satisfied when $s$ holds
\be
&& s \in  G = {\cal G}_{M+\frac{1}{2}} := \{ t \in {\mathbb N}_0 \ | \ t \ {\rm mod} \ (2 M + 2) \in {\cal A}^{(3)}_{M+\frac{1}{2}} \}, \nl
&& {\cal A}^{(3)}_{M+\frac{1}{2}} := \{0, \pm (M + \tfrac{1}{2}) \} \subset {\cal M}, \qquad (M \in {\mathbb N} + \tfrac{1}{2}) 
\label{eq:s_cond_nonzero_halfM}
\ee
where $\bmod (2M+2)$ is defined to project onto  ${\cal M}$.
Thus, one can find the selection rule as
\be
   {\cal S}_1 = {\mathbb N}_0/{\cal G}_{M+\frac{1}{2}} \cong {\mathbb Z}_{2M+2}/{\cal A}^{(3)}_{M+\frac{1}{2}} \ \ \text{for} \ \ M \in {\mathbb N} + \frac{1}{2}.
\ee
Here, we define a \textit{formal} quotient set ${\mathbb Z}_{2M+2}/{\cal A}^{(3)}_{M+\frac{1}{2}}$, interpreted as a partition of symmetry sectors under the ${\mathbb Z}_{2M+2}$ Symanzik rotation, modulo identification by the subset ${\cal A}^{(3)}_{M+\frac{1}{2}}$\footnote{Equivalently, we introduce an equivalence relation on $\mathbb{N}_0$ by $n \sim n'$ iff $(n-n') \bmod (2M+2) \in {\cal A}^{(3)}_{M+\frac{1}{2}}$, and take the set of equivalence classes. This plays the role analogous to the subgroup quotient in the $M \in {\mathbb N} + 1$ case.}.
Unlike the case for $M \in \mathbb{N} + 1$, the set ${\mathbb Z}_{2M+2}/{\cal A}^{(3)}_{M+\frac{1}{2}}$ does not carry a group structure, but rather constitutes a quotient set defined with respect to the selection subset ${\cal A}^{(3)}_{M+\frac{1}{2}}$.
Moreover, this selection rule does \textit{not} coincide with the shift symmetry of $\Phi^{(1)}_{2M  +2}$ in Eq.\eqref{eq:Phi_M+1_halfM}.
For example, the ESRs for $M = \frac{3}{2}$ are written down as (cf. Refs.~\cite{Watkins:2011gx,Voros:2022yos})
\subbe
\be
&& (\sqrt{5}-1) \zeta_{{\cal PT}}(1)^2 - (5 + 3\sqrt{5}) \zeta_{{\cal PT}}(2) = 0, \\
&& \frac{3}{4} \zeta_{{\cal PT}}(1)^3 - (6 + 2 \sqrt{5}) \zeta_{{\cal PT}}(1) \zeta_{{\cal PT}}(2) - \frac{10 + 6 \sqrt{5} }{3} \zeta_{{\cal PT}}(3) = 0, \\
&& \frac{5 + \sqrt{5}}{60} \zeta_{{\cal PT}}(1)^5 - \frac{1 + \sqrt{5}}{2} \zeta_{{\cal PT}}(1)^3 \zeta_{{\cal PT}}(2)  + \frac{\sqrt{5} - 1}{3} \zeta_{{\cal PT}}(1)^2 \zeta_{{\cal PT}}(3) + \frac{9 + 5 \sqrt{5}}{4} \zeta_{{\cal PT}}(1) \zeta_{{\cal PT}}(2)^2 \nl
&& \quad - \frac{1 + \sqrt{5}}{2} \zeta_{{\cal PT}}(1) \zeta_{{\cal PT}}(4) 
+( 3 + \sqrt{5} ) \zeta_{{\cal PT}}(2) \zeta_{{\cal PT}}(3) - \frac{20 + 8 \sqrt{5}}{5} \zeta_{{\cal PT}}(5) = 0, \\
&& \frac{3 - \sqrt{5}}{280} \zeta_{\cal PT}(1)^7 - \frac{1 + \sqrt{5}}{40} \zeta_{\cal PT}(1)^5 \zeta_{\cal PT}(2) + \frac{5 + \sqrt{5}}{8} \zeta_{\cal PT}(1)^3 \zeta_{\cal PT}(2)^2 - \frac{3 - \sqrt{5}}{4} \zeta_{\cal PT}(1)^3 \zeta_{\cal PT}(4) \nl
&& \quad + (1 + \sqrt{5}) \zeta_{\cal PT}(1)^2 \zeta_{\cal PT}(2) \zeta_{\cal PT}(3) - \frac{3 + 9 \sqrt{5}}{5} \zeta_{\cal PT}(1)^2 \zeta_{\cal PT}(5) -\frac{3 + 3 \sqrt{5}}{4} \zeta_{\cal PT}(1) \zeta_{\cal PT}(2) \zeta_{\cal PT}(4) \nl
&& \quad - \frac{13 + 5 \sqrt{5}}{8} \zeta_{\cal PT}(1) \zeta_{\cal PT}(2)^3 
+ ( 3 + \sqrt{5} ) \zeta_{\cal PT}(1) \zeta_{\cal PT}(3)^2 + (\sqrt{5} - 1) \zeta_{\cal PT}(1) \zeta_{\cal PT}(6) \nl
&& \quad - (4 + 2 \sqrt{5}) \zeta_{\cal PT}(2)^2 \zeta_{\cal PT}(3) + \frac{57 + 27 \sqrt{5}}{5} \zeta_{\cal PT}(2) \zeta_{\cal PT}(5) - (3 + \sqrt{5}) \zeta_{\cal PT}(3) \zeta_{\cal PT}(4) \nl
&& \quad - \frac{30 + 18 \sqrt{5}}{7} \zeta_{\cal PT}(7) = 0,  \\
\vdots. && \nn
\ee
\subee

The ZGFs, $\bm{\zeta}_n = \bm{\zeta}_n (\bm{\zeta}_1)$, is easily constructable by using Eq.\eqref{eq:Cn_prods_C1_simp_zeta2}.
Unlike the case in Eqs.\eqref{eq:s_vanish}\eqref{eq:zero_omega_mult}, the linear part of $\zeta_1(s)$ in the ZGFs $\bm{\zeta}_{n} = \bm{\zeta}_{n}(\bm{\zeta}_{1})$ does not become zero for any $n \in {\cal K}$.
Hence, $\bm{\zeta}_{n} = \bm{\zeta}_{n}(\bm{\zeta}_{n^\prime })$ for $n, n^\prime \in {\cal K}$ are all invertible, i.e., ${\cal S}_n$ are the same for all $n \in {\cal K}$:
\be
{\cal S}_n = {\mathbb N}_{0}/{\cal G}_{M + \frac{1}{2}} \cong {\mathbb Z}_{2M+2}/{\cal A}^{(3)}_{M+\frac{1}{2}} \ \ \text{for} \ \ n \in {\cal K}, \ M \in {\mathbb N} + \frac{1}{2}.
\ee
For example, the ZGF for $M = \frac{5}{2}$, $\bm{\zeta}_{2} = \bm{\zeta}_{2}(\bm{\zeta}_{1})$, can be written down as
\subbe
\be
\zeta_{{\cal PT}_2}(1) &=& \frac{\csc^2 \frac{\pi }{14}}{4 (1 + 2 \cos \frac{\pi}{7} )} \zeta_{{\cal PT}_1}(1), \\
\zeta_{{\cal PT}_2}(2) &=& \zeta_{{\cal PT}_1}(1)^2 -\frac{\csc^4 \frac{\pi}{14} }{8 \left(1+2 \cos \frac{\pi}{7}\right)^4} \left( 1 - \sin \frac{\pi }{14}  \right) \zeta_{{\cal PT}_1}(2), \\
\zeta_{{\cal PT}_2}(3) &=&
\frac{\csc ^8\frac{\pi}{14}}{256 \left(1 + 2 \cos \frac{\pi}{7}\right)^6} \left[  \left( -5 + 26 \sin \frac{\pi}{14} + 61 \cos \frac{\pi}{7} - 88 \sin \frac{3 \pi}{14} \right) \zeta_{{\cal PT}_1}(1)^3 \right. \nl
  && \left. - 3  \left( - 5 + 14 \sin \frac{\pi}{14} + 23 \cos \frac{\pi}{7}  - 30 \sin \frac{3 \pi}{14} - 5 \right) \zeta_{{\cal PT}_1}(1) \zeta_{{\cal PT}_1}(2) \right. \nl
  && \left. -2 \left(4 - 8 \sin \frac{\pi}{14} -9 \cos \frac{\pi}{7} + 11 \sin \frac{3 \pi}{14} \right) \zeta_{{\cal PT}_1}(3) \right], \\
\vdots. && \nn
\ee
\subee
The structure for ${M} \in {\mathbb N} + \frac{1}{2}$ can be schematically summarized as
\be
\begin{aligned}
& \text{For $M \in {\mathbb N} + \frac{1}{2}$: } &\quad&  {\large
\begin{tikzcd}
  \bm{\zeta}_{n \in {\cal K}} \arrow[leftrightarrow]{rrr}{{\cal S}_n \cong
    {\mathbb Z}_{2M+2}/{\cal A}^{(3)}_{M + \frac{1}{2}}
  }[swap]{\bm{\zeta}_{n} = \bm{\zeta}_{n}(\bm{\zeta}_{n^\prime})} & & & \bm{\zeta}_{n^{\prime} \in {\cal K}}
\end{tikzcd}}
\end{aligned} \label{eq:sch_fig_Mhalf}
\ee
\\ \, \\ \indent
Finally, we comment on the difference from $M \in {\mathbb N} + 1$ in our results.
Since we actually try to characterize the structure of the ESRs and the ZGFs by the selection rule ${\cal S}_n$, we focus on this quantity.
The remarkable point is that it has ${\mathbb Z}_N$ group and coincides with the shift symmetry in $\Phi^{(1,2)}_{2M + 2}$ for $M = {\mathbb N} + 1$ but does not for $M \in {\mathbb N} + \frac{1}{2}$.
This difference can be intuitively explained from symmetry of the ${\cal PT}$-symmetric operator $\widehat{\cal L}_{\cal PT}$ and domains $x_{\cal PT}$ (or states) in Eqs.\eqref{eq:L_PT}\eqref{eq:dom_xPT}.
For $M \in {\mathbb N }+ 1$, the operator $\widehat{\cal L}_{\cal PT}$ is invariant under ${\cal P}$ or ${\cal T}$ but the domains are not closed under the transform, which implies that there are copies in the upper half-plane of the complex $x$-plane, being mapped from the lower half-plane.
Due to the copied domains, taking analytic continuation going around the origin contains doubly counted $C^{(n)}(\omega^k E)$, and $\Phi^{(1)}_{2M+2}$ consequently consists of $\{[-M]_1,[-M+2]_1,\cdots, [M-2]_1,[M]_{1}\}$ by factorization of $[M-1]_{2M}-1$ as Eq.\eqref{eq:gen_iden_PT_pre}. 
The resulting $\Phi^{(1)}_{2M+2}$ is compatible with the modulo identification by $2M + 2$ and keeps the ${\mathbb Z}_N$ symmetric structure which coincides with the shift symmetry, $[k_1,\cdots,k_\ell]_1 \mapsto [k_1+c,\cdots, k_\ell + c]_1$, in the subgroup of the ${\mathbb Z}_{2M+2}$ Symanzik rotation.
This structure is directly transferred to the selection rule ${\cal S}_{n \in {\cal K}}$ by performing Eq.\eqref{eq:Cn_prods_C1_simp_zeta2}.
In contrast, for $M \in {\mathbb N} + \frac{1}{2}$, because of no copied domain, the factorization of $[M-1]_1 -1$ does not work, and the modulo identification enters into $\Phi^{(1)}_{2M + 2}$ in a non-trivial way breaking the shift symmetry.
In addition, in the process to find the ESRs using Eq.\eqref{eq:Cn_prods_C1_simp_zeta2}, the cosine structure of $e^{- 2k \zeta^\prime_1(0)}$ in Eq.\eqref{eq:zeta1_omega}, which is a consequence of the structure of the Chebyshev polynomials, non-trivially combine with the ${\mathbb Z}_{2M + 2}$ Symanzik rotation in $\Omega({\bf a},s)$.
As a result, the selection rules are given by its combined structure as the quotient set ${\mathbb Z}_{2M + 2}/{\cal A}^{(3)}_{M+\frac{1}{2}}$.
This structure does not change through the ZGFs, and thus, ${\cal S}_{n \in {\cal K}}$ are all the same.

\section{Application: the Ai-Bender-Sarkar conjecture for massless QM} \label{sec:ABS}

We show the simple application to the Ai-Bender-Sarkar (ABS) conjecture~\cite{Ai:2022csx} for the massless QM, which corresponds to $M=K + \varepsilon/2 = 2$ in Eq.\eqref{eq:L_PT}.
The ABS conjecture was proposed as a relationship between the $D \ge 1$ dimensional Euclidean partition function of the ${\cal PT}$-symmetric theory and the analytic continuation of the Hermitian (AC) theory, but Refs.~\cite{Lawrence:2023woz,Kamata:2023opn,Chen:2024ynx} reported that the ABS conjecture does not hold in the massless QM\footnote{
In the massive case, Ref.~\cite{Kamata:2023opn} pointed out that the ABS conjecture is broken in the second non-perturbative and higher orders and reformulated the conjecture using the exact WKB.
}.
The partition functions of AC and Hermitian QMs can be easily related to each other by a simple phase rotation, and thus, the problem reduces to constructing a relation between the ${\cal PT}$- and Hermitian partition functions.
Our framework circumvents these analytic difficulties entirely. 
By mapping the problem into constructing a relation of SZFs, the ZGFs provide a direct \textit{algebraic shortcut} that relates the two sectors without requiring explicit tracking of analytic continuation paths.
This demonstrates the robustness of the ZGF approach, particularly in the massless regime where traditional analytic methods become cumbersome.

Let us illustrate the application to the ABS conjecture for the massless QM using the ZGF for $M=2$. 
In order to show the application to the ABS conjecture, we start with the partition function, which is defined as
\be
Z_n(\beta) := {\rm Tr}_n [e^{-\beta \widehat{H}}] = \sum_{\alpha \in {\mathbb N}_0} e^{- \beta E_{n,\alpha}}, \qquad (\beta \in {\mathbb R}_{>0})
\ee
where we assumed that energy-degeneracies do not exist in our setup, $n$ is the fusion label which admits to take $n=1$ and $2$ (${\cal PT}$ and ${\cal H}$), and ${\rm Tr}_n$ denotes trace over the Hilbert space of the $``{\cal C}"{\cal PT}$ or Hermitian state \cite{Mostafazadeh:2003gz,Bender:2004zz}.
The partition function does not directly relates to the SZFs but does by using the Harglotz function defined through the Laplace integral as
\be
 \Psi_{n} (t) &:=& \int_0^{+\infty} e^{\beta t}  Z_n (\beta) d \beta = \int_{0}^{+\infty} \frac{d {\cal N}_n(E)}{E-t},
\ee
where $d{\cal N}_n(E)$ is a measure defined as
\be
d{\cal N}_n(E) := \sum_{\alpha \in {\mathbb N}_0} \delta(E-E_{n,\alpha}) dE.
\ee
The Harglotz function can be exprressed by the SZFs as
\be
\Psi_n(t) &=& \sum_{\alpha \in {\mathbb N}_0} \frac{1
}{E_{n,\alpha} - t} =  \sum_{s \in {\mathbb N}} \zeta_n (s) t^{s-1}, \qquad \zeta_n(s) = \int_{0}^{+\infty} \, E^{-s} d {\cal N}_n(E),
\ee
where the partition function can be reproduced from the boundary value on the positive real axis for the Harglotz function as
\be
&& Z_n(\beta) = - \frac{1}{\pi} {\rm Im} \int_{0}^{+\infty} dE \, e^{-\beta E} \Psi_{n}(E+i 0_+), \\
&& {\rm Im}[\Psi_{n}(E+i 0_+)] = -\pi \sum_{\alpha \in {\mathbb N}_0} \delta(E-E_{n,\alpha}).
\ee
By these transformations, the partition function has one-to-one correspondence to the SZFs via Harglotz function.
In addition to this fact, the ZGFs directly bridges the ${\cal PT}$ and Hermitian SZFs, and the ${\rm AC}$ partition function can be easily found by the complex phase rotation of $\beta$ from the Hermitian one.
Therefore, the ZGFs provide an indirect route for the modified massless ABS conjecture which relates the ${\cal PT}$ and AC QMs.
The structure can be schematically depicted as
\be
\large{
\begin{tikzcd}
  Z_{\cal PT}(\beta) \ar[leftrightarrow,rrr,dashed] \ar[shift left=0.5ex]{dd}{\int e^{\beta t} d\beta} &&& Z_{\cal H}(\beta) \ar[leftrightarrow, rrr, "\beta \mapsto \omega^{\pm 1} \beta"] \ar[shift left=0.5ex]{dd}{} &&& \ar[shift left=0.5ex]{dd}{} \ar[bend right=15,swap,leftrightarrow]{llllll}{\fbox{\text{ABS conj.}}} Z_{{\rm AC}_\pm}(\beta) \\ \\
  \Psi_{\cal PT}(t) \ar[leftrightarrow,rrr,dashed] \ar[shift left=0.5ex]{uu}{\substack{- \frac{1}{\pi}{\rm Im}\int e^{-\beta t} d E}} \ar[shift left=0.5ex]{dd}{\substack{\text{expand as}  \\ t\rightarrow 0_+}} &&& \Psi_{\cal H}(t) \ar[leftrightarrow]{rrr}{\Psi_{\cal H}(t) \mapsto \omega^{\mp 1} \Psi_{\cal H}(\omega^{\mp 1} t)} \ar[shift left=0.5ex]{uu}{} \ar[shift left=0.5ex]{dd}{} &&& \Psi_{{\rm AC}_\pm}(t) \ar[shift left=0.5ex]{uu}{} \ar[shift left=0.5ex]{dd}{} \\ \\
  \{ {\zeta}_{\cal PT}(s) \}_s \ar[shift left=0.5ex]{uu}{\sum_{s} t^{s-1}} \ar[leftrightarrow]{rrr}{\text{\normalsize ZGF}} &&& \{ {\zeta}_{\cal H}(s) \}_s \ar[shift left=0.5ex]{uu}{} \ar[leftrightarrow]{rrr}{\zeta_{\cal H}(s) \mapsto \omega^{\mp s} \zeta_{\cal H}(s)} &&& \{ {\zeta}_{{\rm AC}_\pm}(s) \}_s \ar[ shift left=0.5ex]{uu}{}
\end{tikzcd}
} \label{eq:ABS_comm}
\ee
where the solid arrows denote the mappings explicitly constructed in our framework, while the dashed arrows represent the mappings which are circumvented by our algebraic approach.

The discussion can be immediately extended to larger $M \in {\mathbb N} + 1$.
Since the essential point of this method is a mapping between $\bm{\zeta}_{{\cal PT}_{K}}$ and $\bm{\zeta}_{\cal H}$ instead of directly mapping between the partition functions, one can easily derive the similar relations to Eq.\eqref{eq:ABS_comm} when the invertibility holds.
Notice that, as we discussed in Sec.~\ref{sec:general_cases}, the ZGF from the Hermitian SZFs to those of ${\cal PT}_K$ is constructable for any $M \in {\mathbb N} + 1$, but the inverse process is not always possible when $M \in 2{\mathbb N} + 1$ depending on $K$.

\section{Summary and outlook} \label{sec:summary}
In this work we have investigated the spectral structure of
one-dimensional quantum mechanics with ${\cal PT}$-symmetric and Hermitian potentials defined as
\[
  V_{{\cal PT}}(x) = x^{2K}(ix)^{\varepsilon} 
  \ \  \text{with} \ \ K,\varepsilon \in \mathbb{N},\qquad
  V_{{\cal H}}(x) = x^{2M} \ \ \text{with}\ \  M \in \mathbb{N} + 1.
\]
For each \textit{fusion} label $n$ we have introduced spectral zeta functions (SZFs)
$\zeta_{n}(s) = \sum_{\alpha \in \mathbb{N}_{0}} E_{n,\alpha}^{-s}$ and
constructed two complementary structures:
exact sum rules (ESRs), which relate zeta values at fixed $n$, and zeta generating formulas (ZGFs), which map spectral data
between different fusion labels.
The underlying guidance comes from the fusion relations of the
$A_{2M-1}$ T-system in the ODE/IM correspondence, with
$M=K+\varepsilon/2$, but our final formulations are entirely spectral
in nature.

Our main results can be summarized as follows:
\begin{itemize}
\item \textbf{ESRs at fixed fusion label.}
The ESRs give algebraic relations among SZFs
$\zeta_{n}(s)$ at integer arguments for each fixed fusion label $n$
(which coincides with the ${\cal PT}$ sector label).
At specific values of $s$, the quantity $\zeta_{n}(s)$ can be written
as a polynomial in lower-order zeta values $\zeta_{n}(s')$ with
$s' < s$.
For $M=2$ in the Hermitian case this reproduces the known ESR obtained
via Borel resummation in exact WKB analysis, while our construction
extends such identities systematically to ${\cal PT}$-symmetric sectors and to
general half-integer $M$.

\item \textbf{ZGFs as inter-sector mappings.}
The ZGFs provide explicit, non-recursive formulas expressing
$\zeta_{n}(s)$ in terms of zeta values in a fixed \textit{source} sector
$n'$.
Thus they implement a direct transfer of spectral information between
fusion labels, and make precise to what extent the spectrum in one
sector determines the spectrum in another.

\item \textbf{Selection rules and information hierarchy.}
The ESRs and ZGFs are controlled by a selection rule ${\cal S}_{n}$, which
specifies the set of integer $s$ yielding non-trivial ESRs and encodes
structural properties of the ZGFs such as accessibility and
invertibility.
Algebraically, ${\cal S}_{n}$ can be described in terms of quotient sets
generated by the Chebyshev structure of the fusion relations together
with the $\mathbb{Z}_{2M+2}$ Symanzik rotational symmetry.
For odd $M$ we have found a structural, purely algebraic information-loss phenomenon: when mapping from \textit{odd} fusion sectors to \textit{even} ones,
certain coefficients in the ZGFs vanish identically due to exact
destructive interference of the Chebyshev phases, and the map cannot be
inverted.
In other words, even-sector spectral data carry strictly less
information than odd-sector data, yielding a concrete \textit{no-go}
statement for inverse spectral reconstruction from restricted sectors.
For even and half-integer $M$ the situation is different: all sectors in the relevant set are mutually invertible, and the selection rule organizes them into
an information-equivalent family.

\item \textbf{Application to the Ai--Bender--Sarkar conjecture.}
As a concrete test, we reformulated the massless Ai--Bender--Sarkar
conjecture using the ZGFs.
Rather than relating ${\cal PT}$-symmetric and Hermitian partition functions
directly, we connect them indirectly via Laplace transforms and Taylor
expansions of the SZFs.
This gives a purely spectral-theoretic route to the conjectured
relation between ${\cal PT}$-symmetric and Hermitian spectra.
\end{itemize}

Conceptually, the present work shifts emphasis within the ODE/IM
correspondence from quantization conditions towards spectral data:
fusion hierarchies on the integrable-model side acquire a spectral-zeta
interpretation on the ODE side.
Crucially, this interpretation reveals that the algebraic structure of
the fusion relations imposes strict hierarchical constraints, distinct
from analytic ambiguities, on the recoverability of spectral
information from restricted sector data.
The ESRs match non-perturbative identities familiar from exact WKB analysis (where they arise via Borel resummation), while the ZGFs capture how spectral data are redistributed across sectors in a way reminiscent of analytic continuation between Stokes sectors.
These connections suggest a structural bridge between the spectral zeta framework, the ODE/IM correspondence, and resurgence analysis.
\\ \, \\ \indent
Several directions for future work are natural.
On the mathematical side, it would be interesting to extend the present
construction to higher-rank T-/Y-systems and to clarify more
systematically how selection rules and information-loss phenomena are
encoded in the representation theory of the underlying integrable
structures.
On the analytical side, one may seek a more explicit link between the
ESR/ZGF framework and exact WKB quantities such as Voros symbols and
Stokes automorphisms, in particular for potentials with poles or branch
cuts, where modified quantization conditions are required.
More broadly, we hope that thinking in terms of spectral zeta data will provide a useful perspective for organizing non-perturbative information in ${\cal PT}$-symmetric and Hermitian quantum mechanics, and for exploring how notions of integrability manifest directly at the level of spectral functions.

\acknowledgments
We would like to express sincere gratitude to the organizers of the international workshop \textit{``Non-Hermitian Physics: from Classical to Quantum and Beyond''} (London, 2024), which provided a stimulating environment and many fruitful discussions that greatly influenced the development of this work.
In particular, we are grateful to Sarben Sarkar and Wen-Yuan Ai for their warm hospitality and insightful conversations during the workshop.
We are particularly thankful to Paul~Romatschke for suggesting questions and discussions that motivated this paper.
We thank to Hongfei~Shu for helpful discussions about the ODE/IM correspondence.
S.~K is supported by JSPS KAKENHI Grant Nos.~22H05118 and 25K07298.

\appendix


\section{Review of the fusion relations and the spectral zeta functions} \label{app:review}

This appendix outlines the foundational elements necessary for our analysis.
Sec.~\ref{sec:ODE_IM} provides a brief overview of the fusion relations associated with the QCs derived from the ODE/IM correspondence.
We review the spectral zeta functions (SZFs) in Sec.~\ref{sec:SZF} and mention the technical assumptions through this paper in Sec.~\ref{app:tec_assum}.
We refer the reader to, e.g., Refs.~\cite{Dorey:2007zx,ito2025ode,Voros:1986vw,voros1992spectral} for further background on the ODE/IM correspondence and the SZFs.

\subsection{The fusion relations of the Stokes multipliers} \label{sec:ODE_IM}

In this appendix we recall the derivation of the fusion relations for the Stokes multipliers
$C^{(n)}(E)$ and the definition of the SZFs. These are the building blocks
used in Sec.~\ref{sec:preparation} for the construction of the ESRs and ZGFs.
In our analysis, we practically need the fusion relations only and do not use the ODE/IM correspondence itself.
In this part, we briefly review the derivation of the fusion relations in the ODE side.
See Refs.~\cite{Dorey:2007zx} in detail, for example.

Suppose a solution of the Schr\"{o}dinger equation \eqref{eq:L_Herm} on $S_0$, denoted by $y(x,E)$, which asymptotically behaves as
\be
y(x,E) \sim \frac{x^{-M/2}}{ \sqrt{2 i}} \exp \left[ - \frac{x^{M+1}}{M + 1}\right], \qquad |x| \rightarrow \infty \ \ \text{with} \ \ | \arg x | < 2 \mu.  \label{eq:yxE_asym}
\ee
The following wavefunctions are also solutions of the same equation:
\be
y_{k}(x,E) := w^{k/2} y(\omega^{-k} x, \omega^{2k} E), \qquad  (k \in {\mathbb Z}) \label{eq:yk_bases}
\ee
where $\omega$ is defined as
\be
\omega := e^{\frac{\pi i}{M+1}} = e^{2 i \mu}. \label{eq:def_omega}
\ee
These solutions are well-defined on each Stokes sector, $S_{k \in \{ 0,\cdots,2M+1 \}}$, and we use them as bases for analytic continuations.
It is remarkable that our problem \eqref{eq:LPTtoLH} has the ${\mathbb Z}_{2M+2}$ rotational symmetry, so called as the Symanzik rotation, that can be seen from Eq.\eqref{eq:yk_bases}. 
In the ODE/IM correspondence, it is common to use the Wronskian for analytic continuations on the complex $x$-plane, which is defined as
\be
&& W_{k_1,k_2}(E) := y_{k_1}(x,E) \pd_x y_{k_2}(x,E) - y_{k_2}(x,E) \pd_x y_{k_1}(x,E).
\ee
The Wronskian has the properties as
\be
&& W_{k_1,k_2}(E) = - W_{k_2,k_1}(E), \qquad W_{k_1+\ell,k_2+\ell}(E) = W_{k_1,k_2}(\omega^{2 \ell} E). \label{eq:W_iden}
\ee
Then, we perform the analytic continuation across the Stokes sectors from $S_{-1}$ to $S_{+1}$ and express the wavefunction using the bases in Eq.\eqref{eq:yk_bases}.
It can be generally written as
\be
y_{-1} (x,E) = C^{(1)}(E) y_0(x,E) + \widetilde{C}^{(1)}(E) y_1(x,E).
\ee
These coefficients, $C^{(1)}(E)$ and $\widetilde{C}^{(1)}(E)$, are called as \textit{Stokes multipliers} and can be written by the Wronskian as
\be
C^{(1)}(E) = \frac{W_{-1,1} (E)}{W_{0,1}(E)}, \qquad \widetilde{C}^{(1)}(E) = - \frac{W_{-1,0}(E)}{W_{0,1}(E)} = -1.
\ee
The fusion relations can be derived by repeating the similar procedure.
Define
\be
C^{(n)}_k(E) := W_{k-1,k+n}(E), \qquad \widetilde{C}^{(n)}_k(E) := - W_{k-1,k+n-1}(E), \label{eq:def_Cnk_tilCnk}
\ee
and consider the two analytic continuations, $S_{k-1} \rightarrow S_{k+n}$ and $S_{k} \rightarrow S_{k+n}$.
These can be expressed by the matrix form as
\be
\begin{pmatrix}
  y_{k-1} \\
  y_{k}
\end{pmatrix}
= {\bf C}^{(n)}_k
\begin{pmatrix}
  y_{k+n-1} \\
  y_{k+n}
\end{pmatrix}, \qquad
{\bf C}^{(n)}_k :=
\begin{pmatrix}
C^{(n)}_k &  \widetilde{C}^{(n)}_k \\
C^{(n-1)}_{k+1} &  \widetilde{C}^{(n-1)}_{k+1} 
\end{pmatrix},
\ee
where
\be
{\bf C}^{(0)}_k = {\mathbb I}_2, \qquad
{\bf C}^{(1)}_k =
\begin{pmatrix}
C^{(1)}_k & -1 \\
1 & 0
\end{pmatrix}. \label{eq:ini_Cmat}
\ee
Since the analytic continuation taking $S_{k(-1)} \rightarrow S_{k+n(-1)} \rightarrow S_{k+n+r(-1)}$ should be the same to $S_{k(-1)} \rightarrow S_{k+n+r(-1)}$, the product of ${\bf C}^{(n)}_{k}$ and ${\bf C}_{k+n}^{(r)}$ can be expressed using another matrix ${\bf C}_{k}^{(n+r)}$ as
\be
   {\bf C}_{k}^{(n)} {\bf C}_{k+n}^{(r)} = {\bf C}_{k}^{(n+r)}. \label{eq:CC_C}
\ee
For Eq.\eqref{eq:CC_C}, we individually consider (i) $n=1$ and (ii) $r=-n$.
From Eqs.\eqref{eq:W_iden}\eqref{eq:ini_Cmat} and fixing $k=0$, one can obtain the fusion relations as
\subbe
\be
\text{(i)} &:& C^{(1)}(E)  C^{(n)}(\omega^{n+1} E) = C^{(n-1)}(\omega^{n+2} E) + C^{(n+1)}(\omega^{n} E), \label{eq:fusion_1-app} \\
\text{(ii)} &:& C^{(n)}(\omega^{-1} E)  C^{(n)}(\omega E) = 1 + C^{(n-1)}(E) C^{(n+1)}(E), \label{eq:fusion_2-app}
\ee
where $\widetilde{C}_{k}^{(n)} = -C^{(n-1)}_k$, and 
\be
C^{(-1)}(E) = 0, \qquad C^{(0)}(E) = 1, \qquad C^{(n \in {\mathbb N})}(E) = W_{-1,n}(\omega^{-n+1} E).
\ee
Here, we set $C^{(n)}(E):= C_0^{(n)}(\omega^{-n+1} E)$.
Due to the asymptotic behaviors of the wavefunction which are $y_{k+2M+2}(x,E) = - y_{k}(x,E)$ in Eqs.\eqref{eq:yxE_asym}\eqref{eq:yk_bases} and the ${\mathbb Z}_{2M+2}$ Symanzik rotation as $\omega^{2M+2} = 1$, one can find that $C^{(2M+1)}(E) = 0$ and $C^{(0)}(E) = C^{(2M)}(E) = 1$, and thus, the fusion relations \eqref{eq:fusion_1}\eqref{eq:fusion_2} are truncated up to $n=2M-1$.
In addition, from the definition of $C^{(n)}_k(E)$ in Eq.\eqref{eq:def_Cnk_tilCnk}, one can directly show the symmetry that
\be
C^{(n)}(E) = C^{(2M - n)}(E). \label{eq:Cn_C2Mn-app}
\ee
\subee
This type of the fusion relations obtained is known as the $A_{2M-1}$ T-system\cite{reshetikhin1987spectrum,Kuniba:1993,Bazhanov:1994ft,Ito:2015nla}.
The resulting $C^{(n)}(E)$ directly correspond to QCs and spectral determinants in the QMs \eqref{eq:LPTtoLH}.
This fact can be seen by reminding Eq.\eqref{eq:CC_C} with $k=0$, which is explicitly written as
\be
y_{-1}(x,E) &=& C^{(n)}(\omega^{n-1} E) y_{n-1}(x,E) - C^{(n-1)}(\omega^{n-2}E) y_{n}(x,E).
\ee
Similar to Eq.\eqref{eq:yxE_asym}, $y_k(x,E)$ converges to zero as $|x| \rightarrow \infty$ with $| \arg \omega^{-k} x | < 2 \mu$.
Hence, if one considers the analytic continuation, $S_{-1} \rightarrow S_{n}$, then the condition that $C^{(n)}(\omega^{n-1} E) = 0$ has to be imposed to make the analytic continued wavefunction, $y_{-1}(x,E)$, to be convergent at $|x| = \infty$ on $S_{n}$.
(See also Ref.~\cite{Dorey:2007zx}, for example.)
From the analytic continuations in Eq.\eqref{eq:PT_Stokes_Kod}, the ${\cal PT}_K$ and Hermitian QCs can be obtained as
\subbe
\be
{\cal PT}_{K< \lfloor M + \frac{1}{2}\rfloor} &:& W_{- \lfloor \frac{K+2}{2} \rfloor, \lfloor \frac{K+1}{2} \rfloor }(E) \mid_{E = e^{2 i \theta^*} E_{\cal PT}} = C^{(K)}(\omega^{(K+1) \bmod 2} E) \mid_{E = e^{2 i \theta^*} E_{\cal PT}} = C^{(K)}(- E_{\cal PT}) = 0, \label{eq:QC_PT-app} \\
 \text{Hermitian} &:& W_{0,M+1}(E) \mid_{E=E_{\cal H}} = C^{(M)}(- E_{\cal H}) = 0, \label{eq:QC_H-app}
\ee
\subee
where we used Eq.\eqref{eq:E_PT_H}.
Solving these conditions with respect to $E_{{\cal PT},{\cal H}}$ yields the energy spectra.

We make a brief comment on Borel resummation.
From the perspective of exact WKB analysis underlying Borel resummation theory, methodology for finding these QCs is the same.
When one considers a bounded potential, its QC is determined from a component in the monodromy matrix by imposing normalizability to the wavefunctions at $|x| = \infty$.
Those QCs consist of cycle integrations going around two turning points, that are called as \textit{Voros symbols (multipliers)}.
Hence, the Stokes multiplier $C^{(n)}(E)$ corresponds to the QC constructed by (a combination of) the Voros multipliers in exact WKB analysis~\cite{Voros1983,Kamata:2024tyb}.

\subsection{The spectral zeta functions} \label{sec:SZF}
In this part, we introduce the spectral zeta functions (SZFs) \cite{Voros:1986vw,voros1992spectral}.
We start with a (normalized) QC, denoted by ${\frak D}(E)$.
Suppose that it can be expressed by a spectral determinant as
\be
   {\frak D}(E) = \prod_{\alpha \in {\mathbb N}_0} (E_\alpha - E) = \exp \left[ \sum_{\alpha \in {\mathbb N}_0} \log (E_\alpha - E) \right], \label{eq:QC_DE}
\ee
with an energy spectrum labeled by $\alpha$, $\{E_\alpha\}_{\alpha \in {\mathbb N}_0}$, where $E_\alpha \in {\mathbb R}_{\ge 0}$, $E \in {\mathbb R}\setminus \{E_\alpha \}_{\alpha \in {\mathbb N}_0}$.
Trivially, solving ${\frak D}(E) = 0$ gives $E=E_\alpha$.
We rewrite Eq.\eqref{eq:QC_DE} using the Hurwitz-type SZF defined as
\be
&& \zeta(s,E) := \sum_{\alpha \in {\mathbb N}_0} (E_\alpha - E)^{-s}. \qquad (s \in {\mathbb R})
\label{eq:zeta_s_E}
\ee
Performing Taylor expansion around $E=0$ to the exponential part of Eq.\eqref{eq:QC_DE} leads to
\be
\sum_{\alpha \in {\mathbb N}_0} \log (E_\alpha - E) &=& - \zeta^{\prime}(0) - \sum_{s \in {\mathbb N}} \frac{\zeta(s)}{s}  E^{s},
\ee 
where $\zeta(s)$ and $\zeta^{\prime}(s)$ are the SZFs and its derivative of $s$ defined from Eq.\eqref{eq:zeta_s_E} as
\be
&& \zeta(s) := \zeta(s,0) = \sum_{\alpha \in {\mathbb N}_0}E_\alpha^{-s}, \qquad \zeta^{\prime}(0) := \left. \frac{\pd \zeta(s)}{\pd s} \right|_{s = 0} = - \sum_{\alpha \in {\mathbb N}_0} \log E_\alpha. \label{eq:def_zetas}
\ee
Therefore, using the SZFs $\zeta(s)$ with $s \in {\mathbb N}$ and its derivative $\zeta^\prime(0)$, one can write down the QC \eqref{eq:QC_DE} as
\be
 {\frak D}(E) &=& \exp \left[ - \zeta^{\prime}(0) - \sum_{s \in {\mathbb N}} \frac{\zeta(s)}{s} E^{s} \right] \nl
 &=& e^{-\zeta^{\prime}(0)} + e^{-\zeta^{\prime}(0)} \sum_{m \in {\mathbb N}} \sum_{\substack{{\bf t} \in {\mathbb N}_0^m  \\ |{\bf t}| > 0}}
 (-1)^{|{\bf t}|} \left[ \prod_{s=1}^{m} \frac{\zeta(s)^{t_s}}{t_s ! s^{t_s}} \right] E^{({\bf m}, {\bf t})}, \qquad ({\bf m}, {\bf t}) := \sum_{s = 1}^m s t_s, \label{eq:D_to_zeta}
\ee
where $|{\bf t}| := \sum_{m=1}^{{\rm dim} ({\bf t})} t_m$.
As one can see from Eq.\eqref{eq:D_to_zeta}, the SZFs appear in coefficients of the expanded form of the QC and keep information of the energy spectrum as summation of the inverse energy, which means that searching all the SZFs (and $\zeta^\prime(0)$) is essentially the same to finding the energy spectrum.

In the main text, the SZFs $\zeta_n(s)$ used in Sec.~\ref{sec:preparation} are precisely the quantities entering
the expansion~\ref{eq:D_to_zeta} for the QCs $C^{(n)}(E)$ associated with each ${\cal PT}_K$/H sector.

\bibliographystyle{utphys}
\bibliography{IM_Herm.bib}

\end{document}